\documentclass[fleqn,usenatbib]{mnras}

\usepackage{newtxtext,newtxmath}

\usepackage[T1]{fontenc}

\DeclareRobustCommand{\VAN}[3]{#2}
\let\VANthebibliography\thebibliography
\def\thebibliography{\DeclareRobustCommand{\VAN}[3]{##3}\VANthebibliography}

\usepackage{graphicx}	
\usepackage{amsmath}	

\title[Ly$\alpha$ emission at $z\simeq5-6$]{Ly$\alpha$ emission in galaxies at $z\simeq5-6$: new insight from {\it JWST} into the statistical distributions of Ly$\alpha$ properties at the end of reionization\thanks{Some of the data presented herein were obtained at Keck Observatory, which is a private 501(c)3 non-profit organization operated as a scientific partnership among the California Institute of Technology, the University of California, and the National Aeronautics and Space Administration. The Observatory was made possible by the generous financial support of the W. M. Keck Foundation.}}

\author[M. Tang et al.]{
Mengtao Tang$^{1}$\thanks{E-mail: tangmtasua@arizona.edu},
Daniel P. Stark$^{1}$, 
Richard S. Ellis$^{2}$, 
Fengwu Sun$^{1}$, 
Michael Topping$^{1}$, 
Brant Robertson$^{3}$, \newauthor
Sandro Tacchella$^{4,5}$, 
Santiago Arribas$^{6}$, 
William M. Baker$^{4,5}$, 
Rachana Bhatawdekar$^{7}$, 
Kristan Boyett$^{8,9}$, \newauthor
Andrew J. Bunker$^{10}$, 
St\'{e}phane Charlot$^{11}$, 
Zuyi Chen$^{1}$, 
Jacopo Chevallard$^{10}$, 
Gareth C. Jones$^{10}$, \newauthor
Nimisha Kumari$^{12}$, 
Jianwei Lyu$^{1}$, 
Roberto Maiolino$^{4,5,2}$, 
Michael V. Maseda$^{13}$, 
Aayush Saxena$^{10,2}$, \newauthor
Lily Whitler$^{1}$, 
Christina C. Williams$^{14}$, 
Chris Willott$^{15}$ and 
Joris Witstok$^{4,5}$ 
\\
\\
$^{1}$ Steward Observatory, University of Arizona, 933 N Cherry Ave, Tucson, AZ 85721, USA \\
$^{2}$ Department of Physics and Astronomy, University College London, Gower Street, London WC1E 6BT, UK \\
$^{3}$ Department of Astronomy and Astrophysics, University of California, Santa Cruz, 1156 High Street, Santa Cruz, CA 95064, USA \\
$^{4}$ Kavli Institute for Cosmology, University of Cambridge, Madingley Road, Cambridge, CB3 0HA, UK \\
$^{5}$ Cavendish Laboratory, University of Cambridge, 19 JJ Thomson Avenue, Cambridge, CB3 0HE, UK \\
$^{6}$ Centro de Astrobiolog\'{i}a (CAB), CSIC-INTA, Cra. de Ajalvir Km.~4, 28850- Torrej\'{o}n de Ardoz, Madrid, Spain \\
$^{7}$ European Space Agency (ESA), European Space Astronomy Centre (ESAC), Camino Bajo del Castillo s/n, 28692 Villanueva de la Ca\~{n}ada, Madrid, Spain \\
$^{8}$ School of Physics, University of Melbourne, Parkville 3010, VIC, Australia \\
$^{9}$ ARC Centre of Excellence for All Sky Astrophysics in 3 Dimensions (ASTRO 3D), Australia \\
$^{10}$ Department of Physics, University of Oxford, Denys Wilkinson Building, Keble Road, Oxford OX1 3RH, UK \\
$^{11}$ Sorbonne Universit\'{e}, CNRS, UMR 7095, Institut d'Astrophysique de Paris, 98 bis bd Arago, 75014 Paris, France \\
$^{12}$ AURA for European Space Agency, Space Telescope Science Institute, 3700 San Martin Drive. Baltimore, MD, 21210 \\
$^{13}$ Department of Astronomy, University of Wisconsin-Madison, 475 N. Charter St., Madison, WI 53706 USA \\
$^{14}$ NSF's National Optical-Infrared Astronomy Research Laboratory, 950 North Cherry Avenue, Tucson, AZ 85719, USA \\
$^{15}$ NRC Herzberg, 5071 West Saanich Rd, Victoria, BC V9E 2E7, Canada \\
}

\date{Accepted XXX. Received YYY; in original form ZZZ}

\pubyear{2024}

\begin{document}
\label{firstpage}
\pagerange{\pageref{firstpage}--\pageref{lastpage}}
\maketitle

\begin{abstract}
{\it JWST} has recently sparked a new era of Ly$\alpha$ spectroscopy, delivering the first measurements of the Ly$\alpha$ escape fraction and velocity profile in typical galaxies at $z\simeq6-10$. 
These observations offer new prospects for insight into the earliest stages of reionization. 
But to realize this potential, we need robust models of Ly$\alpha$ properties in galaxies at $z\simeq5-6$ when the IGM is mostly ionized. 
Here we use new {\it JWST} observations from the JADES and FRESCO surveys combined with VLT/MUSE and Keck/DEIMOS data to characterize statistical distributions of Ly$\alpha$ velocity offsets, escape fractions, and EWs in $z\simeq5-6$ galaxies. 
We find that galaxies with large Ly$\alpha$ escape fractions ($>0.2$) are common at $z\simeq5-6$, comprising $30$~per~cent of Lyman break selected samples. 
Comparing to literature studies, our census suggests that Ly$\alpha$ becomes more prevalent in the galaxy population toward higher redshift from $z\sim3$ to $z\sim6$, although we find that this evolution slows considerably between $z\sim5$ and $z\sim6$, consistent with modest attenuation from residual H~{\small I} in the mostly ionized IGM at $z\simeq5-6$. 
We find significant evolution in Ly$\alpha$ velocity profiles between $z\simeq2-3$ and $z\simeq5-6$, likely reflecting the influence of resonant scattering from residual  intergalactic H~{\small I} on the escape of Ly$\alpha$ emission near line center.
This effect will make it challenging to use Ly$\alpha$ peak offsets as a probe of Lyman continuum leakage at $z\simeq5-6$. 
We use our $z\simeq5-6$ Ly$\alpha$ distributions to make predictions for typical Ly$\alpha$ properties at $z\gtrsim8$ and discuss implications of a recently-discovered Ly$\alpha$ emitter at $z\simeq8.5$ with a small peak velocity offset ($156$~km~s$^{-1}$).
\end{abstract}





\section{Introduction} \label{sec:introduction}

Understanding the reionization of the hydrogen in the intergalactic medium (IGM) offers key clues to investigating the early history of structure formation. 
Over the past two decades, plenty of observational efforts have been made to constrain the history of reionization (see \citealt{Stark2016,Robertson2022,Fan2023} for reviews). 
Measurements of the electron scattering optical depth faced by cosmic microwave background photons imply a mid-point of reionization at $z\simeq7.7$ \citep{Planck2020}. 
Observations of quasar absorption spectra suggest that the IGM is partially neutral at $z\gtrsim7$ \citep[e.g.,][]{Banados2018,Davies2018,Wang2020,Yang2020a,Greig2022} and indicate that the reionization of the intergalactic hydrogen likely comes to an end fairly late at $z\simeq5.5$ or below \citep[e.g.,][]{Worseck2014,Becker2021,Bosman2022,Zhu2023}. 

Lyman-alpha (Ly$\alpha$, rest-frame $\lambda=1215.67$~\AA) emission from high-redshift galaxies provides another useful probe of the timeline of cosmic reionization (see \citealt{Dijkstra2014} and \citealt{Ouchi2020} for reviews). 
Because the cross section for scattering by neutral hydrogen (H~{\small I}) is large, Ly$\alpha$ photons from galaxies in the reionization era should be strongly suppressed by the damping wings of the partially neutral IGM \citep[e.g.,][]{Miralda-Escude1998,Garel2021}. 
Spectroscopic observations have shown that strong Ly$\alpha$ emitting galaxies (with Ly$\alpha$ equivalent width EW $\gtrsim25$~\AA) become much rarer from $z\simeq6$ to $z\gtrsim7$ \citep[e.g.,][]{Stark2010,Ono2012,Treu2013,Schenker2014,Pentericci2018,Kusakabe2020,Jones2024}. 
Narrowband surveys have also demonstrated that the Ly$\alpha$ luminosity function declines between $z\simeq6$ and $z\simeq7$ \citep[e.g.,][]{Ouchi2010,Santos2016,Ota2017,Konno2018}. 
These observations suggest that the IGM transfers from significantly neutral (with neutral fraction $x_{\rm HI}\gtrsim0.5$) at $z\gtrsim7$ \citep[e.g.,][]{Mason2018a,Hoag2019,Whitler2020,Morales2021,Bolan2022} to highly ionized at $z\simeq5-6$, consistent with the evolution implied from quasar spectra. 

Attention has recently been focusing on the small subset of galaxies with Ly$\alpha$ detections at $z\gtrsim7$ \citep[e.g.,][]{Oesch2015,Zitrin2015,Roberts-Borsani2016,Stark2017,Larson2022}. 
Galaxies situated in ionized bubbles will have their Ly$\alpha$ emission redshifted significantly before encountering neutral hydrogen, greatly reducing the attenuation provided by the IGM.
As such, Ly$\alpha$ emitters (LAEs) at $z\gtrsim7$ are thought to provide signposts of early ionized regions in the mostly neutral IGM \citep[e.g.,][]{Wyithe2005,Furlanetto2006}. 
The larger the ionized bubble, the more that Ly$\alpha$ will be transmitted through the IGM \citep[e.g.,][]{Dijkstra2007,Mason2020}. 
Systematic searches for Ly$\alpha$ over wide ($>1$~deg$^2$) areas have been conducted at $z\gtrsim7$, either spectroscopically following up ultraviolet (UV) continuum selected galaxies \citep[e.g.,][]{Endsley2021b,Endsley2022a,Cooper2023} or via narrowband filters that efficiently pick up line emitters \citep[e.g.,][]{Zheng2017,Itoh2018,Goto2021}. 
The results have uncovered a variety of likely ionized regions \citep[e.g.,][]{Endsley2021b,Jung2022,Cooper2023}. 
Perhaps the most compelling of these surrounds an overdensity of galaxies spanning $11\times15$~arcmin$^2$ in the $1.5$~deg$^2$ Cosmic Evolution Survey (COSMOS) field. 
Spectroscopic follow-up has revealed that $9$ of $10$ $z\simeq7$ galaxies exhibit strong Ly$\alpha$ emission \citep{Endsley2022a}, well above the success rate typically found at these redshifts. 
These results are consistent with expectations for an ionized bubble spanning a radius of $\simeq3$~physical~Mpc (pMpc), carved out by an abundant population of faint galaxies.

{\it JWST} \citep{Gardner2023} has now ushered in a new era of Ly$\alpha$ observations. 
Initial results have revealed Ly$\alpha$ out to $z\simeq11$ \citep{Bunker2023a}, while also confirming the downturn in Ly$\alpha$ emission at $z\gtrsim7$ \citep[e.g.,][]{Chen2024,Jones2024,Nakane2024}. 
Among the most exciting early results have been detections of extremely strong Ly$\alpha$ emission at $z\simeq7-8$ where the IGM is expected to be significantly neutral \citep{Fujimoto2023,Saxena2023,Chen2024}. 
The rest-frame EWs of Ly$\alpha$ in these galaxies ($\simeq200-400$~\AA) are close to the maximum intrinsic values expected for star forming galaxies (see \citealt{Chen2024}), suggesting that these galaxies must have large ($\gtrsim1$~pMpc) ionized sightlines that allow the majority of their Ly$\alpha$ emission to be transmitted through the IGM. 
Recent work has begun to identify large galaxy overdensities in $\gtrsim1$~arcmin$^2$ areas around the strong Ly$\alpha$ emitters \citep[e.g.,][]{Endsley2023b,Chen2024,Whitler2024}, consistent with the source requirements for carving out such a large ionized bubble.

The spectroscopic capabilities of {\it JWST} with the Near Infrared Spectrograph (NIRSpec; \citealt{Jakobsen2022}) or the Near Infrared Camera (NIRCam; \citealt{Rieke2023a}) grisms \citep{Greene2017} have also introduced new methods for characterizing Ly$\alpha$ in early galaxies. 
Rest-frame optical spectra enable detection of hydrogen Balmer lines (i.e., H$\alpha$, H$\beta$) from which the intrinsic Ly$\alpha$ luminosity can be predicted under nominal recombination assumptions. 
Comparison to the observed Ly$\alpha$ flux yields the escape fraction of Ly$\alpha$. 
We expect this quantity to decrease at redshifts where the IGM is significantly neutral (see \citealt{Chen2024}). 
The emission line spectra also constrain the systemic redshift, allowing the velocity profile of Ly$\alpha$ to be computed. 
Typically Ly$\alpha$ emission emerges redshifted from the line center, reflecting transfer through outflowing neutral gas \citep[e.g.,][]{Verhamme2006,Dijkstra2014}. 
As the IGM becomes neutral, the Ly$\alpha$ profiles will be further altered as the H~{\small I} damping wing will preferentially scatter photons near the line center. 
As a result, we should only expect to see Ly$\alpha$ emerging near the systemic redshift if galaxies are situated in very large bubbles \citep{Saxena2023}. 
Ly$\alpha$ velocity profiles thus provide an independent method for mapping ionized bubbles across {\it JWST} deep imaging fields \citep{Lu2024}. 

The results described above underscore the potential of {\it JWST} for advancing our understanding of reionization. 
But if we are to realize this potential, we must extract the imprint of reionization on the distribution of Ly$\alpha$ strengths and velocity profiles. 
This task requires a robust ``intrinsic\footnote{Here we refer to intrinsic as the Ly$\alpha$ emission that would emerge from galaxies if surrounded by the ionized IGM at the end of reionization. This includes processing through the circumgalactic medium and the ionized IGM.} model'' of Ly$\alpha$ in galaxies just after reionization (i.e., at $z\simeq5-6$), when the IGM is mostly ionized. 
Over the past decade, large spectroscopic campaigns with Keck and Very Large Telescope (VLT) have taken steps toward establishing an intrinsic model for Ly$\alpha$ emission in star forming galaxies \citep[e.g.,][]{Vanzella2009,Stark2010,Stark2011,Bacon2017,Pentericci2018,Urrutia2019}. 
In addition to measuring the EW distribution of Ly$\alpha$ near the end of reionization, these observations have quantified how the intrinsic Ly$\alpha$ distributions are likely to change as galaxy properties (i.e., dust attenuation, stellar population age) evolve at $z\gtrsim6$. 
With these datasets in hand, Ly$\alpha$ measurements at $z\gtrsim7$ have been effectively mapped to constraints on both IGM neutral fractions \citep[e.g.,][]{Mesinger2015,Mason2018a,Bolan2022,Jones2024,Nakane2024} and ionized bubble sizes in the vicinity of known Ly$\alpha$ emitters \citep[e.g.,][]{Tilvi2020,Jung2022,Leonova2022,Tang2023,Chen2024,Whitler2024,Witstok2024a}. 

Unfortunately, the intrinsic Ly$\alpha$ models developed over the last decade are not equipped to interpret the large body of data that {\it JWST} is now providing at $z\gtrsim7$. 
Ly$\alpha$ escape fractions and velocity profiles are now routinely measured at $z\gtrsim7$ \citep[e.g.,][]{Bunker2023a,Jung2023,Tang2023,Chen2024,Saxena2024}, but we currently have no statistical knowledge of the distribution of either quantity in galaxies at $z\simeq5-6$. 
While the Ly$\alpha$ datasets are sufficiently deep at these redshifts, measurements of velocity profiles (requiring systemic redshifts) and Ly$\alpha$ escape fractions (requiring hydrogen Balmer lines) have never been possible prior to {\it JWST}. 
Even interpretation of Ly$\alpha$ EWs faces challenges. 
The current intrinsic $z\simeq5-6$ models have largely been derived from bright galaxies ($H<27$) for which continuum measurements were possible with the {\it Hubble Space Telescope} ({\it HST}). 
However, {\it JWST} spectroscopic measurements at $z\gtrsim7$ are rapidly pushing to fainter galaxies ($H=27-30$), with several of these showing extremely intense Ly$\alpha$ emission that likely indicates location in a large ionized bubble \citep{Saxena2023,Chen2024}. 
We expect to see enhanced Ly$\alpha$ emission in the vicinity of these strong Ly$\alpha$ emitters, with the IGM transmission increasing in lockstep with the bubble radius. 
But without knowledge of the Ly$\alpha$ EW distribution in similarly low luminosity samples at $z\simeq5-6$, it will be impossible to reliably use $z\gtrsim7$ measurements to compute the size of ionized regions around strong Ly$\alpha$ emitters. 

In this paper, we use new {\it JWST} data to improve our understanding of the Ly$\alpha$ properties in galaxies at $z\simeq5-6$, with the ultimate goal of developing the intrinsic models necessary to interpret the Ly$\alpha$ measurements now being obtained at $z\gtrsim7$ with {\it JWST}. 
Our parent sample is based on deep Ly$\alpha$ spectroscopy that has been conducted in the Great Observatories Origins Deep Survey (GOODS; \citealt{Giavalisco2004}) North \citep{Stark2010,Stark2011} and South \citep{Bacon2017,Bacon2023,Urrutia2019} fields. 
We use the deep {\it JWST}/NIRCam imaging in these fields from the {\it JWST} Advanced Deep Extragalactic Survey (JADES; \citealt{Eisenstein2023a}) to characterize the physical properties of sources with Ly$\alpha$ spectroscopic constraints. 
The deep NIRCam photometry also crucially provides the underlying continuum required for extending the Ly$\alpha$ EW distributions to faint galaxies. 
In addition, we utilize H$\alpha$ measurements obtained in the two GOODS fields from the First Reionization Epoch Spectroscopically Complete Observations (FRESCO) survey \citep{Oesch2023}. 
FRESCO utilizes {\it JWST}/NIRCam Wide Field Slitless Spectroscopy (WFSS; \citealt{Greene2017}), delivering H$\alpha$ emission line measurements for all galaxies in its footprint at $z\simeq5-6$. 
The FRESCO H$\alpha$ spectra provide the systemic redshifts necessary for mapping the Ly$\alpha$ detections into the rest-frame, delivering our first look at the Ly$\alpha$ velocity profiles in faint galaxies at $z\simeq5-6$. 
The H$\alpha$ line is also key for predicting the intrinsic Ly$\alpha$ luminosities required for Ly$\alpha$ escape fraction measurements. 
We compute Ly$\alpha$ escape fractions of $z\simeq5-6$ galaxies using the H$\alpha$ flux from FRESCO and that inferred from color excesses in JADES spectral energy distributions (SEDs; e.g., \citealt{Simmonds2024}). 
With this new observational database, we derive statistical distributions of Ly$\alpha$ properties (Ly$\alpha$ EW, Ly$\alpha$ escape fraction) in $z\simeq5-6$ Ly$\alpha$ break selected (hereafter Lyman break selected) samples for use in interpreting emerging $z\gtrsim7$ measurements.

The organization of this paper is as follows. 
In Section~\ref{sec:data}, we describe a sample of Ly$\alpha$ emitters with H$\alpha$ emission line measurements at $z\simeq5-6$. 
We characterize the Ly$\alpha$ properties (EW, escape fraction, and velocity profile) of sources in our Ly$\alpha$ emitter sample and discuss the impact of neutral hydrogen and dust in Section~\ref{sec:lae_property}. 
We then introduce a more general, Lyman break selected galaxy population at $z\simeq5-6$ and derive the Ly$\alpha$ EW and Ly$\alpha$ escape fraction distributions in Section~\ref{sec:lya_lbg}, seeking to explore how frequently the typical $z\simeq5-6$ galaxies show large Ly$\alpha$ EWs and Ly$\alpha$ escape fractions. 
Using our $z\simeq5-6$ sample as a baseline, we discuss the implications for early ionized regions around Ly$\alpha$ emitters in the reionization era in Section~\ref{sec:discussion}. 
Finally, we summarize our conclusions in Section~\ref{sec:summary}. 
Throughout the paper we adopt a $\Lambda$-dominated, flat universe with $\Omega_{\Lambda}=0.7$, $\Omega_{\rm{M}}=0.3$, and $H_0=70$~km~s$^{-1}$~Mpc$^{-1}$. 
All magnitudes are quoted in the AB system \citep{Oke1983} and all EWs are quoted in the rest frame.


\section{Data and analysis} \label{sec:data}

In this section, we assemble and analyze a sample of Ly$\alpha$ emitters with {\it JWST} measurements of the H$\alpha$ strength and redshift. 
We describe the construction of our sample in Section~\ref{sec:lae}, and measurements of Ly$\alpha$ emission line flux in Section~\ref{sec:lya_flux}. 
We fit the NIRCam-based SEDs of the Ly$\alpha$ emitters in Section~\ref{sec:sed}. 
Utilizing the new H$\alpha$ measurements to constrain intrinsic Ly$\alpha$ luminosities and systemic redshifts, we characterize the Ly$\alpha$ escape fraction and Ly$\alpha$ velocity offset. 
We also use the continuum constrained by NIRCam photometry to derive the Ly$\alpha$ EW. 
The methodology of these measurements is described in Section~\ref{sec:lya_measure}. 

\subsection{Selection of Ly$\alpha$ emitter sample at $z\simeq5-6$} \label{sec:lae}

Our sample consists of Ly$\alpha$ emitting galaxies in the two GOODS fields. 
We describe the Ly$\alpha$ emitters in GOODS-South in Section~\ref{sec:muse} and GOODS-North in Section~\ref{sec:deimos}. 

\subsubsection{Ly$\alpha$ emitter sample in GOODS-South} \label{sec:muse}

There have been numerous Ly$\alpha$ surveys in GOODS-South. 
In GOODS-South, our analysis centers on the publicly available MUSE-Wide\footnote{\url{https://musewide.aip.de/}} \citep{Herenz2017b,Urrutia2019} and MUSE-Deep\footnote{\url{https://amused.univ-lyon1.fr/}} \citep{Bacon2017,Bacon2023,Inami2017} surveys. 
These programs identify Ly$\alpha$ emitters using the integral field spectrograph Multi Unit Spectroscopic Explorer (MUSE; \citealt{Bacon2010}) at VLT. 
The Wide Field Mode was used in the surveys, which has a $1$~arcmin $\times1$~arcmin field of view. 
MUSE covers the wavelength in optical from $4750$~\AA\ to $9350$~\AA, with an average spectral resolution of $\sim2.5$~\AA\ (corresponding to a velocity resolution of $\simeq100$~km~s$^{-1}$ at $\lambda\simeq7000$~\AA). 
The wavelength range of MUSE allows the detection of Ly$\alpha$ emission in galaxies at $2.9<z<6.7$. 

The MUSE-Wide survey provides a relatively shallower dataset covering a wider area than MUSE-Deep. 
The current MUSE-Wide survey covers an area of $\sim44$~arcmin$^2$ with $44$ pointings, with $1$ hour exposure time on each pointing. 
This results in a $5\sigma$ emission line detection limit of $\simeq2\times10^{-18}$~erg~s$^{-1}$~cm$^{-2}$ for point sources. 
The MUSE-Deep survey focuses on the {\it Hubble} Ultra Deep Field (HUDF; \citealt{Beckwith2006}), providing much deeper data over a smaller area. 
MUSE-Deep has released three datasets (all of which we use): a 9-pointing campaign with $10$-hour exposure times (MOSAIC; \citealt{Bacon2017}), a single $31$-hour pointing (UDF-10; \citealt{Bacon2017}), and a new, deeper $141$-hour pointing MUSE eXtremely Deep Field (MXDF; \citealt{Bacon2023}). 
The $5\sigma$ detection limits for point sources are $5\times10^{-19}$, $2.5\times10^{-19}$, and $1\times10^{-19}$~erg~s$^{-1}$~cm$^{-2}$ in MOSAIC, UDF-10, and MXDF, respectively. 
We refer readers to \citet{Urrutia2019} for a full description of the data reduction process for MUSE-Wide, and \citet{Bacon2017,Bacon2023} for MUSE-Deep. 
In total, there are $479$ and $1308$ Ly$\alpha$ emitting galaxies at $2.9<z<6.7$ with Ly$\alpha$ line S/N $>5$ in the MUSE-Wide and Deep catalogs, respectively. 
The MUSE-Wide survey mainly identifies the more rare luminous Ly$\alpha$ emitters with Ly$\alpha$ luminosity brighter than the $L^*$ of the Ly$\alpha$ luminosity function at $3<z<6$ ($L^*\sim10^{42.2}$~erg~s$^{-1}$; \citealt{Herenz2019}). 
On the other hand, the MUSE-Deep survey primarily is comprised of the abundant population of sub-$L^*$ LAEs.

In this paper, our focus is on the Ly$\alpha$ properties of galaxies at redshifts at the tail end of reionization ($z\simeq5-6$), limiting us to a subset of the total MUSE samples. 
Rest-frame optical emission lines are important to quantify Ly$\alpha$ properties including the Ly$\alpha$ escape fraction ($f_{{\rm esc,Ly}\alpha}$) and the Ly$\alpha$ velocity offset ($\Delta v_{{\rm Ly}\alpha}$). 
We use the available {\it JWST}/NIRCam F444W grism spectra obtained from the FRESCO survey to measure the rest-frame optical emission lines of MUSE Ly$\alpha$ emitters. 
Given the wavelength coverage of F444W grism ($3.8-5.1\ \mu$m), we focus on the redshift range at $4.9<z<6.5$ to allow H$\alpha$ detection.
Among the $479$ and $1308$ Ly$\alpha$ emitters in the MUSE-Wide and MUSE-Deep surveys, there are $69$ and $289$ galaxies at $4.9<z<6.5$, respectively. 
We find $3$ duplicates between MUSE-Wide and MUSE-Deep sources, leaving $69+289-3=355$ MUSE identified Ly$\alpha$ emitters at $4.9<z<6.5$ in total. 

We now identify a Ly$\alpha$ selected sample with H$\alpha$ line detections in GOODS-South at $4.9<z<6.5$. 
We cross-match the MUSE Ly$\alpha$ emitter catalog to the NIRCam grism emission line catalog based on a joint analysis of FRESCO and JADES data (Sun et al. in prep.).
We visually inspect the MUSE Ly$\alpha$ narrowband image and NIRCam F444W image (Sun et al. in prep.) of each Ly$\alpha$ emitter and identify the matched NIRCam source. 
Among the $355$ MUSE Ly$\alpha$ emitters at $4.9<z<6.5$, there are $82$ sources with H$\alpha$ emission line detections (here an H$\alpha$ detection refers to an H$\alpha$ line detected in the NIRCam grism spectrum). 
For the remaining $273$ Ly$\alpha$ emitters, the H$\alpha$ emission lines are not detected either because the expected position of H$\alpha$ is out of the individual F444W grism spectra or the H$\alpha$ fluxes are below the FRESCO detection limit ($5\sigma$ emission line sensitivity $\sim2\times10^{-18}$~erg~s$^{-1}$~cm$^{-2}$ for a point source; \citealt{Oesch2023}). 
In order to derive the Ly$\alpha$ properties, we also need to measure the Ly$\alpha$ flux and Ly$\alpha$ redshift accurately. 
Therefore, we removed $18$ galaxies whose Ly$\alpha$ emission lines are contaminated by sky line residuals. 
We also removed four objects identified as active galactic nuclei (AGN) \citep{Lyu2024,Matthee2024}, leaving a final sample containing $60$ galaxies with both Ly$\alpha$ and H$\alpha$ detections at $4.9<z<6.5$ in GOODS-South. 
In Section~\ref{sec:lae_property} we will primarily focus on this sample, but we will also discuss the potential bias of this sample relative to the entire Ly$\alpha$ emitting galaxy sample (i.e., including both sources with and without H$\alpha$ detection) therein. 

\subsubsection{Ly$\alpha$ emitter sample in GOODS-North} \label{sec:deimos}

In addition to the Ly$\alpha$ emitters identified from the public VLT/MUSE catalogs targeting the GOODS-South field, we also include Ly$\alpha$ emitters at $4.9<z<6.5$ in the GOODS-North field identified from a large spectroscopic survey taken with the DEep Imaging Multi-Object Spectrograph (DEIMOS; \citealt{Faber2003}) at the Keck II telescope. 
We direct readers to \citet{Stark2010} and \citet{Stark2011} for detailed descriptions of the survey. 
Below we briefly summarize the DEIMOS spectra. 
Our Keck/DEIMOS survey targeted $B$-, $V$-, and $i$-band dropouts (i.e., $z\sim4-6$ Lyman break galaxies). 
The dropouts were identified in \citet{Stark2009} utilizing the standard color selection criteria. 
The follow-up DEIMOS spectroscopic observations were performed between 2008 and 2015 with eight multi-object slitmasks. 
The slit width is $1.0$~arcsec. 
Seven masks primarily targeting $B$- and $V$-band dropouts ($z_{\rm phot}\sim4-5$) were observed using the $600$ line/mm grating blazed at $7500$~\AA, covering wavelength $4850-10150$~\AA\ with a resolution of $3.5$~\AA\ (velocity resolution $\simeq140$~km~s$^{-1}$). 
The remaining mask primarily targets $i$-band dropouts ($z_{\rm phot}\sim6$), and was observed using the $830$ line/mm grating blazed at $8640$~\AA\ covering $6800-10100$~\AA\ (spectral resolution $=2.5$~\AA, corresponding to velocity resolution $\simeq87$~km~s$^{-1}$). 
The on-target integration time of each mask is between $3$ and $12.5$ hours with an average seeing $=0.5-1.1$~arcsec, resulting in $5\sigma$ Ly$\alpha$ line flux limits of $0.3-2.0\times10^{-17}$~erg~s$^{-1}$~cm$^{-2}$. 
The DEIMOS spectra were reduced following the methodology described in \citet{Stark2010}. 
We summarize the DEIMOS observations in Table~\ref{tab:deimos_obs}.


\begin{table*}
\centering
\begin{tabular}{cccccccc}
\hline
Mask ID & R.A. & Decl. & Date & Exposure Time & N$_{\rm dropout}$ & Grating & Seeing \\
 & (hh:mm:ss) & (dd:mm:ss) & & (s) & & (line/mm) & (arcsec) \\
\hline
gn\_A & 12:37:06.02 & +62:16:33.2 & April 2008 & 21600 & 94 & 600 & 1.0 \\
gn\_B & 12:37:16.83 & +62:15:00.4 & April 2008 & 21600 & 107 & 600 & 0.8 \\
gn\_C & 12:37:16.29 & +62:15:04.9 & April 2008 & 20400 & 100 & 600 & 1.1 \\
kcGNv1B & 12:37:02.14 & +62:13:48.1 & March 2009 & 18000 & 108 & 600 & 1.0 \\
kcGNv2B & 12:36:55.53 & +62:14:24.9 & March 2009 & 25200 & 79 & 600 & 0.5 \\
GNm1v5 & 12:36:54.84 & +62:14:11.6 & April 2010 & 45000 & 23 & 830 & 0.8 \\
Bdrop & 12:36:46.16 & +62:13:26.3 & June 2012 & 16800 & 73 & 600 & 1.0 \\
GN\_AZ & 12:37:15.18 & +62:14:19.0 & March 2015 & 10800 & 79 & 600 & 1.0 \\
\hline
\end{tabular}
\caption{Summary of Keck/DEIMOS observations of $z\simeq4-6$ dropouts in the GOODS-North field. Observations were taken in between 2008 and 2015. N$_{\rm dropout}$ represents the number of $B$-, $V$-, and $i$-band dropouts (primary targets) placed on each mask.}
\label{tab:deimos_obs}
\end{table*}

We visually inspect the DEIMOS spectrum of each galaxy to search for Ly$\alpha$ emission lines. 
We identify Ly$\alpha$ emission in $220$ galaxies at $3<z<6.5$, including $32$ Ly$\alpha$ emitters at $4.9<z<6.5$. 
Similar to our MUSE sample, we cross-match these $32$ galaxies to the NIRCam grism emission line catalog in Sun et al. (in prep.) in GOODS-North and find H$\alpha$ detections in $29$ of them. 
For the remaining three systems without an H$\alpha$ detection, two of them have H$\alpha$ flux below the detection limit, and the expected H$\alpha$ position of another one is shifted out of its F444W grism spectrum. 
Again, we removed $9$ sources for which the Ly$\alpha$ emission lines are contaminated by sky line residuals, and one AGN identified in \citet{Matthee2024}. 
This leaves a sample of $19$ galaxies with both Ly$\alpha$ and H$\alpha$ detections at $4.9<z<6.5$ in GOODS-North. 
Combining with the sample in GOODS-South, we have identified $79$ Ly$\alpha$ emitting galaxies with H$\alpha$ detections at $4.9<z<6.5$. 
We list these $79$ systems in Table~\ref{tab:lya_properties}.

\subsection{Measurements of Ly$\alpha$ flux} \label{sec:lya_flux}

For the $79$ galaxies with Ly$\alpha$ and H$\alpha$ detections at $4.9<z<6.5$ in our sample, we measure their Ly$\alpha$ emission line fluxes from MUSE or DEIMOS spectra in a self-consistent way. 
Before analyzing the Ly$\alpha$ fluxes of our sources, we first consider the potential aperture loss for Ly$\alpha$ emission. 
We then present the measurements of Ly$\alpha$ fluxes.

For the $60$ VLT/MUSE sources in our $4.9<z<6.5$ sample, we take advantage of the 1D MUSE spectra extracted and published by the MUSE-Wide \citep{Urrutia2019} and MUSE-Deep \citep{Bacon2023} teams. 
For objects in MUSE-Wide, we use the point spread function (PSF; full width at half maximum FWHM $\simeq0.7$~arcsec for MUSE) weighted 1D spectra extracted with the \texttt{LSDCat} software \citep{Herenz2017a} from the 3D MUSE data cube by \citet{Urrutia2019}. 
For objects in MUSE-Deep, we use the reference spectra provided by \citet{Bacon2023}. 
Among the extractions used in \citet{Bacon2023}, the \texttt{ORIGIN} algorithm \citep{Mary2020} is primarily favored for extracting spectra of Ly$\alpha$ emitters. 
The reference spectra of the most of our MUSE-Deep sources ($29$ out of $36$) are extracted with \texttt{ORIGIN}. 
Source blending increases in deep exposures, and the spectra of blended sources are better extracted with the de-blended algorithm \citep{Bacon2023}. 
There are $5$ blended sources (MUSE-547, MUSE-2071, MUSE-6294, MUSE-6462, and MUSE-7125) in our MUSE-Deep sample, and their reference spectra are extracted with the de-blended algorithm \texttt{ODHIN}. 
For galaxies with very bright Ly$\alpha$ emission, the spectra are best extracted with the \texttt{NBEXT} algorithm \citep{Bacon2023}.
MUSE-68 and MUSE-7605 are the two bright Ly$\alpha$ emitters in our MUSE-Deep sample, and their spectra are extracted with \texttt{NBEXT}. 
Comparing the MUSE-Wide and MUSE-Deep spectra for duplicates, we find that the Ly$\alpha$ emission line profiles extracted from MUSE-Wide and MUSE-Deep are consistent. 
The difference of Ly$\alpha$ fluxes measured between the MUSE-Wide and MUSE-Deep spectra of duplicates is less than $10$~per~cent. 

We consider the effects of aperture loss of Ly$\alpha$ fluxes measured by MUSE. 
It has been established that Ly$\alpha$ emitting galaxies are commonly surrounded by extended Ly$\alpha$ halos \citep[e.g.,][]{Matsuda2012,Hayes2013,Momose2014,Matthee2016,Wisotzki2016,Leclercq2017,Wu2020,Guo2023,Zhang2024}. 
From the segmentation maps (used to extract 1D spectra) published by \citet{Urrutia2019} and \citet{Bacon2023}, we find that the MUSE spectra extraction apertures of our $z\simeq5-6$ Ly$\alpha$ emitters have a median diameter $=1.5$~arcsec. 
The Ly$\alpha$ line flux measured within the above mentioned MUSE aperture (as well as slit spectrographs including Keck/DEIMOS and the {\it JWST}/NIRSpec) could miss a portion of the flux from the Ly$\alpha$ halo. 
Here we estimate the ratio of Ly$\alpha$ flux recovered by MUSE measurements to the total Ly$\alpha$ flux. 
To do this, we generate a Ly$\alpha$ surface brightness profile for a typical galaxy at $z\simeq5-6$ based on the Ly$\alpha$ halo measurements in \citet{Leclercq2017} and estimate the fraction of the Ly$\alpha$ flux within the MUSE aperture. 
\citet{Leclercq2017} fit individual Ly$\alpha$ surface brightness profile with a two-component model (a core and a halo), each described by an exponentially decreasing distribution. 
About $65$~per~cent of the total Ly$\alpha$ flux comes from the halo. 
The median scale length of the halo is $3.8$~kpc at $z\simeq5-6$, while the average scale length of the core is $\simeq1/10$ of the halo scale length. 
Then we create the Ly$\alpha$ surface brightness profile based on the above parameters, assuming exponential decreasing distributions for both the core and the halo component. 
After convolving with MUSE PSF, we estimate that about $75$~per~cent of the total Ly$\alpha$ flux for a $z\simeq5-6$ galaxy will be recovered when extracting the line flux using the MUSE aperture.

Similarly, the Keck/DEIMOS slit spectra should also miss a portion of the total Ly$\alpha$ flux due to the slit loss. 
We estimate the DEIMOS slit loss for Ly$\alpha$ flux following the same procedures for estimating MUSE aperture loss. 
We note that our DEIMOS sources are brighter than our MUSE sources (median absolute UV magnitude M$_{\rm UV}=-19.6$ for DEIMOS vs. M$_{\rm UV}=-18.8$ for MUSE). 
However, \citet{Leclercq2017} show that the Ly$\alpha$ halo scale length does not change significantly with M$_{\rm UV}$. 
Therefore, we use the same Ly$\alpha$ surface brightness profile model generated utilizing the parameters measured in \citet{Leclercq2017} to estimate the fraction of in-slit light. 
We find that $\simeq57$~per~cent of the total Ly$\alpha$ flux will be recovered within the $1.0$~arcsec width DEIMOS slit. 
Here we note that the MUSE or DEIMOS aperture loss of Ly$\alpha$ flux is estimated based on the median parameters of Ly$\alpha$ surface brightness profile measurements at $z\simeq5-6$, and thus it could vary among individual sources. 
Taking the Ly$\alpha$ halo scale length range of $z\simeq5-6$ Ly$\alpha$ emitters ($1-10$~kpc) reported in \citet{Leclercq2017}, we estimate that the fraction of total Ly$\alpha$ flux recovered by MUSE ($1$~arcsec width DEIMOS slit) measurement ranges from $61$ to $86$~per~cent ($44$ to $63$~per~cent).
This suggests the true total flux may vary by $0.8-1.1\times$ relative to that derived in this paper. 

We now derive the Ly$\alpha$ fluxes of the $79$ Ly$\alpha$ emitters at $4.9<z<6.5$ in our sample. 
Because the Ly$\alpha$ emission line profile at high redshift can be complex, we compute the line flux by directly integrating the flux between rest-frame $1212$~\AA\ and $1220$~\AA\ ($\simeq\pm1000$~km~s$^{-1}$ in velocity space). 
This wavelength window captures the total Ly$\alpha$ flux for Ly$\alpha$ emitting galaxies \citep[e.g.,][]{Du2020,Matthee2021}. 
We next examine the impact of the underlying continuum to the integrated Ly$\alpha$ line flux. 
For all the $79$ galaxies, the continua measured from spectra are faint with relatively low S/N ($<5$). 
Therefore, we estimate the underlying continuum flux densities using the available {\it JWST}/NIRCam or {\it HST} photometry (Section~\ref{sec:lya_measure}). 
We find that the continua make negligible changes ($<10$~per~cent) to the integrated Ly$\alpha$ fluxes of these $79$ Ly$\alpha$ emitters. 
Motivated by these, we do not subtract the underlying continua from the integrated Ly$\alpha$ fluxes for our Ly$\alpha$ emitters. 

The observed Ly$\alpha$ fluxes of the $60$ MUSE sources in our sample range from $1.2\times10^{-18}$ to $5.5\times10^{-17}$~erg~s$^{-1}$~cm$^{-2}$. 
These Ly$\alpha$ fluxes are similar to those presented in the MUSE-Wide and Deep data catalogs in \citet{Urrutia2019} and \citet{Bacon2023}, with differences $<0.1$~dex for all the $60$ MUSE sources. 
To ensure the Ly$\alpha$ fluxes of MUSE and DEIMOS sources are measured in a self-consistent way, we will adopt our own Ly$\alpha$ flux measurements throughout the paper. 
For the $19$ DEIMOS sources, the observed Ly$\alpha$ fluxes are from $2.8\times10^{-18}$ to $3.5\times10^{-17}$~erg~s$^{-1}$~cm$^{-2}$. 
In order to be consistent with the MUSE measurements, we convert the Ly$\alpha$ fluxes measured from the DEIMOS slit for the MUSE aperture. 
Using the aperture losses estimated for DEIMOS $1.0$~arcsec slit and MUSE aperture, we multiply the observed DEIMOS Ly$\alpha$ fluxes by a factor of $0.75/0.57=1.3$. 
After conversion, the Ly$\alpha$ fluxes of the $19$ DEIMOS Ly$\alpha$ emitters are from $3.6\times10^{-18}$ to $4.6\times10^{-17}$~erg~s$^{-1}$~cm$^{-2}$. 
To evaluate the uncertainties of Ly$\alpha$ fluxes, we resample the flux densities of each spectrum $1000$ times by taking the observed flux densities as mean values and the errors as standard deviations. 
We measure the Ly$\alpha$ fluxes from the resampled spectra of each source using the same methods and take the standard deviation as the uncertainty.
In Table~\ref{tab:lya_properties}, we list the Ly$\alpha$ fluxes of the $79$ galaxies with both Ly$\alpha$ and H$\alpha$ detections at $4.9<z<6.5$. 

One of the primary goals of this paper is to provide a baseline for interpreting the Ly$\alpha$ measurements at $z\gtrsim7$ with {\it JWST}/NIRSpec. 
Therefore, we also need to consider the possible aperture loss for NIRSpec micro-shutter assembly (MSA; \citealt{Ferruit2022}) observations, and how this compares to MUSE aperture loss. 
To estimate the NIRSpec MSA aperture loss for Ly$\alpha$ emission, we generate the same Ly$\alpha$ surface brightness profile based on the measurements in \citet{Leclercq2017} as we did for MUSE and DEIMOS aperture loss estimation. 
We assume the source is centered in the NIRSpec micro-shutter and convolve the Ly$\alpha$ surface brightness profile with the NIRSpec PSF using the \texttt{WebbPSF} package \citep{Perrin2014}. 
We estimate that $\simeq50$~per~cent of the total Ly$\alpha$ flux at $z\simeq5-6$ will be recovered by NIRSpec MSA measurement. 
Considering the {\it JWST} data reduction pipeline\footnote{\url{https://github.com/spacetelescope/jwst}} \citep{Bushouse2024} will perform an aperture correction assuming a point source, the pipeline corrected Ly$\alpha$ flux will recover $\simeq60$~per~cent of the total Ly$\alpha$ flux. 
Comparing to the fraction of Ly$\alpha$ flux recovered by MUSE aperture ($\simeq75$~per~cent), the typical Ly$\alpha$ flux measured with a NIRSpec MSA shutter will be $0.6/0.75=80$~per~cent of the flux measured with MUSE. 
We note that if the source is not centered in the MSA shutter, there will be a $10-20$~per~cent systematic uncertainty as the NIRSpec PSF depends on the position of the target in the shutter \citep[e.g.,][]{deGraaff2024}.
We test the aperture losses by cross-matching $z\simeq5-6$ Ly$\alpha$ emitters identified in MUSE surveys to the public NIRSpec survey \citep{Bunker2023b,Saxena2024,Witstok2024a}. 
There are three matched Ly$\alpha$ emitters (all in the HUDF field): MUSE-852 (ID 16625 in \citealt{Saxena2024,Witstok2024a}), MUSE-3089 (ID 9365), and MUSE-6231 (ID 14123). 
We find that the NIRSpec Ly$\alpha$ fluxes of these sources are $\simeq70-80$~per~cent of the fluxes measured from MUSE, which is consistent with the NIRSpec to MUSE flux ratio estimated from aperture loss ($\simeq80$~per~cent). 
In the future, a larger reference sample is required to estimate the aperture correction between Ly$\alpha$ measured from IFU and slit spectroscopy.

\subsection{SED fitting of Ly$\alpha$ emitter sample} \label{sec:sed}

To derive the physical properties of the $79$ galaxies with Ly$\alpha$ and H$\alpha$ detections in our $4.9<z<6.5$ sample, we fit their available SEDs with stellar population and photoionization models. 
We utilize the {\it JWST}/NIRCam imaging taken as a part of the JADES observations targeting the two GOODS fields. 
The current JADES NIRCam observations utilizes nine NIRCam filters (F090W, F115W, F150W, F200W, F277W, F335M, F356W, F410M, and F444W), covering a wavelength range of $0.8-5.0\ \mu$m. 
The JADES NIRCam data reduction is introduced in literature \citep{Eisenstein2023a,Rieke2023b,Robertson2023,Tacchella2023a}, and will be fully described in Tacchella et al. (in prep.). 
Among the $79$ Ly$\alpha$ emitters with H$\alpha$ detections at $4.9<z<6.5$ in our sample, $61$ lie in the JADES NIRCam footprint ($52/60$ in GOODS-South and $9/19$ in GOODS-North). 
For these $61$ Ly$\alpha$ emitters, we visually inspect the MUSE Ly$\alpha$ narrowband images or the {\it HST} images used to select DEIMOS targets (\citealt{Stark2010}; see also Section~\ref{sec:deimos}) with JADES NIRCam images. 
We have identified JADES NIRCam counterparts for all these $61$ galaxies. 
For each of these $61$ galaxies, we take advantage of the F090W to F444W photometry provided in the JADES photometry catalog (\citealt{Rieke2023b}; Robertson et al. in prep.). 

We use the JADES $0.2$~arcsec diameter circular aperture (``CIRC1'') fluxes to compute the NIRCam colors, which reduces the background noise associated with larger apertures \citep{Hainline2024b}. 
We also use the CIRC1 fluxes to derive the UV slopes of our sources. 
The UV slope is computed by fitting a power law ($f_{\lambda}\propto\lambda^{\beta}$) to the JADES NIRCam broadband fluxes at rest-frame wavelengths $1250-2600$~\AA\ \citep{Calzetti1994}. 
Then we use the Kron \citep{Kron1980} aperture ($k=2.5$) fluxes, which represent the total fluxes, for SED fitting but keep the CIRC1 colors. 
We modify the Kron aperture fluxes using the following procedures. 
For each object, we compute the median ratio of fluxes measured within $k=2.5$ Kron apertures to CIRC1 apertures in NIRCam F115W, F150W, and F200W filters. 
The Kron aperture fluxes used here are corrected to the total fluxes using the NIRCam PSFs \citep{Rieke2023b}. 
Next, we multiply all the NIRCam CIRC1 fluxes by this factor to obtain the modified Kron aperture fluxes. 

We fit the modified JADES NIRCam Kron aperture photometry of the $61$ galaxies using the Bayesian galaxy SED modelling and interpreting tool BayEsian Analysis of GaLaxy sEds (\textsc{beagle}, version 0.23.0; \citealt{Chevallard2016}). 
The \textsc{beagle} setup and SED fitting procedures follow the description in \citet{Tang2023}, and we summarize these below. 
Models used in \textsc{beagle} combine the latest version of the \citet{Bruzual2003} stellar population synthesis models and the \citet{Gutkin2016} photoionization models of star-forming galaxies with the {\small CLOUDY} code \citep{Ferland2013}. 
The redshift is fixed to the systemic redshift measured from H$\alpha$ emission lines (Sun et al. in prep.). 
We assume a constant star formation history (CSFH), allowing the galaxy age to vary between $1$~Myr and the age of the Universe at the given redshift with a log-uniform prior. 
We assume a \citet{Chabrier2003} initial mass function (IMF) with a stellar mass range of $0.1-300\ M_{\odot}$. 
We allow the metallicity to vary in the range $-2.2\le\log{(Z/Z_{\odot})}\le0.25$ ($Z_{\odot}=0.01524$; \citealt{Caffau2011}) and the dust-to-metal mass ratio ($\xi_{\rm d}$) to span the range $\xi_{\rm d}=0.1-0.5$. 
The interstellar metallicity is set to be equal to the stellar metallicity. 
The ionisation parameter $U$ is adjusted in the range $-4.0\le \log{U}\le-1.0$. 
We adopt log-uniform priors for metallicity and ionization parameter, and a uniform prior for dust-to-metal mass ratio. 
We assume the Small Magellanic Cloud (SMC) extinction curve \citep{Pei1992} to account for the dust attenuation, allowing the $V$-band optical depth $\tau_{V}$ to vary between $0.001$ and $5$ with a log-uniform prior. 
Finally, we adopt the prescription of \citet{Inoue2014} to include the absorption of IGM. 
When fitting the SEDs, we remove fluxes in filters that lie blueward of Ly$\alpha$ to avoid introducing the uncertain flux contribution from Lyman series emission and absorption. 

From the \textsc{beagle} models we derive the median values from the posterior probability distributions and the marginalized $68$~per~cent credible intervals. 
For the $61$ Ly$\alpha$ emitters at $4.9<z<6.5$ with JADES photometry, their SEDs suggest that the light is dominated by very young stellar populations, with luminosity-weighted age ranging from $2$~Myr to $263$~Myr (median age $=14$~Myr) and specific star formation rate (sSFR) $=4-575$~Gyr$^{-1}$ (median sSFR $=69$~Gyr$^{-1}$) assuming CSFH. 
These reflect that our Ly$\alpha$ emitters are likely dominated by a recent burst or upturn in star formation history. 
The stellar masses derived from \textsc{beagle} CSFH models occupy the relatively low-mass space, with $1.2\times10^7-9.3\times10^8\ M_{\odot}$ (median $=4.7\times10^7\ M_{\odot}$). 
We note that the stellar masses derived from CSFH models correspond to the very young stellar populations which dominate the rest-frame UV to optical SEDs. 
Older stellar populations can be easily outshined by the light of young stars, but increasing the stellar mass up to over an order of magnitude \citep[e.g.,][]{Roberts-Borsani2020,Laporte2021,Tacchella2022,Tacchella2023b,Tang2022,Whitler2023}. 
However, this effect will not strongly impact the main results presented in this paper (Section~\ref{sec:lae_property}) because they do not depend on the stellar mass. 

We characterize the [O~{\small III}]+H$\beta$ EWs using the JADES NIRCam SEDs. 
From \textsc{beagle} models, we derive the rest-frame [O~{\small III}]+H$\beta$ EW $=249-5206$~\AA\ for the $61$ Ly$\alpha$ emitters at $4.9<z<6.5$, with a median of $1283$~\AA. 
This median [O~{\small III}]+H$\beta$ EW is nearly $2$ times larger than the average EW of galaxies in the reionization era ($\simeq700-800$~\AA; e.g., \citealt{Endsley2023a,Endsley2023b}), similar to that in emission line selected samples ($\simeq1000$~\AA; e.g., \citealt{Matthee2023,Rinaldi2023,Tang2023}). 
This is consistent with the very young luminosity-weighted ages inferred from models for our Ly$\alpha$ emitters at $4.9<z<6.5$ \citep[e.g.,][]{Chevallard2018,Tang2019,Tang2023}. 

The NIRCam SEDs also constrain the hydrogen ionizing photon production efficiency ($\xi_{\rm ion}$). 
Throughout this paper we use the most commonly definition of $\xi_{\rm ion}$ in literature: $\xi_{\rm ion}$ is the hydrogen ionizing photon production rate per unit intrinsic UV luminosity density at rest-frame $1500$~\AA\ ($L_{\rm UV}$), where $L_{\rm UV}$ is the observed UV luminosity (including both stellar and nebular continuum) corrected for dust attenuation (see \citealt{Chevallard2018} for definitions of various $\xi_{\rm ion}$). 
We find large ionizing photon production efficiencies for our Ly$\alpha$ emitters at $4.9<z<6.5$ (median $\xi_{\rm ion}\simeq10^{25.6}$~erg$^{-1}$~Hz), comparable to the $z\gtrsim6$ population (median $\xi_{\rm ion}\simeq10^{25.5-25.7}$~erg$^{-1}$~Hz; e.g., \citealt{Endsley2023a,Simmonds2023,Tang2023}). 
The results indicate that the majority of our Ly$\alpha$ emitters at $4.9<z<6.5$ have hard ionizing spectra. 
We find that assuming a different SFH (e.g., a burst on top of more evolved stellar population or a non-parametric SFH) in SED fitting does not change the derived $\xi_{\rm ion}$ significantly. 
In Table~\ref{tab:sed} we present the galaxies properties inferred from \textsc{beagle} models for the $61$ Ly$\alpha$ emitters with H$\alpha$ detection at $4.9<z<6.5$. 

\subsection{Characterization of Ly$\alpha$ emission line properties} \label{sec:lya_measure}

Deep {\it JWST}/NIRCam imaging observations allow us to constrain the underlying continuum that is essential to derive the Ly$\alpha$ EW. 
In addition, the H$\alpha$ emission lines measured from NIRCam grism spectra provide constraints to the intrinsic Ly$\alpha$ luminosity and systemic redshift, enabling us to quantify the Ly$\alpha$ escape fraction and Ly$\alpha$ velocity offset. 
These properties allow us to investigate the neutral hydrogen and dust distribution of galaxies as well as the ionization state of the surrounding IGM. 
In the following we derive the Ly$\alpha$ EWs, Ly$\alpha$ escape fractions, and Ly$\alpha$ velocity offsets for the $79$ galaxies with Ly$\alpha$ and H$\alpha$ detections at $4.9<z<6.5$ in our sample. 

\begin{figure}
\begin{center}
\includegraphics[width=\linewidth]{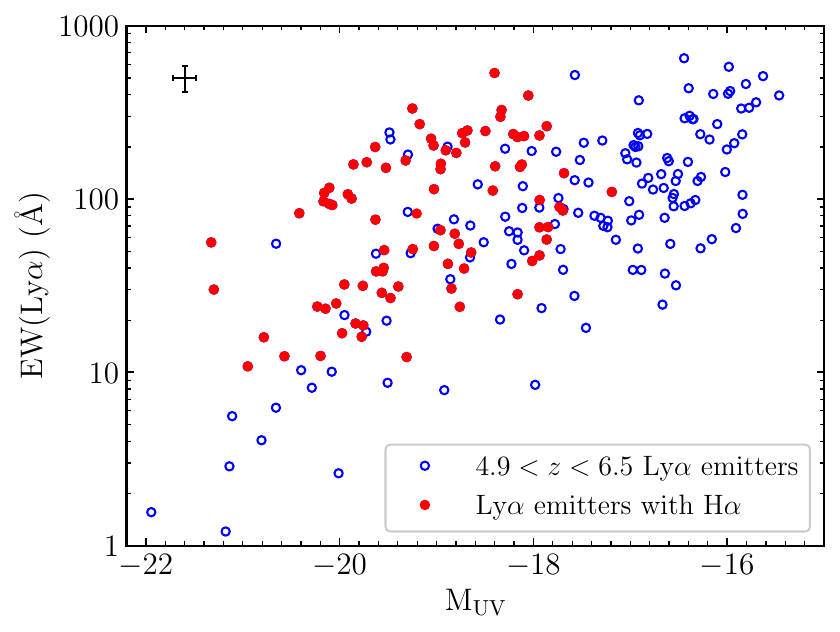}
\caption{Ly$\alpha$ EW versus absolute UV magnitude (M$_{\rm UV}$) of all the Ly$\alpha$ emitting galaxies at $4.9<z<6.5$ identified from VLT/MUSE (Section~\ref{sec:muse}) and Keck/DEIMOS (Section~\ref{sec:deimos}) observations (open blue circles). The typical uncertainties are shown as the error bars at the upper left of the figure. We have identified H$\alpha$ detections in $79$ galaxies to derive Ly$\alpha$ properties including Ly$\alpha$ escape fraction and Ly$\alpha$ velocity offset. These $79$ sources with both Ly$\alpha$ and H$\alpha$ detections are marked by red circles.}
\label{fig:lyaew_muv_all}
\end{center}
\end{figure}

We compute the Ly$\alpha$ emission line EWs using the aperture corrected Ly$\alpha$ fluxes derived in Section~\ref{sec:lya_flux} and the underlying continuum flux densities. 
For the $61$ galaxies with JADES NIRCam observations (Section~\ref{sec:sed}), we estimate the continuum from NIRCam photometry. 
For the other $18$ galaxies we utilize the {\it HST} broadband photometry obtained from the {\it Hubble} Legacy Field (HLF; \citealt{Whitaker2019}) archive, which includes all the {\it HST} imaging data in the two GOODS fields \citep[e.g.,][]{Ellis2013,Illingworth2013}. 
We visually inspect the Ly$\alpha$ images of these $18$ Ly$\alpha$ emitters with {\it HST} images, and we find {\it HST} counterparts for all of them. 
Then we fit the photometry from filters covering rest-frame $1250-2600$~\AA\ with a power law $f_{\lambda}\propto\lambda^{\beta}$. 
From the best-fit $f_{\lambda}(\lambda)$ relation we derive the average continuum flux density at rest-frame $1225-1250$~\AA\ \citep{Kornei2010,Stark2010}. 
The derived Ly$\alpha$ EWs of the $60$ MUSE Ly$\alpha$ emitters with H$\alpha$ detections at $4.9<z<6.5$ are from $12$~\AA\ to $534$~\AA\ (rest-frame), with a median value of $111$~\AA. 
The absolute UV magnitudes of our MUSE sources at rest-frame $1500$~\AA\ (M$_{\rm UV}$) range from $-21.3$ to $-17.2$, with a median M$_{\rm UV}=-18.7$. 
For the $19$ DEIMOS sources, we compute the Ly$\alpha$ EWs from $11$~\AA\ to $212$~\AA\ with a median EW $=76$~\AA\ after converting the Ly$\alpha$ fluxes measured with DEIMOS slits to the MUSE aperture. 
The absolute UV magnitudes of these $19$ systems are from $-21.0$ to $-18.4$ with a median M$_{\rm UV}=-19.6$. 
Comparing to our MUSE sources, our DEIMOS sample extends to lower Ly$\alpha$ EW and brighter M$_{\rm UV}$. 
The uncertainties of Ly$\alpha$ EWs are estimated by adding the uncertainties of Ly$\alpha$ fluxes (Section~\ref{sec:lya_flux}) and the errors of continua in quadrature. 
We summarize the Ly$\alpha$ EWs in Table~\ref{tab:lya_properties}. 
In Fig.~\ref{fig:lyaew_muv_all}, we show the M$_{\rm UV}$ and Ly$\alpha$ EW for the $79$ galaxies with Ly$\alpha$ and H$\alpha$ detections (open red circles). 


To compute the Ly$\alpha$ escape fractions, we use the NIRCam F444W grism spectra obtained from FRESCO observations and processed by Sun et al. (in prep.) to measure the H$\alpha$ luminosities and hence derive the intrinsic Ly$\alpha$ luminosities. 
For each object, the H$\alpha$ flux was derived by fitting a Gaussian profile to the 1D NIRCam grism spectrum. 
The 1D spectrum was optimally extracted \citep{Horne1986} from the 2D grism spectrum using the 2D emission line profile (Sun et al. in prep.). 
The typical aperture size for extracting 1D NIRCam grism spectra is $0.75$~arcsec, which is similar to the MUSE aperture size after deconvolving the MUSE PSF (Section~\ref{sec:lya_flux}). 
It has been noticed in Sun et al. (in prep.) that the H$\alpha$ fluxes measured from NIRCam grism spectra are slightly lower (by a factor of $0.82$) than those inferred from NIRCam imaging. 
This may be explained if the extended H$\alpha$ emission line component is missed in the over-subtraction of sky continuum (Sun et al. in prep.). 
To examine this discrepancy, we compare the H$\alpha$ fluxes measured from NIRCam grism spectra to those inferred from \textsc{beagle} models (Section~\ref{sec:sed}) for our objects. 
We find that the grism measured H$\alpha$ fluxes are on average $0.82\times$ of the fluxes inferred from SED fitting, consistent with the result found in Sun et al. (in prep.). 
Because the NIRCam grism spectra allow direct measurement of H$\alpha$ lines that are free of nearby emission lines such as [N~{\small II}]~$\lambda\lambda6548,6584$, we still use the spectral measured H$\alpha$ luminosity to infer the intrinsic Ly$\alpha$ luminosity. 
To correct the spectral measured H$\alpha$ flux to the total flux, we multiply with a factor derived from the comparison between spectral flux and SED inferred value: $1/0.82=1.22$. 

We then correct the observed H$\alpha$ luminosities for dust attenuation. 
For the $61$ galaxies with JADES NIRCam photometry measurements, we derive the dust attenuation to the H$\alpha$ emission line ($A_{{\rm H}\alpha}$) from \textsc{beagle} models (Table~\ref{tab:sed}). 
The dust attenuation is low for the $z\simeq5-6$ galaxies with both Ly$\alpha$ and H$\alpha$ detections in our sample, with a median value of $A_{{\rm H}\alpha}=0.007$~mag. 
For the $18$ galaxies without JADES NIRCam SEDs, we apply the above median dust attenuation inferred from \textsc{beagle} models of the $61$ galaxies with JADES SEDs. 

The dust-corrected H$\alpha$ luminosity ($L_{{\rm H}\alpha}$) is converted to the intrinsic Ly$\alpha$ luminosity ($L_{{\rm Ly}\alpha,{\rm int}}$) assuming case B recombination. 
We apply $L_{{\rm Ly}\alpha,{\rm int}}=8.7\times L_{{\rm H}\alpha}$ (see the discussion on the $8.7$ factor in \citealt{Hayes2015} and \citealt{Henry2015}) assuming an electron temperature $T_{\rm e}=10000$~K and an electron density $n_{\rm e}=100$~cm$^{-3}$. 
The $25-50-75$~per~centile values of the computed Ly$\alpha$ escape fractions of our $79$ sources are $0.14$, $0.26$, and $0.54$. 
We note that varying the electron temperature and electron density in $T_{\rm e}=5000-20000$~K and $n_{\rm e}=100-1000$~cm$^{-3}$ does not impact the $L_{{\rm Ly}\alpha}/L_{{\rm H}\alpha}$ ratio and hence the calculated $f_{{\rm esc,Ly}\alpha}$ significantly. 
We also note that assuming case B recombination, which is valid for optically-thick H~{\small II} regions, may not always be applicable. 
If a galaxy is leaking a large fraction of Ly$\alpha$ emission through optically-thin H~{\small I} gas, which might be the case for a subset of our sources with very large Ly$\alpha$ escape fractions (assuming case B recombination), case A recombination may be a better approximation. 
Assuming case A recombination, we utilize $L_{{\rm Ly}\alpha,{\rm int}}=11.4\times L_{{\rm H}\alpha}$ \citep{Osterbrock2006}. 
The resulting Ly$\alpha$ escape fractions are $1.3$ times lower than the values derived assuming case B recombination, with $25-50-75$~percentile values $=0.10$, $0.20$, and $0.41$. 
We list Ly$\alpha$ escape fractions assuming both case B and case A recombination in Table~\ref{tab:lya_properties}. 
In order to ensure robust comparison with values in the literature \citep[e.g.,][]{Hayes2010,Matthee2016,Yang2017,Chen2024,Saxena2024}, we will adopt Ly$\alpha$ escape fractions assuming case B recombination in the following sections. 

We note that the Ly$\alpha$ escape fractions computed here include the contribution from extended Ly$\alpha$ halos. 
The extended Ly$\alpha$ emission is not only produced by recombinations inside galaxies but also other origins (cooling radiation, Ly$\alpha$ fluorescence, satellite galaxies; e.g., \citealt{Leclercq2017}; \citealt{Kusakabe2020,Leclercq2020,HerreroAlonso2023}). 
Although it is difficult to quantify the contribution of each origin, recent simulations suggest that the scattering of Ly$\alpha$ emission produced by recombinations inside the galaxy dominates the inner region ($r<7$~kpc) at $3<z<6$ \citep{Mitchell2021}. 
Therefore, the Ly$\alpha$ emission covered by our MUSE apertures (radius $\simeq0.75$~arcsec, corresponding to $r\lesssim4.5$~kpc at $z\simeq5-6$) is not likely significantly impacted by that from non-recombination origins or nearby sources. 

Finally, we quantify the Ly$\alpha$ velocity offset of the $79$ galaxies in our $4.9<z<6.5$ Ly$\alpha$-selected sample. 
The systemic redshifts are derived by fitting the H$\alpha$ emission lines in NIRCam grism spectra with Gaussian profiles. 
We measure the Ly$\alpha$ redshifts ($z_{{\rm Ly}\alpha}$) from the peak of the Ly$\alpha$ emission lines identified from VLT/MUSE and Keck/DEIMOS spectra. 
In order to be consistent with {\it JWST} measurements, we convert the wavelengths of MUSE and DEIMOS spectra from air to vacuum using the formula in \citet{Ryabchikova2015}\footnote{\url{https://www.astro.uu.se/valdwiki/Air-to-vacuum\%20conversion}}, and the Ly$\alpha$ redshifts are derived based on vacuum wavelengths. 
For the $79$ galaxies with Ly$\alpha$ and H$\alpha$ detections in our sample, their Ly$\alpha$ velocity offsets are from $+61$~km~s$^{-1}$ to $+725$~km~s$^{-1}$. 
We estimate the uncertainty of Ly$\alpha$ velocity offset following the similar way in estimating the Ly$\alpha$ flux uncertainty. 
For each galaxy, we resample the flux densities of its spectrum $1000$ times and take the standard deviation of Ly$\alpha$ velocity offsets derived from the $1000$ resampled spectra as the uncertainty. 
We note that the wavelength calibration for NIRCam grism spectra is also subject to an uncertainty of $\sim10-20$~\AA\ (Sun et al. in prep.), corresponding to a velocity uncertainty of $\simeq100$~km~s$^{-1}$ for F444W grism. 
This uncertainty randomly scatters the derived velocity offsets but does not systematically shift the velocity offsets to one direction. 
We summarize the Ly$\alpha$ velocity offsets of the $79$ Ly$\alpha$ emitters in our sample in Table~\ref{tab:lya_properties}. 
In the following section we will discuss the Ly$\alpha$ properties of systems in our Ly$\alpha$-selected sample at $4.9<z<6.5$.


\section{The Properties of Ly$\alpha$ Emitters at $z\simeq5-6$} \label{sec:lae_property}

Before considering the Ly$\alpha$ properties of the more general (Lyman break selected) population of $z\simeq5-6$ galaxies (Section~\ref{sec:lya_lbg}), we first use new {\it JWST} observations to investigate the subset known to show Ly$\alpha$ emission. 
We characterize the relationship of Ly$\alpha$ EW and Ly$\alpha$ escape fraction in our Ly$\alpha$ emitter sample in Section~\ref{sec:lya_ew_escape}, with particular interest in the systems with the largest escape fractions ($>0.5$) and largest Ly$\alpha$ EWs ($>100$~\AA). 
In Section~\ref{sec:lya_offset} and Section~\ref{sec:blue_peak}, we quantify the velocity structure of Ly$\alpha$ lines relative to the systemic redshift (determined from new H$\alpha$ redshifts) and discuss the potential impact of the IGM on the line profiles at $z\simeq5-6$.

\subsection{Galaxies with large Ly$\alpha$ EWs and Ly$\alpha$ escape fractions} \label{sec:lya_ew_escape}


\begin{figure*}
\begin{center}
\includegraphics[width=0.9\linewidth]{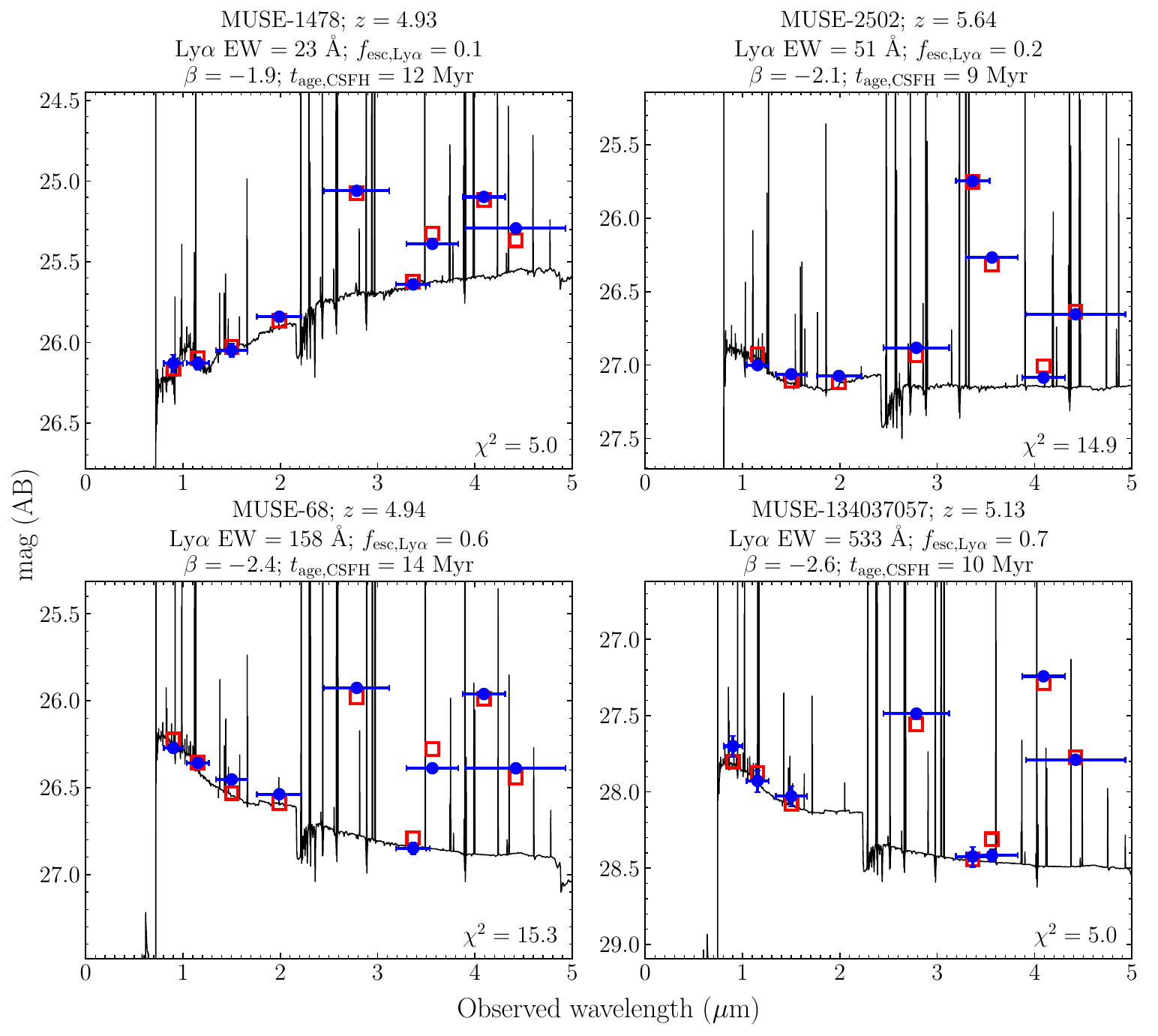}
\caption{JADES SEDs of example galaxies in the four Ly$\alpha$ EW bins (top two panels: Ly$\alpha$ EW $\simeq20$~\AA, $50$~\AA; bottom two panels: $150$~\AA, and $500$~\AA). Observed {\it JWST}/NIRCam photometry is shown by blue circles. The spectra (black lines) and synthetic photometry (red squares) are derived from the posterior median of \textsc{beagle} models. All the SEDs show large flux excesses with respect to the continuum in some of the NIRCam long wavelength filters (F277W, F335M, F356W, F410M, F444W), indicating intense rest-frame optical line emission (e.g., [O~{\scriptsize III}]+H$\beta$, H$\alpha$+[N~{\scriptsize III}]) often associated with very young stellar populations (CSFH age $\simeq10$~Myr). All these four galaxies present very efficient ionizing (and hence Ly$\alpha$) photon production inferred from \textsc{beagle} ($\xi_{\rm ion}\gtrsim10^{25.6}$~erg$^{-1}$~Hz), while the Ly$\alpha$ escape fraction varies with Ly$\alpha$ EW, with $f_{{\rm esc,Ly}\alpha}=0.1$ seen in the Ly$\alpha$ EW $\simeq20$~\AA\ system to $f_{{\rm esc,Ly}\alpha}=0.7$ at Ly$\alpha$ EW $\simeq500$~\AA.}
\label{fig:sed}
\end{center}
\end{figure*}


\begin{figure*}
\begin{center}
\includegraphics[width=\linewidth]{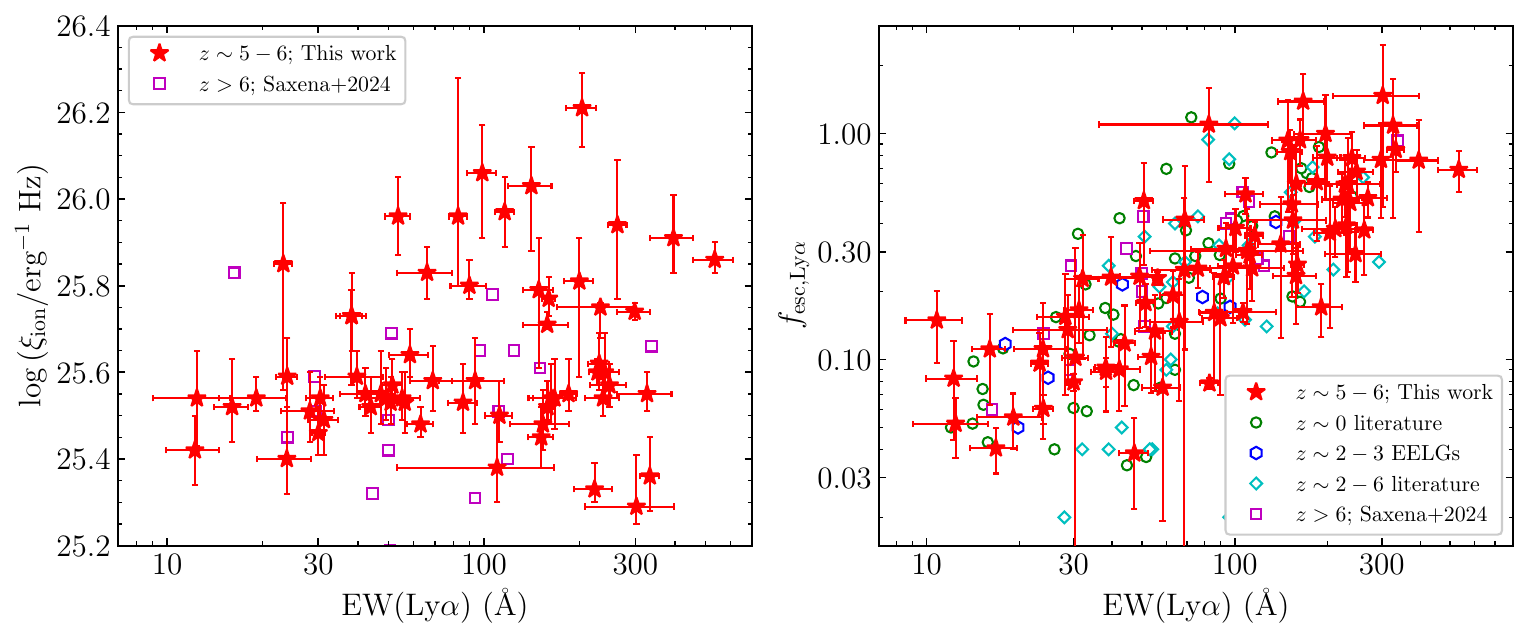}
\includegraphics[width=\linewidth]{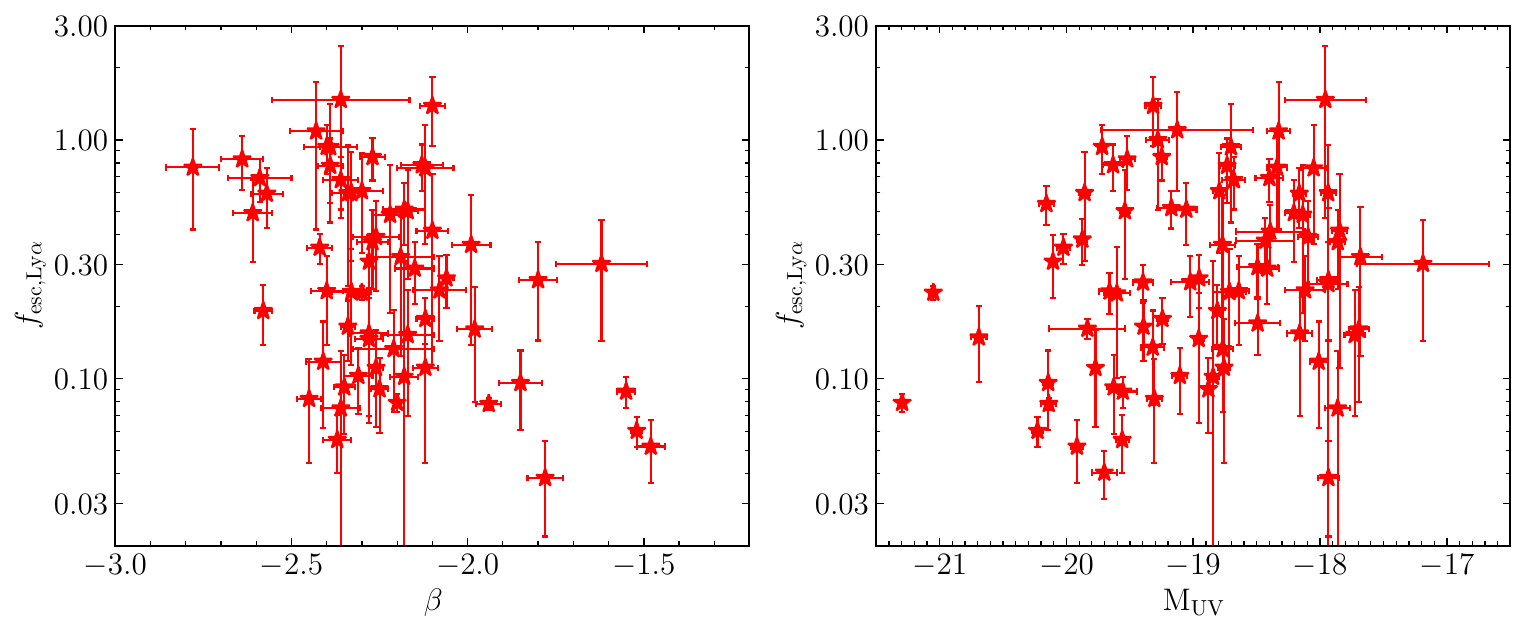}
\caption{Top left panel: ionizing photon production efficiency ($\xi_{\rm ion}$) inferred from \textsc{beagle} models versus Ly$\alpha$ EW for the $61$ Ly$\alpha$ emitters with H$\alpha$ detections and JADES NIRCam SEDs at $4.9<z<6.5$ (red stars). We find galaxies that are efficient in producing ionizing photons ($\xi_{\rm ion}>10^{25.6}$~erg$^{-1}$~Hz, hence with very large intrinsic Ly$\alpha$ EWs $>300$~\AA) span a wide range of observed Ly$\alpha$ EW from $\simeq20$~\AA\ to $400$~\AA. Top right panel: Ly$\alpha$ escape fraction (assuming case B recombination) as a function of Ly$\alpha$ EW for the $79$ Ly$\alpha$ emitters in our sample (red stars). We find a tight correlation between Ly$\alpha$ EW and Ly$\alpha$ escape fraction. As a comparison, we overplot $z\simeq0$ Green peas \citep{Yang2017,Jaskot2019} or metal-poor star-forming galaxies \citep{Izotov2024} (open green circles), extreme emission line galaxies (EELGs) at $z\sim2-3$ (open blue hexagons; \citealt{Tang2021,Tang2024}), Ly$\alpha$ emitters at $z\sim2-6$ (open cyan diamonds; \citealt{Matthee2021,Roy2023}), and Ly$\alpha$ emitters at $z\gtrsim6$ (open magenta squares; \citealt{Saxena2024}). Bottom left and right panels: Ly$\alpha$ escape fraction versus UV slope and M$_{\rm UV}$ for the $79$ Ly$\alpha$ emitters in our sample. Galaxies with larger Ly$\alpha$ escape fractions tend to be bluer systems.}
\label{fig:lyaew_xi_fesc}
\end{center}
\end{figure*}

Galaxies with extremely large Ly$\alpha$ EWs ($>100-300$~\AA) have recently been discovered at $z\gtrsim7$ \citep{Saxena2023,Chen2024}, providing signposts of ionized bubbles in the mostly neutral universe. 
If we are to link these sources to useful information about the early IGM, we will need to understand what separates the most extreme Ly$\alpha$ emitters (EW $>100-300$~\AA) from the more commonly-studied population with moderate-EW Ly$\alpha$ (EW $=10-50$~\AA). 
Our Ly$\alpha$ emitter sample spans from EW $=10$ to $500$~\AA\ (Fig.~\ref{fig:lyaew_muv_all}), allowing us to investigate this question at $z\simeq5-6$ where the impact of the IGM damping wing on Ly$\alpha$ is minimal.

Prior to interaction with the IGM, the Ly$\alpha$ EW is largely regulated by a combination of ionizing photon production (which sets the intrinsic luminosity of the line) and transmission of line photons through the interstellar medium (ISM) and the circumgalactic medium (CGM). 
On one hand, we may expect that the strongest Ly$\alpha$ emitters are simply those that are able to transmit most of their Ly$\alpha$ photons through the ISM and CGM.
Alternatively, the galaxies with Ly$\alpha$ EW $>100$~\AA\ may stand out as those with an extreme population of ionizing sources that are boosting the strength of the line relative to other galaxies \citep[e.g.,][]{Maseda2020,Maseda2023}. 
According to the stellar population synthesis models used in this paper (see Section~\ref{sec:sed}), the intrinsic Ly$\alpha$ EW will increase by a factor of $6$ between an age of $300$~Myr and $2$~Myr (here assuming CSFH). 
This change reflects the difference in the ionizing photon production efficiency (parameterized as $\xi_{\rm ion}$, the hydrogen ionizing photon production rate per unit intrinsic luminosity density at rest-frame $1500$~\AA) in these stellar populations, ranging from $\xi_{\rm ion}=10^{25.3}$~erg$^{-1}$~Hz at $300$~Myr to $\xi_{\rm ion}=10^{26.1}$~erg$^{-1}$~Hz at $2$~Myr. 
In this case, we would expect that the most extreme Ly$\alpha$ emitters stand out as the sources dominated by the youngest stellar populations.

We investigate these possibilities in the $61$ galaxies in our sample with NIRCam-based SEDs (which constrain age) and grism-based H$\alpha$ measurements (which constrain the Ly$\alpha$ escape fraction). 
To illustrate the trends, we first show four representative sources spanning the full range in Ly$\alpha$ EW, from $23$~\AA\ to $533$~\AA\ (see Fig.~\ref{fig:sed}). 
It is immediately apparent that the inferred age does not vary with the Ly$\alpha$ EW. 
In all four galaxies, we see young ages ($\simeq10$~Myr) linked with efficient ionizing production. 
In contrast, the Ly$\alpha$ escape fraction shows a clear trend, increasing from relatively modest transmission ($f_{{\rm esc,Ly}\alpha}=0.1-0.2$) in the two moderate-EW Ly$\alpha$ emitters to near-unity ($f_{{\rm esc,Ly}\alpha}=0.6-0.7$) in the two most extreme line emitters (EW $=150-500$~\AA). 
For these four galaxies, variations in line transmission (and not production efficiency) are what separates galaxies with moderate and extreme Ly$\alpha$ emission.

The full Ly$\alpha$ selected sample with H$\alpha$ detections shows a similar picture. 
We find a positive correlation between the Ly$\alpha$ EW and the Ly$\alpha$ escape fraction (top right panel of Fig.~\ref{fig:lyaew_xi_fesc}), with Ly$\alpha$ emitters with EW $=20$~\AA\ generally showing $f_{{\rm esc,Ly}\alpha}=0.1$, and those with EW $\simeq100-500$~\AA\ having $f_{{\rm esc,Ly}\alpha}=0.6$. 
This trend is qualitatively consistent with relations found in other samples at a variety of redshifts \citep[e.g.,][]{Yang2017,Jaskot2019,Izotov2024,Saxena2024} and indicates that the transmission of Ly$\alpha$ is likely playing a dominant role in regulating the Ly$\alpha$ EW. 
In Fig.~\ref{fig:lyaew_xi_fesc}, we show how galaxy properties vary with Ly$\alpha$ EW. 
In the top left panel, we see the derived ionizing photon production efficiency as a function of Ly$\alpha$ EW, with the results showing extremely young galaxies ($\lesssim20$~Myr) with Ly$\alpha$ EW spanning the full range in our sample ($20-500$~\AA). 
\citet{Saxena2024} also shows $z\gtrsim6$ galaxies with large ionizing photon production efficiencies (often associated with very young ages) spanning a wide range of Ly$\alpha$ EW ($\simeq20-300$~\AA). 
While young galaxies will have enhanced Ly$\alpha$ production, this does not guarantee they will be observed with the largest Ly$\alpha$ EWs. 

Our $z\simeq5-6$ sample includes $23$ galaxies with extremely large Ly$\alpha$ escape fractions ($f_{{\rm esc,Ly}\alpha}>0.5$). 
These systems appear to have blue UV slopes ($\beta<-2.1$), with a median $\beta=-2.4$ and the bluest UV slope down to $\beta=-2.8$ (bottom left panel of Fig.~\ref{fig:lyaew_xi_fesc}). 
Since UV slope is a tracer of dust attenuation \citep[e.g.,][]{Calzetti1994,Meurer1999}, the above finding indicates that Ly$\alpha$ emitters in our sample with enhanced Ly$\alpha$ transmission are likely associated with low dust content \citep[e.g.,][]{Verhamme2008,Hayes2011,Matthee2016,Lin2024}. 
The UV slope is also impacted by a variety of other properties (i.e., age, metallicity, ionizing photon escape), so it is possible that the blue colors are additionally driven by these factors \citep{Bouwens2010b,Chisholm2022,Topping2022,Topping2024}. 
We may also expect galaxies with large Ly$\alpha$ escape fractions to be fainter systems with very young stellar ages (e.g., $\lesssim10$~Myr). 
This is because large gaseous disks may have not yet developed in faint galaxies \citep[e.g.,][]{Erb2014} and the strong feedback associated with intense bursts can clear the pathway \citep[e.g.,][]{Kimm2019,Ma2020,Kakiichi2021}, both are conducive to the leakage of Ly$\alpha$ photons. 
While we do see that some of the galaxies with $f_{{\rm esc,Ly}\alpha}>0.5$ in our Ly$\alpha$ selected sample are faint (M$_{\rm UV}=-18$) with young CSFH ages ($2-10$~Myr), we also find $f_{{\rm esc,Ly}\alpha}>0.5$ among relatively bright (M$_{\rm UV}=-20$) systems (bottom right panel of Fig.~\ref{fig:lyaew_xi_fesc}) and galaxies with more evolved stellar populations (CSFH age $\simeq100$~Myr). 
We will discuss the dependence of Ly$\alpha$ transmission on galaxy properties for the entire Lyman break selected sample in Section~\ref{sec:lya_lbg} to explore whether very large Ly$\alpha$ escape fractions ($f_{{\rm esc,Ly}\alpha}>0.5$) are more frequent in a subset of systems (e.g., with bluer UV slopes, fainter M$_{\rm UV}$, or younger CSFH ages).

\subsection{Ly$\alpha$ peak velocity offsets of LAEs at $z\simeq5-6$} \label{sec:lya_offset}


\begin{figure}
\begin{center}
\includegraphics[width=\linewidth]{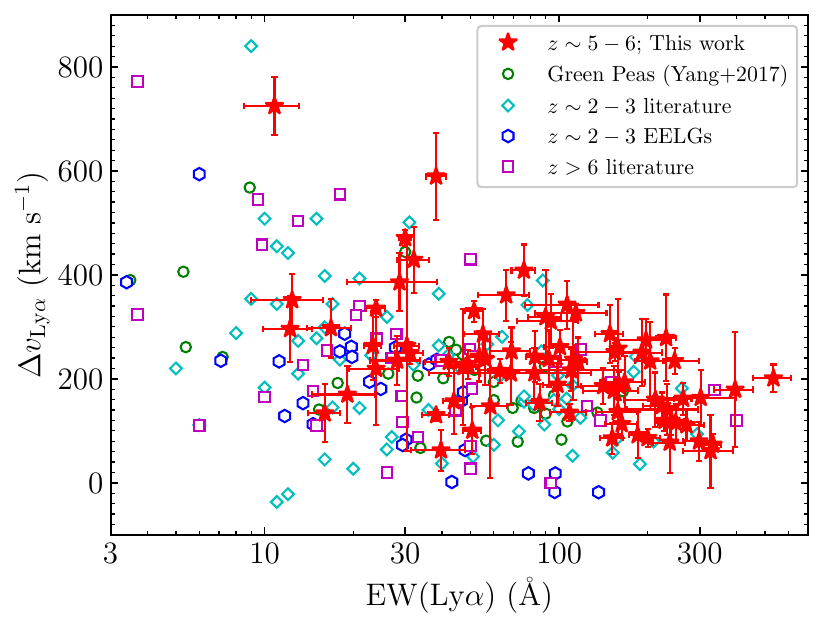}
\caption{Ly$\alpha$ velocity offset versus Ly$\alpha$ EW for the $79$ Ly$\alpha$ emitters with H$\alpha$ detections at $4.9<z<6.5$ (red stars). While our strong Ly$\alpha$ emitters at $4.9<z<6.5$ tend to have lower Ly$\alpha$ velocity offsets comparing to those with lower Ly$\alpha$ EWs, there are EW $>100$~\AA\ Ly$\alpha$ emitters presenting relatively large Ly$\alpha$ velocity offsets ($\gtrsim200$~km~s$^{-1}$) indicating dense neutral hydrogen along the sightlines. For comparison, we overplot literature data at lower and higher redshifts including Green Peas (open green circles; \citealt{Yang2017}), Ly$\alpha$ emitters at $z\simeq2-3$ (open cyan diamonds; \citealt{Finkelstein2011,McLinden2011,McLinden2014,Hashimoto2013,Erb2014,Matthee2021}), EELGs at $z\simeq2-3$ (open blue hexagons; \citealt{Tang2021,Tang2024}), and $z>6$ galaxies (open magenta squares; \citealt{Maiolino2015,Stark2015,Stark2017,Willott2015,Inoue2016,Pentericci2016,Bradac2017,Laporte2017,Mainali2017,Hashimoto2019,Hutchison2019,Endsley2022b,Bunker2023a,Tang2023,Saxena2024}).}
\label{fig:vo_lyaew}
\end{center}
\end{figure}


\begin{figure}
\begin{center}
\includegraphics[width=\linewidth]{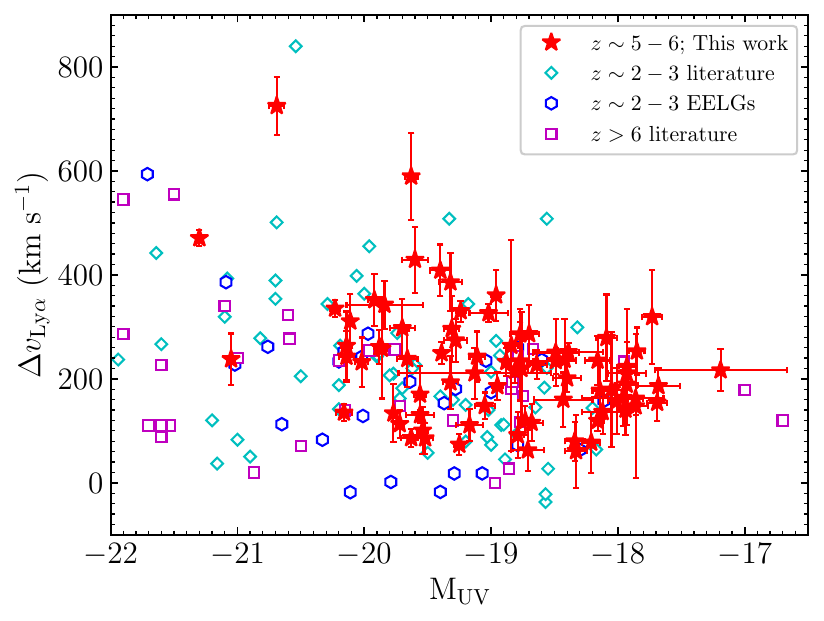}
\caption{Ly$\alpha$ velocity offset versus absolute UV magnitude for the $79$ Ly$\alpha$ emitters with H$\alpha$ detections at $4.9<z<6.5$ (red stars). We overplot literature data as a comparison, including Ly$\alpha$ emitters at $z\simeq2-3$ (open cyan diamonds; \citealt{Erb2014,Matthee2021}) and EELGs at $z\simeq2-3$ (open blue hexagons; \citealt{Tang2021,Tang2024}), as well as $z>6$ galaxies (open magenta squares; \citealt{Maiolino2015,Stark2015,Stark2017,Willott2015,Inoue2016,Pentericci2016,Bradac2017,Laporte2017,Mainali2017,Hashimoto2019,Hutchison2019,Endsley2022b,Bunker2023a,Tang2023,Saxena2024}).}
\label{fig:vo_muv}
\end{center}
\end{figure}

The Ly$\alpha$ velocity profile provides insight to the escape of Ly$\alpha$ through the ISM and CGM \citep[e.g.,][]{Dijkstra2017,Blaizot2023,Almada-Monter2024}. 
Trends between the Ly$\alpha$ velocity offset and galaxy properties have been established in Ly$\alpha$ emitter samples at $z\simeq0.3-2$ \citep[e.g.,][]{Erb2014,Hashimoto2015,Yang2017,Tang2024}. 
Galaxies with the largest EW Ly$\alpha$ emission at these redshifts are generally found with the smallest Ly$\alpha$ velocity offsets \citep[e.g.,][]{Erb2014,Tang2021,Tang2024}, likely reflecting lower H~{\small I} column densities which in turn allow Ly$\alpha$ to escape without significant diffusion to large velocities \citep[e.g.,][]{Verhamme2006,Verhamme2015}. 
When Ly$\alpha$ is seen to emerge near the line center ($\lesssim100$~km~s$^{-1}$), it may point to situations where Ly$\alpha$ emerges through low density channels that are optically thin to Lyman continuum (LyC) emission \citep[e.g.,][]{Behrens2014,Verhamme2015,Dijkstra2016,Naidu2017,Rivera-Thorsen2017,Izotov2021,Choustikov2024b}. 
Prior to the launch of {\it JWST}, Ly$\alpha$ velocity profiles at $z\gtrsim5$ could only be constrained in a few sources with systemic redshifts measured from non-resonant UV lines (i.e., C~{\small III}]) or far-infrared lines ([C~{\small II}]~$158\ \mu$m) \citep[e.g.,][]{Stark2017,Cassata2020,Matthee2020,Endsley2022b}. As such samples are small, it has long been unclear how common low velocity offsets are in $z\simeq5-6$ galaxies.

We explore the velocity profiles at $z\simeq5-6$ using the $79$ galaxies in our sample with Ly$\alpha$ and H$\alpha$ measurements. 
As expected, the velocity offset decreases with increasing Ly$\alpha$ EW (Fig.~\ref{fig:vo_lyaew}). 
In the weakest Ly$\alpha$ emitters in our sample (EW $=10-30$~\AA), we find large velocity offsets (median $\Delta v_{{\rm Ly}\alpha}=300$~km~s$^{-1}$). 
These decrease to a median $\Delta v_{{\rm Ly}\alpha}=180$~km~s$^{-1}$ in the strongest Ly$\alpha$ emitters (EW $>100$~\AA). 
Notably absent in our $z\simeq5-6$ sample are galaxies with very low Ly$\alpha$ velocity offsets. 
Among the $39$ strongest Ly$\alpha$ emitters (EW $>100$~\AA) in our sample, only $7$ galaxies present low Ly$\alpha$ velocity offsets with $\Delta v_{{\rm Ly}\alpha}<100$~km~s$^{-1}$ and none of them shows $\Delta v_{{\rm Ly}\alpha}<50$~km~s$^{-1}$. 
On the contrary, the strongest Ly$\alpha$ emitters (EW $\gtrsim100$~\AA) at $z\simeq2-3$ occasionally show Ly$\alpha$ peaks closer to the line center ($\Delta v_{{\rm Ly}\alpha}<100$~km~s$^{-1}$; e.g., \citealt{Erb2014,Rivera-Thorsen2017,Matthee2021,Tang2024}).
Among galaxies with large [O~{\small III}]+H$\beta$ EWs ($>600$~\AA), existing data suggests the median Ly$\alpha$ peak velocity offset of large EW Ly$\alpha$ emitters at $z\simeq2-3$ is $20$~km~s$^{-1}$ \citet{Tang2024}, considerably lower than that of strong Ly$\alpha$ emitters at $z\simeq5-6$ ($\Delta v_{{\rm Ly}\alpha}=180$~km~s$^{-1}$) with the same [O~{\small III}]+H$\beta$ EWs.
This may indicate significant evolution in Ly$\alpha$ peak velocities at $2<z<6$, with the small offset sources linked to LyC leakage (though with scatter; \citealt{Pahl2024}) disappearing as we enter the reionization era. 

Previous work has demonstrated that Ly$\alpha$ velocity offsets are often largest in the most massive galaxies \citep[e.g.,][]{Erb2014,Shibuya2014,Stark2017,Mason2018a,Endsley2022b}, likely reflecting larger reservoirs of neutral gas through which Ly$\alpha$ photons must escape. 
This trend is also seen when looking at absolute UV magnitude. 
Luminous galaxies have been found with very large velocity offsets ($\Delta v_{{\rm Ly}\alpha}=300-800$~km~s$^{-1}$) at lower redshifts ($z\simeq2-3$; e.g., \citealt{Erb2014}) and at $z>6$ \citep[e.g.,][]{Endsley2022b,Bunker2023a,Tang2023}. 
In Fig.~\ref{fig:vo_muv}, we plot the Ly$\alpha$ velocity offset versus absolute UV magnitude for both our Ly$\alpha$ selected sample at $4.9<z<6.5$ and Ly$\alpha$ emitters over different redshifts from literature. 
While luminous galaxies (M$_{\rm UV}<-20.5$) are few (three galaxies) in our sample, they appear to show the largest velocity offsets (median $\Delta v_{{\rm Ly}\alpha}=470$~km~s$^{-1}$). 
On the other hand, the less luminous systems (M$_{\rm UV}=-20$ to $-18$) have lower Ly$\alpha$ velocity offsets with a median value of $\Delta v_{{\rm Ly}\alpha}=220$~km~s$^{-1}$, consistent with the $\Delta v_{{\rm Ly}\alpha}-$ M$_{\rm UV}$ trend seen in literature \citep[e.g.,][]{Prieto-Lyon2023}. 
Based on this $z\simeq5-6$ baseline sample, we may expect luminous galaxies to be easier to detect in Ly$\alpha$ at $z\gtrsim7$ as the large velocity Ly$\alpha$ emission faces less attenuation from the neutral IGM \citep[e.g.,][]{Stark2017,Mason2018b,Hashimoto2019,Endsley2022b}. 

To better demonstrate the shift in Ly$\alpha$ velocity profiles over $2\lesssim z \lesssim 6$, we create a composite Ly$\alpha$ spectrum of our $z\simeq5-6$ galaxies. 
We stack the individual spectra following the procedures described in \citet{Tang2024} which we summarize below. 
Our goal is to identify the velocity profile associated with the strongest Ly$\alpha$ emitters, so we create a composite for those systems in our sample with Ly$\alpha$ EW $>40$~\AA. 
To create the composite spectrum, we shift individual spectra (after converting air wavelengths to vacuum) to the rest-frame using the systemic redshifts inferred from H$\alpha$ or [O~{\small III}]~$\lambda5007$ lines. 
Then we interpolate each spectrum to a common wavelength scale with a bin size of $0.2$~\AA\ in rest-frame, which is larger than the wavelength bin size of each individual spectrum. 
We next normalize each individual spectrum using its measured Ly$\alpha$ flux. 
The individual spectra are stacked by median-combining the individual flux densities in each wavelength bin. 
Finally, we convert the rest-frame wavelengths to the velocity space ($\Delta v=c(\lambda_{\rm rest}-\lambda_{{\rm Ly}\alpha})/\lambda_{{\rm Ly}\alpha}$, where $c$ is the speed of light and $\lambda_{{\rm Ly}\alpha}=1215.67$~\AA) to illustrate the velocity profile of Ly$\alpha$ emission line. 
We compare this to line profiles of $z\simeq2-3$ galaxies matched in [O~{\small III}]+H$\beta$ EW and described in \citep{Tang2024}.
That paper presents resolved ($R\sim3900$) Ly$\alpha$ spectroscopy of $z\simeq2-3$ galaxies with [O~{\small III}]+H$\beta$ EW $>600$~\AA\ using Binospec \citep{Fabricant2019} on the MMT. 
For details, the reader is directed to \citet{Tang2024}. 

\begin{figure}
\begin{center}
\includegraphics[width=\linewidth]{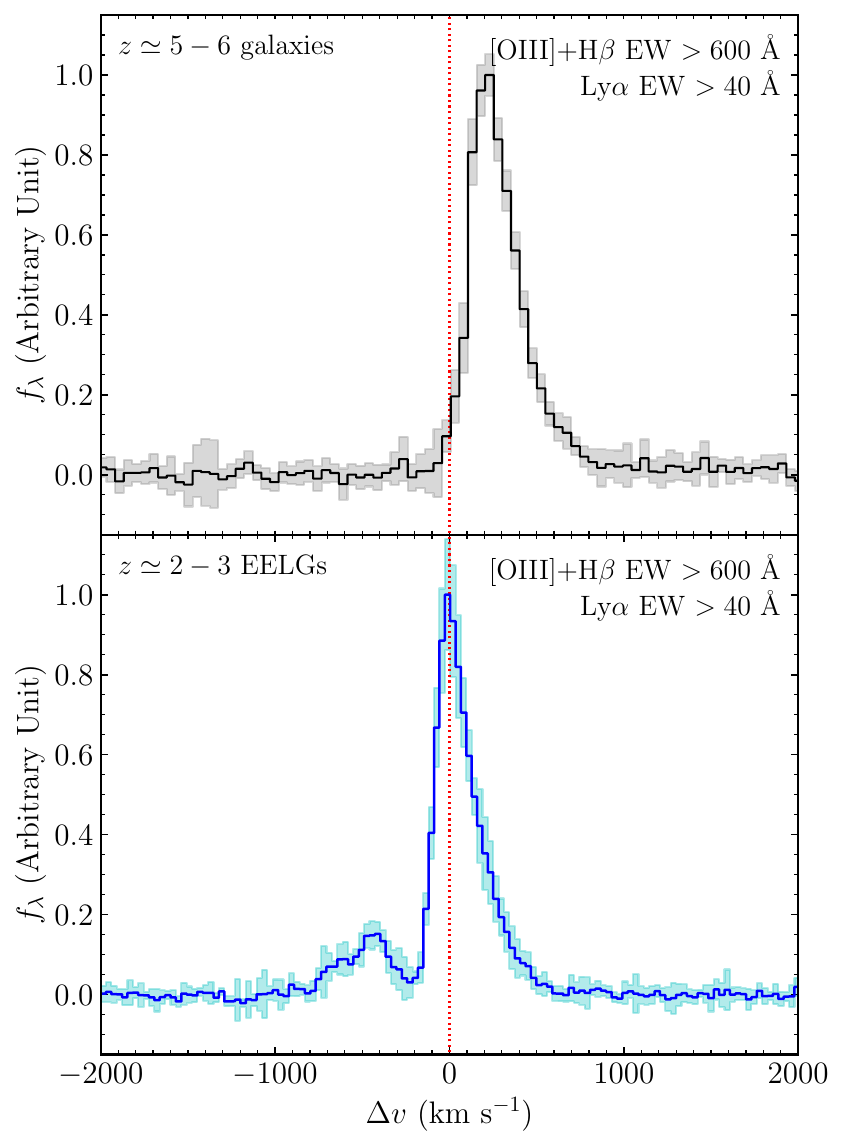}
\caption{Comparison of composite Ly$\alpha$ profile between $z\simeq5-6$ galaxies (top panel, black line) and $z\simeq2-3$ galaxies (bottom panel, blue line; \citealt{Tang2024}). The $1\sigma$ variations are shown as the shaded regions. We compare galaxies selected with the same criteria ([O~{\scriptsize III}]+H$\beta$ EW $>600$~\AA\ and Ly$\alpha$ EW $>40$~\AA; \citealt{Tang2024}) in both samples. The red dotted line shows $\Delta v_{{\rm Ly}\alpha}=0$~km~s$^{-1}$ derived using the systemic redshifts inferred from H$\alpha$ or [O~{\scriptsize III}]~$\lambda5007$ lines. While the composite Ly$\alpha$ profile at $z\simeq2-3$ presents a significant fraction of Ly$\alpha$ flux emitted at the line center ($\sim50$~per~cent) and a blue peak, the bulk of the Ly$\alpha$ flux of $z\simeq5-6$ galaxies is shifted to redder wavelengths with a peak at $\Delta v_{{\rm Ly}\alpha}\simeq230$~km~s$^{-1}$.}
\label{fig:comp_spec}
\end{center}
\end{figure}

The composite Ly$\alpha$ profile of our $z\simeq5-6$ Ly$\alpha$ emitters is shown in the top panel of Fig.~\ref{fig:comp_spec} as the black solid line. 
We see its Ly$\alpha$ peak at a relatively large velocity offset ($\Delta v_{{\rm Ly}\alpha}=230$~km~s$^{-1}$), similar to that of many of the individual spectra shown in Fig~\ref{fig:vo_lyaew}. 
We find only a small portion of emission at the line center -- the fraction of Ly$\alpha$ flux emitted within $\pm100$~km~s$^{-1}$ of the systemic redshift (the so-called ``Ly$\alpha$ central escape fraction''; e.g., \citealt{Naidu2022}) is $9$~per~cent. 
The result suggests that the strongest Ly$\alpha$ emitters (EW $>40$~\AA) at $z\simeq5-6$ may be uniformly covered by dense neutral hydrogen ($N_{\rm HI}\simeq10^{19}-10^{20}$~cm$^{-2}$ assuming an expanding shell model, e.g., \citealt{Hashimoto2015,Verhamme2015}, or equivalently the total H~{\small I} column density $\simeq10^{19}-10^{20}$~cm$^{-2}$ assuming a clumpy slab model, e.g., \citealt{Li2022b}). 
Such H~{\small I} column densities are optically thick to LyC emission \citep[e.g.,][]{Verhamme2015}, indicating that strong Ly$\alpha$ emitters at $z>5$ may have neutral gas distributions that do not facilitate large LyC escape fractions. 
This is significantly different from lower redshifts where strong Ly$\alpha$ emitters are closely linked to LyC leakage \citep[e.g.,][]{Steidel2018,Fletcher2019,Flury2022,Naidu2022,Pahl2023} and have Ly$\alpha$ profiles that reveal low H~{\small I} column densities \citep[e.g.,][]{Erb2014,Rivera-Thorsen2017,Vanzella2018,Jaskot2019,Izotov2021,Matthee2021}. 
This is clearly seen in the bottom panel of Fig.~\ref{fig:comp_spec}, where the composite shows strong Ly$\alpha$ peaking near the line center, with half of the flux emitted within $\pm100$~km~s$^{-1}$ of the systemic redshift \citep{Tang2024}.


Naively we may interpret the disappearance of strong Ly$\alpha$ emitters with small peak velocity offsets ($<100$~km~s$^{-1}$) and large Ly$\alpha$ central escape fractions ($\gtrsim20$~per~cent) as evidence that large LyC escape fractions ($f_{\rm esc,LyC}\gtrsim0.2$; e.g., \citealt{Naidu2022,Choustikov2024b}) are becoming less common as we enter the reionization era. 
But more likely we are seeing the imprint of the IGM on Ly$\alpha$ profiles at $z\simeq5-6$. 
Even if the IGM is mostly ionized at these redshifts, the IGM density is high enough for the residual neutral hydrogen ($x_{\rm HI}\gtrsim10^{-5}-10^{-4}$; e.g., \citealt{Yang2020b,Bosman2022}) to efficiently scatter Ly$\alpha$ photons near the line center. 
The transmission ($\mathcal{T}$) at $\Delta v_{{\rm Ly}\alpha}=0$~km~s$^{-1}$ is expected to be negligible ($\mathcal{T}\simeq0.16$ at $z\sim5$ and $\mathcal{T}\simeq0.01$ at $z\sim6$; e.g., \citealt{Inoue2014,Becker2015,Bosman2018,Eilers2018,Yang2020b,Bosman2022}), effectively attenuating any Ly$\alpha$ peaking near systemic. 
If the IGM is infalling onto galaxies, there is likely further scattering of Ly$\alpha$ photons within $\sim100$~km~s$^{-1}$ on the red side of the line center \citep[e.g.,][]{Santos2004,Dijkstra2007,Laursen2011,Mason2018a}. 
We note that if $z>5$ galaxies occasionally reside in underdense regions or quasar proximity zones the transmission near the line center may be boosted \citep[e.g.,][]{Bosman2020,Mason2020}. 
But in general the IGM effects will make it difficult at $z\simeq5-6$ to recover Ly$\alpha$ emission with velocity profiles linked to LyC leakage (i.e., those with small velocity offsets).

At lower redshifts, Ly$\alpha$ emission provides one of the best indicators of LyC emission. 
While it has always been clear that the IGM damping wing will make the connection between 
Ly$\alpha$ and LyC emission less useful at $z\gtrsim7$, the results presented in this paper suggest that the utility of Ly$\alpha$ as a probe of ionizing photon leakage is likely to also be limited at $z\simeq5-6$. 
If galaxies at these redshifts with the largest escape fractions have ionized channels that facilitate direct escape of Ly$\alpha$ at the line center (or at small positive velocity offsets), the IGM will strongly scatter their Ly$\alpha$ emission, making them unlikely to enter Ly$\alpha$ emitter samples. 
They will be identifiable in continuum-selected samples, likely with properties similar to those of leakers at lower redshifts (e.g., blue UV colors, low masses; e.g., \citealt{Chisholm2022,Flury2022,Kim2023,Pahl2023}). 
But because their Ly$\alpha$ is heavily attenuated by the IGM, we will need to rely on other techniques to more clearly reveal this population as strong LyC leakers. 
Efforts to detect Mg~{\small II} emission \citep[e.g.,][]{Henry2018,Chisholm2020,Xu2022,Xu2023} and low ionization absorption lines \citep[e.g.,][]{Heckman2001,Shapley2003,Erb2015,Reddy2016,Chisholm2018,Steidel2018,Saldana-Lopez2022} are challenging, but they provide the best path to selecting galaxies with conditions that are conducive to escape of ionizing radiation. 
UV slopes provide another promising indicator \citep[e.g.,][]{Bouwens2010b,Ono2010,Raiter2010,Robertson2010,Chisholm2022,Topping2022,Kim2023}. 
Recent cosmological hydrodynamical simulation also suggests composite indicators for predicting LyC escape fraction \citep{Choustikov2024a}. 

We next consider implications of the velocity offsets in the strong Ly$\alpha$ emitters at $z\simeq5-6$. 
The presence of prominent Ly$\alpha$ emission (EW $>100$~\AA) in galaxies with large velocity offsets ($>200$~km~s$^{-1}$) suggests there must be a population that is able to transmit a large fraction of Ly$\alpha$ emission through fairly large column densities of neutral hydrogen. 
We suggest this may be possible in cases where the dust content is low enough for scattered Ly$\alpha$ emission to emerge through H~{\small I} without being absorbed \citep[e.g.,][]{Stark2010,Hayes2011,Matthee2016}. 
Given the low stellar masses (median $4\times10^7 M_{\odot}$) and very blue UV slopes (median $\beta=-2.3$) in the EW $>100$~\AA\ Ly$\alpha$ emitters, it seems plausible that we may be seeing such an effect in the $z\simeq5-6$ sample. 
This could be further amplified if the CGM is compact, allowing more of the Ly$\alpha$ to appear centrally concentrated and higher in surface brightness. 
Deep rest-frame UV spectroscopy have hinted that neutral gas may indeed be more compact at higher redshifts \citep{Jones2012}. 
There are examples of strong Ly$\alpha$ emitters with large velocity offsets at lower redshifts (e.g., XLS-6 and XLS-24 in \citealt{Matthee2021}), and indeed they tend to have very blue UV slopes ($\beta \lesssim -2.5$) suggesting minimal absorption of Ly$\alpha$ photons. 
Such galaxies should become more common at $z\simeq5-6$ as the population shifts toward lower mass galaxies with bluer UV slopes \citep[e.g.,][]{Topping2024}. 

Finally we investigate the Ly$\alpha$ velocity offset distribution of strong Ly$\alpha$ emitters (EW $>50$~\AA) at $z\simeq5-6$. 
We fit the probability density function of the Ly$\alpha$ velocity offsets of the $57$ Ly$\alpha$ emitters with EW $>50$~\AA\ in our sample with a truncated Gaussian distribution, which accounts for the disappearance of galaxies with very low Ly$\alpha$ velocity offsets ($\Delta v_{{\rm Ly}\alpha}\lesssim60$~km~s$^{-1}$) at $z\simeq5-6$:
\begin{eqnarray*}
p(\Delta v) &=& A\cdot \exp{[-(\Delta v-\mu)^2/(2\sigma^2)]}\ \ ({\rm when}\ \Delta v\ge\Delta v_{\rm cut}) \\
p(\Delta v) &=& 0\ \ ({\rm when}\ \Delta v<\Delta v_{\rm cut}).
\end{eqnarray*}
Where $A$, $\mu$, and $\sigma$ are the amplitude, mean, and standard deviation of the Gaussian distribution, and $\Delta v_{\rm cut}$ is the minimum Ly$\alpha$ velocity offset at $z\simeq5-6$. 
We fit the distribution using a Bayesian approach, considering uniform priors: $\mu=50-500$~km~s$^{-1}$, $\sigma=20-200$~km~s$^{-1}$, and $\Delta v_{\rm cut}=0-100$~km~s$^{-1}$. 
Then we derive the posterior probability distributions of the above four parameters using the \textsc{emcee} package \citep{Foreman-Mackey2013}. 
We find that the Ly$\alpha$ velocity offset distribution of strong Ly$\alpha$ emitters at $z\simeq5-6$ can be described by a mean value of $\Delta v_{{\rm Ly}\alpha}=199$~km~s$^{-1}$, a standard deviation $\sigma=82$~km~s$^{-1}$, and a minimum offset at $\Delta v_{\rm cut}=61$~km~s$^{-1}$. 
This demonstrates that only $33$~per~cent of the EW $>50$~\AA\ Ly$\alpha$ emitters at $z\simeq5-6$ show relatively small Ly$\alpha$ velocity offsets with $\Delta v_{{\rm Ly}\alpha}=60-150$~km~s$^{-1}$. The Ly$\alpha$ velocity offset distribution at $z\simeq5-6$ will be useful for predicting how frequently we might expect to find Ly$\alpha$ emitters with low velocity offsets in ionized bubbles at $z\gtrsim7$. 
We note this distribution would likely be different in more luminous galaxies at this redshift \citep{Endsley2022b}. 
We will discuss this in Section~\ref{sec:discussion}.

As a comparison with our $z\gtrsim 5$ measurements, we derive the Ly$\alpha$ velocity offset distribution of $z\simeq2$ Ly$\alpha$ emitters from \citet{Matthee2021} and \citet{Tang2024}, requiring the same Ly$\alpha$ EWs ($>50$~\AA) and [O~{\small III}]+H$\beta$ EWs ($>600$~\AA) as our $z\simeq5-6$ sample. 
Following the approach described above, we find that the mean and standard deviation of the Ly$\alpha$ velocity offset distribution at $z\simeq2$ are $116$~km~s$^{-1}$ and $100$~km~s$^{-1}$, respectively. 
This indicates that $63$~per~cent of the EW $>50$~\AA\ Ly$\alpha$ emitters at $z\simeq2$ show small Ly$\alpha$ velocity offsets with $\Delta v_{{\rm Ly}\alpha}<150$~km~s$^{-1}$ and even $25$~per~cent show $\Delta v_{{\rm Ly}\alpha}$ close to the systemic redshift ($<50$~km~s$^{-1}$). 
This distribution suggests the disappearance of galaxies with very small Ly$\alpha$ velocity offsets from $z\simeq2$ to $z\simeq5-6$, consistent with the trend shown in Fig.~\ref{fig:vo_lyaew}. 
Further data at $z\simeq2-3$ targeting extreme line emitters will be useful for better establishing the evolution in Ly$\alpha$ line profiles. 

\subsection{The strength of blue peaks in Ly$\alpha$ emission at $z\simeq5-6$} \label{sec:blue_peak}


\begin{figure}
\begin{center}
\includegraphics[width=\linewidth]{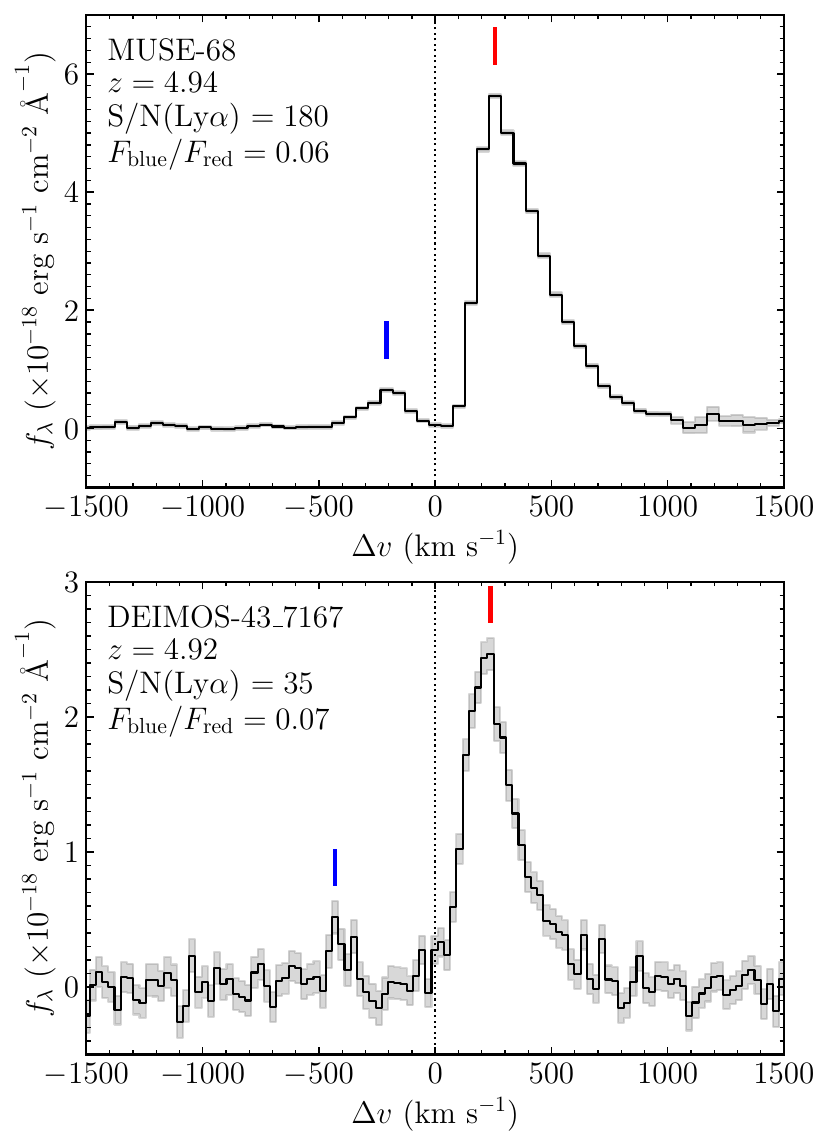}
\caption{Ly$\alpha$ spectra of the two Ly$\alpha$ emitters with blue peaked Ly$\alpha$ emission in our $z\simeq5-6$ sample (top panel: MUSE-68; bottom panel: DEIMOS-43\_7167). The spectra are shown in the velocity space using the systemic redshifts measured from H$\alpha$ emission lines. The error spectra are presented by grey shaded regions. The blue and red peaked Ly$\alpha$ emission are marked by the blue and red lines. These two galaxies are at the lowest redshift among our sample ($z\simeq4.9$) with the highest S/N Ly$\alpha$ detections (S/N $=180$ and $35$). The blue-to-red flux ratios of these two galaxies are $0.06-0.07$.}
\label{fig:blue_peak_lya}
\end{center}
\end{figure}

At lower redshifts ($z\lesssim 5$), Ly$\alpha$ profiles generally show a weak peak blueward of the systemic redshift \citep[e.g.,][]{Jaskot2014,Henry2015,Yang2017,Gazagnes2020,Izotov2020,Hayes2021,Matthee2021}. 
Such blue peak Ly$\alpha$ emission usually originates from Ly$\alpha$ photons that have diffused through the near side of the outflowing gas \citep[e.g.,][]{Verhamme2006,Verhamme2015,Gronke2016,Ouchi2020,Li2022b}. 
At $z>5$, due to the high IGM density, the blue peak Ly$\alpha$ emission should be strongly attenuated by resonant scattering from residual neutral hydrogen even in regions that have been reionized \citep{Gunn1965}. 
However, potential blue peak detections have been presented in four galaxies at $z\gtrsim6$ \citep{Hu2016,Matthee2018,Songaila2018,Bosman2020,Meyer2021}. 
This implies that these systems may be surrounded by intergalactic gas with a higher ionized hydrogen fraction than is typical, perhaps due to hard local radiation fields \citep[e.g.,][]{Mason2020,Torralba-Torregrosa2024}. 
Another potential explanation is that these galaxies have atypically strong blue Ly$\alpha$ emission due to inflowing gas \citep[e.g.,][]{Trebitsch2016,Ao2020,Li2022a}. 
How commonly blue-sided Ly$\alpha$ emission appears at $z\simeq5-6$ (when the IGM is mostly ionized) is not yet known, making it challenging to physically interpret any detections that emerge at $z\gtrsim6$.

We characterize blue peaks in the $79$ Ly$\alpha$ emitters with H$\alpha$-based systemic redshifts at $z\simeq5-6$.
Here we aim to constrain the strength of Ly$\alpha$ emission on the blue side of the line center. 
To do this, we quantify the ratio of Ly$\alpha$ that is blueward of the systemic redshift to that redward (hereafter the blue-to-red flux ratio), by integrating the Ly$\alpha$ flux at velocities $-1000$ to $0$~km~s$^{-1}$ of the Ly$\alpha$ resonance and at $0$ to $+1000$~km~s$^{-1}$.
The integration range is chosen to be similar with the literature \citep[e.g.,][]{Matthee2018,Matthee2021,Hayes2021}. 

Most of our galaxies ($77$ of $79$) present a single Ly$\alpha$ peak with no emission blueward of systemic. 
We derive the upper limits of the blue-to-red Ly$\alpha$ flux ratio for these $77$ single-peaked Ly$\alpha$ emitters, which are uniformly $<0.15$ ($5\sigma$). 
As a comparison, we consider the Ly$\alpha$ profiles of Ly$\alpha$ emitters at $z<0.44$ \citep{Hayes2021} and $z\sim2$ \citep{Matthee2021,Tang2024} where the IGM is mostly transparent to blueshifted Ly$\alpha$ photons. 
For those with similar Ly$\alpha$ EWs ($\simeq20-300$~\AA\ with a median EW $\simeq80$~\AA) to our $z\simeq5-6$ sample, the blue peak Ly$\alpha$ emission at $z\simeq0-2$ is much more prominent, with an average blue-to-red flux ratio $\simeq0.3$. 
This suggests that the galaxies showing strong blue peak Ly$\alpha$ emission at low redshift ($z\simeq0-2$) disappear at the tail of reionization ($z\simeq5-6$), as would be expected from the increasing IGM opacity at earlier times \citep[e.g.,][]{Madau1995,Inoue2014,Eilers2018,Yang2020b}. 

To further illustrate the evolution of blue-sided Ly$\alpha$ emission, we generate a composite Ly$\alpha$ spectrum for all the $79$ Ly$\alpha$ emitters with H$\alpha$ detections in our sample, following the same procedures described in Section~\ref{sec:lya_offset}. 
In the composite, we still do not detect any emission line feature blueward the systemic redshift, with an implied blue-to-red flux ratio $<0.04$ ($5\sigma$). 
Comparing with the average Ly$\alpha$ profiles at $z\simeq0-2$ \citep{Hayes2021,Matthee2021,Tang2024} with matched Ly$\alpha$ EWs, the average blue-to-red flux ratio of our $z\simeq5-6$ sample is less than $13$~per~cent of that at $z\simeq0-2$ ($\simeq0.3$). 
This is consistent with the decline of the IGM transmission at the Ly$\alpha$ resonance with redshift, from $\mathcal{T}=0.9-1$ at $z\simeq0-2$ to just $\mathcal{T}=0.16$ ($z=5$) and $\mathcal{T}=0.01$ ($z=6$) assuming the \citet{Inoue2014} IGM transmission model. 
Given the typically low S/N with which blue peaks are detected at $z\simeq2-3$, it is not surprising that we do not see them at $z\simeq5-6$ if the IGM is only transmitting at most $10$~per~cent of the line emission.

While most of our Ly$\alpha$ emitters show a single, redshifted Ly$\alpha$ peak, we do identify two $z\gtrsim5$ double-peaked Ly$\alpha$ emitting galaxies (MUSE-68 and DEIMOS-43\_7167) with a peak blueward of the line center (Fig.~\ref{fig:blue_peak_lya}). 
These two systems are at the lowest redshift ($z=4.9$) in our sample, where the IGM opacity is likely the lowest. 
This increases the transmission of blue peak Ly$\alpha$ line relative to other sources in our sample, making blue peaks easier to detect.
The blue peaks in both galaxies are much weaker than the red peaks, with blue-to-red flux ratios of $0.06$ (MUSE-68; top panel of Fig.~\ref{fig:blue_peak_lya}) and $0.07$ (DEIMOS-43\_7167; bottom panel of Fig.~\ref{fig:blue_peak_lya}). 
Given typical transmission factors at $z=4.9$ ($\mathcal{T}=0.16$; \citealt{Inoue2014}), we would expect intrinsic blue-to-red flux ratios of $\simeq0.4$ for the two sources. 
These intrinsic values are consistent with average ratios at $z\simeq0-2$, suggesting that these sources are not atypical in their blue peak flux fractions. 
The two galaxies with blue peaks also have the largest S/N among Ly$\alpha$ detections in our sample, with S/N(Ly$\alpha$) $=180$ for MUSE-68 (whose Ly$\alpha$ spectrum is obtained from the deepest MUSE field used in this work, MXDF) and S/N(Ly$\alpha$) $=35$ for DEIMOS-43\_7167. In contrast, most of the other sources in our sample have S/N(Ly$\alpha$) $<25$. 
The visibility of blue Ly$\alpha$ peaks of these two sources may primarily reflect their low redshifts (maximizing IGM transmission) and the S/N of the spectra. 
Finally we note that the two galaxies with blue peaks in our sample have $5-10\times$ lower blue-to-red flux ratios than those presented in the literature at $z\gtrsim6$ ($=0.3-0.7$; \citealt{Hu2016,Matthee2018,Songaila2018,Bosman2020,Meyer2021}). 
The four galaxies with blue peak Ly$\alpha$ detections in the literature at $z\gtrsim6$ are relatively brighter with M$_{\rm UV}<-21$.
The absence of very prominent blue peaks in our sample suggests that these very large blue-to-red flux ratios must be relatively rare in the faint $z\simeq5-6$ galaxies which dominate our sample. 
It is conceivable they become more common in more luminous galaxies where our statistics are currently limited.


\section{L\lowercase{y}$\alpha$ Properties in Lyman Break Galaxies} \label{sec:lya_lbg}

In Section~\ref{sec:lae_property}, we have demonstrated that galaxies with large EW ($>100$~\AA) Ly$\alpha$ emission and high Ly$\alpha$ transmission ($f_{{\rm esc,Ly}\alpha}>0.5$) are common among the Ly$\alpha$ selected sample at $z\simeq5-6$. 
We showed the strongest Ly$\alpha$ emitters tend to have relatively large velocity offsets, reflecting significant column densities of H~{\small I}.
In this section, we seek to explore how frequently the typical, Lyman break galaxies (LBGs) at $z\sim5-6$ present large Ly$\alpha$ EWs and transmission by establishing the Ly$\alpha$ EW and Ly$\alpha$ escape fraction distributions. 
We present the selection of $z\sim5-6$ galaxies using Lyman-break techniques in Section~\ref{sec:lbg}. 
Then we describe the methodology of constructing Ly$\alpha$ EW and escape fraction distributions in Section~\ref{sec:dist_method}. 
Finally, we present the distributions of Ly$\alpha$ properties of our LBGs at $z\sim5-6$ in Section~\ref{sec:lya_dist}.

\subsection{Identification of Lyman break galaxies at $z\sim5-6$} \label{sec:lbg}

To establish the Ly$\alpha$ EW and Ly$\alpha$ escape fraction distributions at $z\sim5-6$, we need to identify a sample of typical $z\sim5-6$ galaxies. 
We now select star-forming galaxies at $z\sim5-6$ via the standard Lyman-break techniques, using the JADES {\it JWST}/NIRCam imaging (see Section~\ref{sec:sed}) and the complementary {\it HST} Advanced Camera for Surveys (ACS) imaging from the HLF archive. 
Here we focus on the GOODS-South field overlapped with JADES observations as well as the MUSE-Wide and the MUSE-Deep observations. 
Using MUSE exposure maps \citep{Bacon2017,Bacon2023,Urrutia2019}, we only consider the regions that have exposure time $>80$~per~cent of the designed exposure time of each MUSE observation ($1$ hour for MUSE-Wide, $10$ hours for MOSAIC, $31$ hours for UDF-10, and $141$ hours for MXDF) and are overlapped with JADES observations. 
This allows us to derive robust Ly$\alpha$ properties (including both detections and upper limits) for all the LBGs covered by the MUSE field of view. 
We do not apply this study to the GOODS-North field because our Keck/DEIMOS observations only target a portion of LBGs within the DEIMOS field of view. 

The Lyman-break galaxies at $z\sim5$ ($4.5\lesssim z\lesssim5.5$) and $z\sim6$ ($5.5\lesssim z\lesssim6.5$) were identified separately, using the rest-frame UV color computed from JADES CIRC1 aperture ($0.2$~arcsec diameter aperture) fluxes (see descriptions in Section~\ref{sec:sed}). 
At $z\sim5$, the Lyman break is located at $\simeq0.73\ \mu$m, thus LBGs at this redshift range should appear as strong {\it HST} ACS/F606W dropouts. 
We select $z\sim5$ star-forming galaxies using the similar criteria utilized in \citet{Bouwens2021}:
\begin{eqnarray*}
{\rm F606W}-{\rm F814W}&>&1.2 \\
{\rm F090W}-{\rm F150W}&<&0.9 \\
{\rm F606W}-{\rm F814W}&>&1.2\times({\rm F090W}-{\rm F150W})+1.3.
\end{eqnarray*}
For objects with the S/N $<1$ in F606W, we set the F606W flux to its $1\sigma$ upper limit. 
In addition, we require a non-detection (S/N $<2$) in ACS/F435W.

We next select galaxies at $z\sim6$ using the F775W dropout criteria utilized in \citet{Endsley2023b}, which build on the approach of previous studies \citep[e.g.,][]{Bunker2004,Bouwens2015}: 
\begin{eqnarray*}
{\rm F775W}-{\rm F090W}&>&1.2 \\
{\rm F090W}-{\rm F150W}&<&1.0 \\
{\rm F775W}-{\rm F090W}&>&{\rm F090W}-{\rm F150W}+1.2.
\end{eqnarray*}
Again, we set the F775W flux to its $1\sigma$ upper limit if its S/N $<1$. 
We also require a non-detection in F435W with S/N $<2$, and a strong dropout in F606W: ${\rm F606W}-{\rm F090W}>2.7$, or ${\rm F606W}-{\rm F090W}>1.8$ if S/N(F606W) $<2$, where the F606W flux is set to its $1\sigma$ upper limit if S/N(F606W) $<1$. 
If the F775W dropout is extremely strong (${\rm F775W}-{\rm F090W}>2.5$), we will ignore the F435W non-detection and the F606W dropout criteria. 

To ensure robust dropout detection, we put a magnitude limit to the filter at rest-frame UV. 
This is because when the rest-frame UV band (i.e., F814W at $z\sim5$, or F090W at $z\sim6$) is very faint, we are not likely measure the Lyman break given the $1\sigma$ upper limit in the dropout band (i.e., F606W at $z\sim5$, or F775W at $z\sim6$). 
In HUDF, the typical $1\sigma$ depth of F606W (F775W) is $30.2$~AB~mag \citep[e.g.,][]{Bouwens2015}. 
To ensure we are able to measure the Lyman break with ${\rm F606W}-{\rm F775W}>1.2$ at $z\sim5$ (${\rm F775W}-{\rm F090W}>1.2$ at $z\sim6$), we require the galaxies in our $z\sim5$ ($z\sim6$) sample in HUDF to present F814W $<29$ (F090W $<29$). 
In the other regions in GOODS-South, the typical $1\sigma$ depth of F606W (F775W) is $29.7$~AB~mag ($29.2$). 
Therefore, we require our $z\sim5$ ($z\sim6$) galaxies in those regions to present F814W $<28.5$ (F090W $<28$). 
Finally, to ensure the selected sources are real, we also require S/N $>5$ in at least one NIRCam filter and S/N $>3$ in at least three NIRCam filters for all the galaxies in our $z\sim5$ and $z\sim6$ LBG samples \citep{Endsley2023b}. 

Because we aim to derive Ly$\alpha$ property distributions for LBGs at $z\sim5-6$, we must consider the potential bias of LBG selection to Ly$\alpha$ emitting galaxies. 
The presence of strong Ly$\alpha$ emission at the lower bound of redshift range probed by each dropout selection (e.g., $z\simeq4.5$ for F606W dropouts, or $z\simeq5.5$ for F775W dropouts) may boost the dropout filter and thus dilute the Lyman break color \citep[e.g.,][]{Stanway2008}. 
We cross-match the MUSE Ly$\alpha$ emitting galaxy catalog to our F606W and F775W dropout sample, and we find that several strong Ly$\alpha$ emitters (Ly$\alpha$ EW $\gtrsim50$~\AA) at $4.5<z<4.8$ are missed in the F606W dropout selection. 
While there are also strong Ly$\alpha$ emitters at $5.5<z<5.8$ missed in the F775W dropout selection, they are selected as F606W dropouts because the presence of strong Ly$\alpha$ boosts the F814W flux and hence these systems are still included in our total $z\sim5-6$ LBG sample. 
Similarly, strong Ly$\alpha$ emission at $4.5<z<4.8$ can boost the F606W flux and hence the F435W $-$ F606W color. 
Therefore, to include the missing strong Ly$\alpha$ emitters at $4.5<z<4.8$, we also identify a sample of F435W dropout sources using the criteria presented in \citet{Bouwens2021}:
\begin{eqnarray*}
{\rm F435W}-{\rm F606W}&>&1 \\
{\rm F814W}-{\rm F115W}&<&1 \\
{\rm F435W}-{\rm F606W}&>&1.8\times({\rm F814W}-{\rm F115W})+1.
\end{eqnarray*}
We then cross-match the selected F435W dropouts with MUSE Ly$\alpha$ emitters. 
We add Ly$\alpha$ emitters at $z>4.5$ that fall in the F435W dropout sample into our $z\sim5-6$ LBG sample.

We visually inspected the {\it JWST}/NIRCam and {\it HST}/ACS imaging of every selected Lyman-break galaxy at $z\sim5-6$ to remove suspicious objects usually due to diffraction spikes, hot pixels, or diffuse emission from nearby bright low redshift objects. 
We also examined whether there are brown dwarfs selected as F606W or F775W dropouts. 
After cross-matching with the JADES brown dwarf catalog \citep{Hainline2024a}, we find that we are not selecting any brown dwarfs as Lyman break galaxies. 
AGN were removed by cross-matching our dropouts with AGN catalogs in literature (\citealt{Maiolino2023,Lyu2024,Matthee2024}; Sun et al. in prep.)

Our final LBG sample contains $543$ sources at $z\sim5$ (F606W dropout) and $171$ sources at $z\sim6$ (F775W dropout) in the GOODS-South field overlapped with JADES and MUSE observations. 
For LBGs with spectroscopic redshift measurements (i.e., either from MUSE Ly$\alpha$ detection or NIRCam grism H$\alpha$ detection), we move those at $4.5<z_{\rm spec}<5.5$ ($5.5<z_{\rm spec}<6.5$) to the $z\sim5$ ($z\sim6$) sample no matter whether they were selected as F435W, F606W, or F775W dropouts. 
Among the total $714$ LBGs in our $z\sim5-6$ sample, Ly$\alpha$ emission lines are detected in $167$ sources with MUSE observations \citep{Urrutia2019,Bacon2017,Bacon2023}, and H$\alpha$ emission lines are detected in $161$ sources with NIRCam grism observations (Sun et al. in prep.). 
These include $64$ sources with both Ly$\alpha$ and H$\alpha$ emission line detections, comprising all $52$ MUSE-identified galaxies at $4.9<z<6.5$ with Ly$\alpha$ and H$\alpha$ detections (and NIRCam imaging) introduced in Section~\ref{sec:data}. 
The remaining $12$ sources have Ly$\alpha$ emission near skylines. 
In the analysis presented in Section~\ref{sec:lae_property}, we do not include galaxies with Ly$\alpha$ lines obscured by skylines (see Section~\ref{sec:muse}). 
In this section, because we will consider the incompleteness due to skyline obscuration (Section~\ref{sec:dist_method}), we will keep these systems in the LBG sample. 

For the $167$ LBGs with Ly$\alpha$ detections, we derive their Ly$\alpha$ emission line fluxes and EWs from MUSE spectra using the same approach described in Section~\ref{sec:data}. 
For the remaining galaxies without Ly$\alpha$ detection, we put the $5\sigma$ upper limit to their Ly$\alpha$ fluxes. 
The typical $5\sigma$ upper limit of flux for galaxies in the MUSE-Wide field is $2\times10^{-18}$~erg~s$^{-1}$~cm$^{-2}$, and for galaxies in the MUSE-Deep field is $5\times10^{-19}$~erg~s$^{-1}$~cm$^{-2}$ (MOSAIC), $2.5\times10^{-19}$~erg~s$^{-1}$~cm$^{-2}$ (UDF-10), or $1\times10^{-19}$~erg~s$^{-1}$~cm$^{-2}$ (MXDF). 
The upper limits of Ly$\alpha$ EW for LBGs without Ly$\alpha$ emission detection are computed using the $5\sigma$ upper limits on Ly$\alpha$ flux and the underlying continuum flux densities derived from broadband photometry. 

\begin{figure}
\begin{center}
\includegraphics[width=\linewidth]{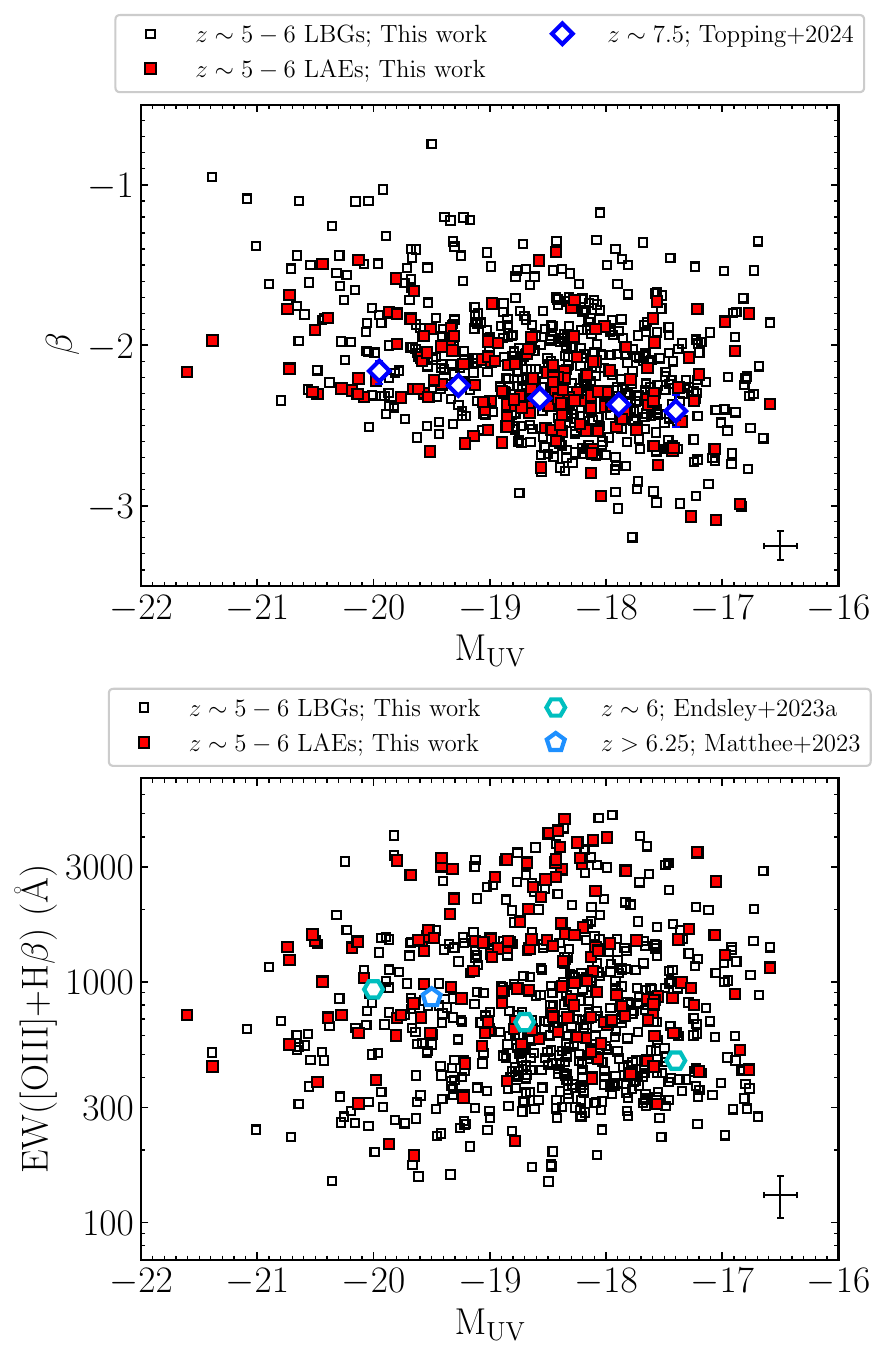}
\caption{Absolute UV magnitude versus UV slope (top panel) and [O~{\scriptsize III}]+H$\beta$ EWs (inferred from \textsc{beagle} models; bottom panel) for the LBGs at $z\sim5-6$ in our sample (open black squares), with those with Ly$\alpha$ detections marked as red squares. The typical uncertainties are shown as the black error bars at the lower right of each panel. As a comparison, we overplot the median UV slope of $z\sim7$ LBGs (open blue diamonds) presented in \citet{Topping2024} in the top panel, and the median [O~{\scriptsize III}]+H$\beta$ EWs of $z\sim5-9$ galaxies (open cyan hexagon; \citealt{Endsley2023b}; open dodger-blue pentagon; \citealt{Matthee2023}) in the bottom panel.} The physical properties of our $z\sim5-6$ LBGs are comparable to the galaxy population in the reionization era.
\label{fig:lbg}
\end{center}
\end{figure}

We fit the NIRCam F090W to F444W SEDs of the $714$ LBGs at $z\sim5-6$ in our sample with \textsc{beagle} models following the same procedures described in Section~\ref{sec:sed}. 
For those without spectroscopic redshifts, we fit the redshift in the range $3\le z\le8$ assuming a uniform prior. 
The absolute UV magnitude of our LBGs span a wide range from $-21.6$ to $-16.6$, and the UV slope varies from $-3.2$ to $-0.7$ (Fig.~\ref{fig:lbg}). 
The [O~{\small III}]+H$\beta$ EWs inferred from \textsc{beagle} models span from $\simeq100$~\AA\ to $\simeq5000$~\AA\ with a median value of $\simeq704$~\AA\ (corresponding to a median CSFH age of $\simeq50$~Myr). 
The properties of these LBGs at $z\sim5-6$ are comparable to the properties of galaxies in the reionization era (e.g., \citealt{Endsley2023b,Topping2024}; Fig.~\ref{fig:lbg}), suggesting that our LBG sample provides an ideal baseline for understanding the Ly$\alpha$ properties of $z\gtrsim7$ galaxies at redshifts where the impact of the neutral IGM is less important.


We then derive the Ly$\alpha$ escape fractions for all the LBGs in our $z\sim5-6$ sample. 
Because not all the LBGs have H$\alpha$ line measurements, we derive the intrinsic Ly$\alpha$ luminosities of all our sources from \textsc{beagle} models instead of from the observed H$\alpha$ luminosity to avoid bias against those without H$\alpha$ measurements. 
From \textsc{beagle} models, we first infer the H$\alpha$ luminosities. 
To examine whether the SED inferred H$\alpha$ luminosities are robust, we compare the H$\alpha$ fluxes measured from NIRCam grism spectra with those inferred from \textsc{beagle} models for the $161$ LBGs with grism H$\alpha$ detections.
We find that on average the grism H$\alpha$ fluxes are $0.82$ times of those inferred from SEDs, likely owing to that a small portion of H$\alpha$ emission is missed in the over-subtraction of sky continuum in NIRCam grism spectra (Sun et al. in prep.). 
This comparison is consistent with the results found in our $z\simeq5-6$ Ly$\alpha$ emitters (Section~\ref{sec:lya_measure}) and in the joint analysis of FRESCO and JADES data by Sun et al. (in prep.).
We then correct the \textsc{beagle} inferred H$\alpha$ luminosities for dust attenuation and convert to Ly$\alpha$ luminosities assuming case B recombination (Section~\ref{sec:lya_measure}).
The Ly$\alpha$ escape fraction of each LBG is computed as the ratio of the observed Ly$\alpha$ luminosity (measured Ly$\alpha$ luminosity for detection, or $5\sigma$ upper limit for non-detection) to the model inferred intrinsic Ly$\alpha$ luminosity. 
We force the Ly$\alpha$ escape fraction to $1$ for sources with computed $f_{{\rm esc,Ly}\alpha}>1$.

\subsection{Methodology for deriving Ly$\alpha$ property distributions} \label{sec:dist_method}

We establish the distributions of Ly$\alpha$ EW and Ly$\alpha$ escape fraction for our LBGs at $z\sim5-6$ using a Bayesian approach \citep{Schenker2014,Endsley2021b,Boyett2022,Chen2024}. 
We assume a log-normal distribution for both quantities \citep{Schenker2014,Endsley2021b,Chen2024}. 
In this subsection we describe the methodology for deriving the log-normal distribution. 

Before deriving the Ly$\alpha$ property distributions, we need to take into account the incompleteness of the Ly$\alpha$ measurements. 
Ly$\alpha$ emission lines with relatively faint fluxes could be hidden by random noise fluctuations. 
And since the Ly$\alpha$ lines are measured from the ground, they could be partially obscured by sky lines. 
Here we estimate the completeness for detecting a Ly$\alpha$ emission line in the MUSE data for a range of redshifts, absolute UV magnitudes, and Ly$\alpha$ EWs. 
We start by creating a MUSE sky line spectrum at $\lambda=4800-9300$~\AA\ (i.e., the wavelength range covered by the MUSE data) using the average MUSE spectrum. 
We consider wide grids of redshift ($z=4.5-6.5$), M$_{\rm UV}$ ($=-22$ to $-16$), and Ly$\alpha$ EW ($=1-1000$~\AA), creating a Ly$\alpha$ line profile for each set of parameters ($z$, M$_{\rm UV}$, and Ly$\alpha$ EW) assuming a FWHM $=230$~km~s$^{-1}$ (i.e., equal to the median FWHM measured for our Ly$\alpha$ emitters at $z\simeq5-6$). 
For each Ly$\alpha$ profile, we insert it into the MUSE sky line spectrum $1000$ times. 
And for each simulated emission line spectrum, we randomly perturb the flux density of each wavelength pixel based on the error spectrum. 
The completeness is calculated as the fraction of realizations in which the emission line is detected at $>5\sigma$ level. 

For either the Ly$\alpha$ EW or the Ly$\alpha$ escape fraction, the distribution is modeled with a set of parameters $\theta=[\mu,\sigma]$, where $\mu$ is the mean of the log-normal distribution and $\sigma$ is the standard deviation. 
For Ly$\alpha$ EW distribution, we consider uniform priors for the model parameters: $\mu=0-6$ (corresponding to mean Ly$\alpha$ EW $\simeq1-400$~\AA\ in linear space) and $\sigma=0.01-3$ \citep{Schenker2014,Endsley2021b}. 
For Ly$\alpha$ escape fraction distribution, we use an uniform prior for $\mu$ ($=-9$ to $0$, corresponding to mean $f_{{\rm esc,Ly}\alpha}\simeq0.0001-1$ in linear space), and a Gaussian prior for $\sigma$ (mean $=0.6$, standard deviation $=0.3$; \citealt{Chen2024}). 

For each set of model parameters $\theta$, the log-normal distribution is given by
\begin{equation}
p(x|\theta) = \frac{A}{\sqrt{2\pi}\sigma\cdot x} \cdot \exp{[-\frac{(\ln{x}-\mu)^2}{2\sigma^2}]},
\end{equation}
where $x$ is the Ly$\alpha$ EW or the Ly$\alpha$ escape fraction, and $A$ is the normalization parameter. 
For the Ly$\alpha$ EW distribution, the normalization parameter $A$ equals $1$. 
For the Ly$\alpha$ escape fraction distribution, because we only consider the range $f_{{\rm esc,Ly}\alpha}=0-1$, the normalization parameter is computed as:
\begin{eqnarray*}
\int^{1}_{0} p(x|\theta)\ dx &=& 1 \\
A &=& \frac{2}{1+{\rm erf}[-\mu/(\sqrt{2}\sigma)]}.
\end{eqnarray*}
And for each galaxy with Ly$\alpha$ detection, we also compute the Gaussian measurement uncertainty as
\begin{equation}
p(x)_{{\rm obs},i} = \frac{1}{\sqrt{2\pi}\sigma_{{\rm obs},i}} \cdot \exp{[-\frac{(x-x_{{\rm obs},i})^2}{2\sigma^2_{{\rm obs},i}}]},
\end{equation}
where $x_{{\rm obs},i}$ and $\sigma_{{\rm obs},i}$ are the observed value and the uncertainty of Ly$\alpha$ EW or Ly$\alpha$ escape fraction for the $i^{\rm th}$ object. 
The likelihood of the entire sample is computed as the product of the individual likelihood of each object in the sample. 
For each galaxy with Ly$\alpha$ detection ($i^{\rm th}$), the individual likelihood for Ly$\alpha$ EW distribution is computed as
\begin{equation}
p({\rm obs},i|\theta)_{\rm det} = \int^{\infty}_{0} p(x)_{{\rm obs},i} \cdot p(x|\theta)\ dx.
\end{equation}
And for Ly$\alpha$ escape fraction distribution the individual likelihood is:
\begin{equation}
p({\rm obs},i|\theta)_{\rm det} = \int^{1}_{0} p(x)_{{\rm obs},i} \cdot p(x|\theta)\ dx.
\end{equation}
For each galaxy without Ly$\alpha$ detection, we write the individual likelihood as
\begin{equation} 
p({\rm obs},i|\theta)_{\rm lim} = p(x<x_{5\sigma}|\theta) + p(x>x_{5\sigma}|\theta) \cdot (1-C)
\end{equation}
The first term $p(x<x_{5\sigma}|\theta)$ considers the likelihood that the Ly$\alpha$ flux is under the $5\sigma$ upper limit and thus is undetected. 
The second term $p(x>x_{5\sigma}|\theta) \cdot (1-C)$ considers the likelihood that the Ly$\alpha$ flux is larger than the $5\sigma$ upper limit, but the emission is not detected due to the impact of skyline obscuration. 
Here $C$ is the completeness (see Section~\ref{sec:lbg}). 

Now we compute the total likelihood for a given set of parameters as
\begin{equation}
p({\rm obs}|\theta) \propto \prod_{i} p({\rm obs},i|\theta)
\end{equation}
Using Bayes' theorem, we write the posterior probability distribution for the model parameters as
\begin{equation}
p(\theta|{\rm obs}) \propto p(\theta) \cdot p({\rm obs}|\theta)
\end{equation}
where $p(\theta)$ is the prior of model parameters. 
Finally, we derive the distribution using a Markov Chain Monte Carlo (MCMC) approach to sample the model parameter space using the \textsc{emcee} package \citep{Foreman-Mackey2013}. 
For each model parameter, we derive its posterior probability distribution and compute the median value and the marginal $68$~per~cent credible interval. 

\subsection{Distributions of Ly$\alpha$ EW and Ly$\alpha$ escape fraction} \label{sec:lya_dist}


\begin{figure*}
\begin{center}
\includegraphics[width=\linewidth]{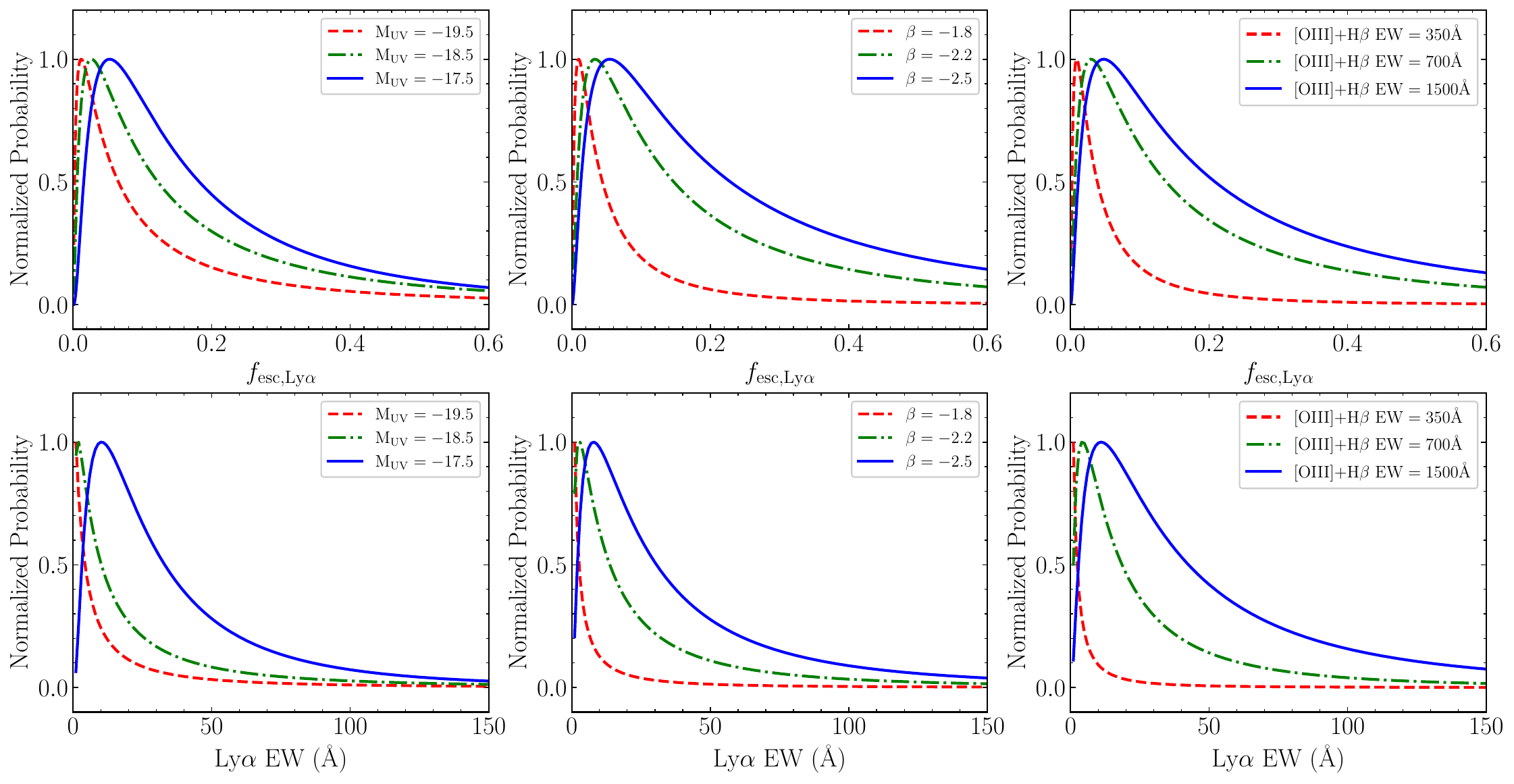}
\caption{The derived log-normal distributions of Ly$\alpha$ escape fraction (top panels) and Ly$\alpha$ EW (bottom panels) for our LBGs at $z\sim5-6$. We show distributions for different groups of LBGs binned by galaxy properies: absolute UV magnitude (left panels), UV slope (middle panels), and [O~{\scriptsize III}]+H$\beta$ EW (right panels). Different colors and line styles represent different properties. We see a trend that the average Ly$\alpha$ escape fraction and EW become higher towards fainter M$_{\rm UV}$, bluer UV slopes, or higher [O~{\scriptsize III}]+H$\beta$ EWs at $z\sim5-6$.}
\label{fig:lya_dist}
\end{center}
\end{figure*}

\subsubsection{Overview of Ly$\alpha$ EW and Ly$\alpha$ escape fraction distributions} \label{sec:lya_dist_all}

Based on the methodology described in the above subsection, we derive the Ly$\alpha$ EW and Ly$\alpha$ escape fraction distributions for the LBGs at $z\sim5-6$. 
We note that our distributions are determined using apertures that are equivalent to the {\it HST} segmentation map convolved with the MUSE PSF, typically corresponding to $1.5$~arcsec diameter for the Keck and VLT spectra (Section~\ref{sec:lya_flux}). 
Comparison with observations with other facilities will require modest flux conversions given the different apertures. 
In particular, we will discuss these conversions when comparing to emerging {\it JWST}/NIRSpec observations in Section~\ref{sec:lya_pred}.

The posterior median values and $68$~per~cent credible intervals of parameters of log-normal EW distributions are shown in Table~\ref{tab:lya_ew_dist} and Table~\ref{tab:lya_fesc_dist}. 
For the entire sample, we infer that the parameters of the Ly$\alpha$ EW distribution are $\mu=2.76^{+0.13}_{-0.14}$ (corresponding to a median Ly$\alpha$ EW $=15$~\AA) and $\sigma=1.48^{+0.11}_{-0.10}$. 
This indicates that strong Ly$\alpha$ emission is common at $z\sim5-6$, with $38^{+4}_{-4}$~per~cent of the LBGs showing Ly$\alpha$ EW $>25$~\AA\ (the so-called ``Ly$\alpha$ fraction''; e.g., \citealt{Stark2010}). 
Focusing on the M$_{\rm UV}$ range that is mostly used in literature ($-20.25<{\rm M}_{\rm UV}<-18.75$), we find that $33^{+6}_{-5}$~per~cent of these systems present Ly$\alpha$ EW $>25$~\AA. More extreme 
Ly$\alpha$ emitters (EW $>50$~\AA) appear less commonly in Lyman break selected samples. Our distributions indicate Ly$\alpha$ fractions of $22^{+3}_{-3}$~per~cent and $11^{+2}_{-2}$~per~cent for EW $>50$~\AA\ and $100$~\AA, respectively. 

From the Ly$\alpha$ escape fraction distribution, we see that many $z\simeq5-6$ LBGs transmit a large fraction of their Ly$\alpha$ emission. 
We find the parameters are $\mu=-2.06^{+0.14}_{-0.14}$ (median $f_{{\rm esc,Ly}\alpha}=0.13$) and $\sigma=1.55^{+0.16}_{-0.15}$. 
This suggests that $32^{+4}_{-4}$~per~cent of the $z\sim5-6$ galaxies show $f_{{\rm esc,Ly}\alpha}>0.2$. This is more than $4$ times the typical Ly$\alpha$ escape fraction of $z\sim2$ galaxies (0.05; \citealt{Hayes2010}), suggesting the transmission of Ly$\alpha$ emission increases between $z\simeq2$ and $z\simeq5-6$.
Very high escape fractions are also seen in our dataset. Our distribution suggests $11^{+1}_{-1}$~per~cent of the UV continuum selected sample at $z\simeq5-6$ have $f_{{\rm esc,Ly}\alpha}>0.5$. 
This baseline value will be particularly useful for predicting how commonly we might expect to detect galaxies with similarly large Ly$\alpha$ transmission at $z\gtrsim7$ with {\it JWST}. 
We will come back to discuss this in Section~\ref{sec:discussion}.

To assess the impact of the assumed distribution, we also fit our data with a commonly used declining exponential distribution $p(x)=A\cdot\exp{(-x/x_0)}$ \citep[e.g.,][]{Dijkstra2011,Jung2018}, where $x$ is Ly$\alpha$ EW or Ly$\alpha$ escape fraction and $x_0$ is the characteristic $e$-folding scale of EW or $f_{{\rm esc,Ly}\alpha}$. 
We find that the declining exponential distributions indicate that the fractions of galaxies showing large Ly$\alpha$ EWs or large Ly$\alpha$ escape fractions are similar to those derived from log-normal distributions. 
Therefore, we argue that choosing different distribution models (log-normal or declining exponential) does not impact our results significantly.


\begin{table}
\centering
\begin{tabular}{cccc}
\hline
Sample & $N_{\rm gal}$ & $e^{\mu}$ (\AA) & $\sigma$ (dex) \\
\hline
All & $714$ & $15^{+2}_{-2}$ & $1.48^{+0.11}_{-0.10}$ \\
Median M$_{\rm UV}=-19.5$ & $238$ & $10^{+2}_{-2}$ & $1.75^{+0.19}_{-0.17}$ \\
Median M$_{\rm UV}=-18.5$ & $238$ & $16^{+3}_{-3}$ & $1.51^{+0.19}_{-0.16}$ \\
Median M$_{\rm UV}=-17.5$ & $238$ & $27^{+5}_{-5}$ & $0.99^{+0.17}_{-0.14}$ \\
Median $\beta=-1.8$ & $238$ & $ 4^{+1}_{-1}$ & $1.99^{+0.26}_{-0.24}$ \\
Median $\beta=-2.2$ & $238$ & $18^{+4}_{-3}$ & $1.40^{+0.16}_{-0.14}$ \\
Median $\beta=-2.5$ & $238$ & $29^{+5}_{-4}$ & $1.16^{+0.14}_{-0.12}$ \\
Median [O~{\scriptsize III}]+H$\beta$ EW $=350$~\AA & $238$ & $ 2^{+1}_{-1}$ & $1.82^{+0.30}_{-0.26}$ \\
Median [O~{\scriptsize III}]+H$\beta$ EW $=700$~\AA & $238$ & $19^{+3}_{-3}$ & $1.24^{+0.15}_{-0.13}$ \\
Median [O~{\scriptsize III}]+H$\beta$ EW $=1500$~\AA & $238$ & $41^{+6}_{-6}$ & $1.15^{+0.13}_{-0.12}$ \\
$z\sim5$, $-20.25<{\rm M}_{\rm UV}<-18.75$ & $138$ & $13^{+4}_{-3}$ & $1.64^{+0.23}_{-0.19}$ \\
$z\sim6$, $-20.25<{\rm M}_{\rm UV}<-18.75$ & $82$ & $ 8^{+4}_{-3}$ & $1.85^{+0.42}_{-0.33}$ \\
\hline
\end{tabular}
\caption{Posterior median values and $68$~per~cent credible intervals of parameters of Ly$\alpha$ EW distributions for our LBGs at $z\sim5-6$. We list the median Ly$\alpha$ EW ($e^{\mu}$) and the standard deviation ($\sigma$) for each subset, as well as the number of galaxies ($N_{\rm gal}$).}
\label{tab:lya_ew_dist}
\end{table}


\begin{table}
\centering
\begin{tabular}{cccc}
\hline
Sample & $N_{\rm gal}$ & $e^{\mu}$ & $\sigma$ (dex) \\
\hline
All & $714$ & $0.13^{+0.02}_{-0.02}$ & $1.55^{+0.16}_{-0.15}$ \\
Median M$_{\rm UV}=-19.5$ & $238$ & $0.10^{+0.02}_{-0.02}$ & $1.46^{+0.19}_{-0.17}$ \\
Median M$_{\rm UV}=-18.5$ & $238$ & $0.14^{+0.03}_{-0.03}$ & $1.30^{+0.18}_{-0.16}$ \\
Median M$_{\rm UV}=-17.5$ & $238$ & $0.16^{+0.03}_{-0.03}$ & $1.05^{+0.19}_{-0.17}$ \\
Median $\beta=-1.8$ & $238$ & $0.05^{+0.01}_{-0.01}$ & $1.28^{+0.20}_{-0.18}$ \\
Median $\beta=-2.2$ & $238$ & $0.16^{+0.03}_{-0.03}$ & $1.26^{+0.19}_{-0.15}$ \\
Median $\beta=-2.5$ & $238$ & $0.24^{+0.06}_{-0.04}$ & $1.22^{+0.19}_{-0.16}$ \\
Median [O~{\scriptsize III}]+H$\beta$ EW $=350$~\AA & $238$ & $0.04^{+0.01}_{-0.01}$ & $1.25^{+0.18}_{-0.17}$ \\
Median [O~{\scriptsize III}]+H$\beta$ EW $=700$~\AA & $238$ & $0.16^{+0.04}_{-0.03}$ & $1.31^{+0.18}_{-0.15}$ \\
Median [O~{\scriptsize III}]+H$\beta$ EW $=1500$~\AA & $238$ & $0.23^{+0.06}_{-0.04}$ & $1.25^{+0.18}_{-0.15}$ \\
$z\sim5$, $-20.25<{\rm M}_{\rm UV}<-18.75$ & $138$ & $0.15^{+0.05}_{-0.03}$ & $1.36^{+0.19}_{-0.18}$ \\
$z\sim6$, $-20.25<{\rm M}_{\rm UV}<-18.75$ & $82$ & $0.09^{+0.03}_{-0.03}$ & $1.15^{+0.21}_{-0.18}$ \\
\hline
\end{tabular}
\caption{Posterior median values and $68$~per~cent credible intervals of parameters of Ly$\alpha$ escape fraction distributions for our LBGs at $z\sim5-6$. We list the median Ly$\alpha$ escape fraction ($e^{\mu}$) and the standard deviation ($\sigma$) for each subset, as well as the number of galaxies ($N_{\rm gal}$).}
\label{tab:lya_fesc_dist}
\end{table}

\subsubsection{Dependence of Ly$\alpha$ distributions on galaxy properties} \label{sec:lya_dist_property}

One of the primary goals of this subsection is to investigate the dependence of Ly$\alpha$ escape fraction and Ly$\alpha$ EW on galaxy properties. 
We first explore how the Ly$\alpha$ strength changes with rest-frame UV luminosity. 
It has been reported that UV-faint galaxies (M$_{\rm UV}\simeq-19$) present stronger Ly$\alpha$ emission than more luminous systems (M$_{\rm UV}\simeq-22$ to $-20$) \citep[e.g.,][]{Shapley2003,Ando2006,Stark2010,DeBarros2017,ArrabalHaro2018}. 
The $25-75$ percentile of the M$_{\rm UV}$ distribution in our sample ranges from $-19.1$ to $-17.8$. The {\it JWST} dataset thus extends to considerably fainter galaxies but does not include the more luminous systems studied previously. 
To explore the luminosity dependence of Ly$\alpha$ EWs, we divide our $z\sim5-6$ LBG sample into three groups with equal number of sources: a bright subset M$_{\rm UV}\lesssim-19$ (median M$_{\rm UV}=-19.5$), a moderately faint subset $-19\lesssim{\rm M}_{\rm UV}\lesssim-18$ (median M$_{\rm UV}=-18.5$), and a very faint subset M$_{\rm UV}\gtrsim-18$ (median M$_{\rm UV}=-17.5$). 
The results are consistent with the trend found previously in the literature, with Ly$\alpha$ EW becoming more prominent at lower luminosities. 
In particular, we find that UV faint galaxies with M$_{\rm UV}\gtrsim-18$ typically show Ly$\alpha$ EWs $2.7\times$ higher than more luminous systems with with M$_{\rm UV}\lesssim-19$ (bottom middle panel of Fig.~\ref{fig:lya_dist}) when comparing the mean values of Ly$\alpha$ EW distributions (Table~\ref{tab:lya_ew_dist}). 
As has been discussed extensively in the literature \citep[e.g.,][]{Shapley2003,Ando2006,Stark2010}, this luminosity trend is likely related to the dependence of dust content on luminosity \citep[e.g.,][]{Bouwens2009,Bouwens2012,Reddy2009}, with the more luminous systems presenting more attenuation to Ly$\alpha$ than less luminous galaxies.

Previous studies have revealed significant trends between Ly$\alpha$ strength and UV slope \citep[e.g.,][]{Shapley2003,Stark2010,Matthee2016,Endsley2021b}. 
We quantify this trend in our Lyman break selected $z\sim5-6$ sample. 
Most earlier results have focused on how the Ly$\alpha$ strength increases between galaxies with significant reddening ($\beta\simeq-1.4$) and those with little sign of reddening ($\beta\simeq-2.0$).
Our sample extends to much bluer colors, with the 25-75th percentile values of the slope distribution ranging from $\beta=-2.4$ to $-1.9$. 
How Ly$\alpha$ properties vary between blue ($\beta\simeq-2.0$) and extremely blue ($\beta\simeq-2.5$) is not known observationally. 
We derive the Ly$\alpha$ escape fraction and Ly$\alpha$ EW distributions for three equally-divided subsets: galaxies with $-2.0\lesssim\beta\lesssim-1.0$ (median $\beta=-1.8$), $-2.3\lesssim\beta\lesssim-2.0$ (median $\beta=-2.2$), and $-3.0\lesssim\beta\lesssim-2.3$ (median $\beta=-2.5$). 
We do find that the bluest subset presents the largest Ly$\alpha$ escape fractions (top middle panel of Fig.~\ref{fig:lya_dist}). In particular, we find that galaxies with $\beta\simeq-2.5$ have Ly$\alpha$ escape fractions that are on average $4.8\times$ higher than those with $\beta\simeq-1.8$ (Table~\ref{tab:lya_ew_dist}). 
However, we see the variation of Ly$\alpha$ escape fraction between the $\beta=-2.2$ and $\beta=-2.5$ subsets is less significant, with $f_{{\rm esc,Ly}\alpha}$ only increasing by $1.6\times$ (Table~\ref{tab:lya_fesc_dist}). 
We find a similar trend in Ly$\alpha$ EW, with the EW modestly increasing (by a factor of $1.6$) from the $\beta=-2.2$ subset to the $\beta=-2.5$ subset (bottom middle panel of Fig.~\ref{fig:lya_dist}). 
This suggests that the increase in Ly$\alpha$ strength toward bluer colors begins to slow for galaxies with $\beta\lesssim-2$. 
Physically this may be expected if the variations of UV slope in this regime are less uniformly linked to dust attenuation. 
It may be other factors less strongly linked to Ly$\alpha$ escape (i.e., gas-phase metallicity, stellar population properties) play a significant role of regulating the UV slope in this blue regime \citep[e.g.,][]{Bouwens2012,Topping2022,Cameron2023b}.

At lower redshifts ($z\simeq2-3$), it has been shown that strong Ly$\alpha$ emission becomes more common among the most extreme [O~{\small III}]+H$\beta$ line emitting galaxies \citep[e.g.,]{Du2020,Tang2021,Tang2024}.
We explore this trend in our dataset at $z\sim5-6$. 
Previous investigations of luminous galaxies ($\gtrsim L^*_{\rm UV}$) at $z\sim7$ have revealed that galaxies with large [O~{\small III}]+H$\beta$ EWs ($>800$~\AA) also present large Ly$\alpha$ EWs \citep{Endsley2021b}. 
Here we extend this study to the more abundant low luminosity population ($\lesssim L^*_{\rm UV}$) at the tail end of reionization using our LBG sample. 
We quantify both the Ly$\alpha$ EW and Ly$\alpha$ escape fraction distributions in three subsets: [O~{\small III}]+H$\beta$ EW $=100-500$~\AA\ (median EW $=350$~\AA), $500-1000$~\AA\ (median EW $=700$~\AA, similar to the average EW of $z>6$ systems; e.g, \citealt{Labbe2013,DeBarros2019,Endsley2023a}), and $>1000$~\AA\ (median EW $=1500$~\AA), where [O~{\small III}]+H$\beta$ EWs are derived from \textsc{beagle} models. 
We find that the most extreme [O~{\small III}] emitters have much stronger Ly$\alpha$ emission (bottom right panel of Fig.~\ref{fig:lya_dist}). 
Comparing to galaxies with [O~{\small III}]+H$\beta$ EW $\simeq350$~\AA, those with [O~{\small III}]+H$\beta$ EW $\simeq1500$~\AA\ show $20\times$ larger Ly$\alpha$ EWs (Table~\ref{tab:lya_ew_dist}). 
The [O~{\small III}]+H$\beta$ EW $\simeq1500$~\AA\ galaxies also show Ly$\alpha$ EWs $2\times$ larger than those with more typical [O~{\small III}]+H$\beta$ EWs ($\simeq700$~\AA) seen at $z>5$. 
We also see that Ly$\alpha$ escape fraction increases with [O~{\small III}]+H$\beta$ EW (bottom right panel of Fig.~\ref{fig:lya_dist}). 
Galaxies with [O~{\small III}]+H$\beta$ EW $\simeq1500$~\AA\ have Ly$\alpha$ escape fraction $5\times$ higher than those with [O~{\small III}]+H$\beta$ EW $\simeq350$~\AA\ (Table~\ref{tab:lya_fesc_dist}). 
This may suggest that the transmission of Ly$\alpha$ through the ISM and the CGM is enhanced in the most extreme [O~{\small III}] emitters, similar to the results found at $z\simeq2-3$ \citep{Tang2021,Tang2024} and at $z\gtrsim6$ \citep[e.g.,][]{Boyett2024}. 
This may be expected if the strong feedback associated with large sSFR disrupts the surrounding gas and boosts the transfer of Ly$\alpha$ photons \citep[e.g.,][]{Kimm2019,Ma2020,Kakiichi2021}. 
It is likely that the Ly$\alpha$ trend is also influenced by the production efficiency of Ly$\alpha$, as the largest sSFR galaxies generally produce more ionizing photons per unit star formation rate and hence have larger intrinsic Ly$\alpha$ EWs \citep[e.g.,][]{Chen2024}. 
These results stress the importance of considering rest-frame optical EWs when assessing the visibility of Ly$\alpha$ emission in the reionization era. In general, [O~{\small III}]+H$\beta$ EWs are the most readily constrained from flux excesses in SEDs.
Of course in very UV-faint (M$_{\rm UV}\gtrsim-17$) galaxies with very low metallicities ($Z\lesssim0.05\ Z_{\odot}$), the Ly$\alpha$ emission may reach to very high EWs ($>100$~\AA) while the [O~{\small III}]+H$\beta$ EWs are low due to the relatively low number of oxygen atoms \citep[e.g.,][]{Endsley2023b,Maseda2023}. 

In the final portion of this section, we investigate the evolution of Ly$\alpha$ fraction with redshift.
To be consistent with previous studies, we focus on systems with $-20.25<{\rm M}_{\rm UV}<-18.75$ and derive the Ly$\alpha$ EW distributions at $z\sim5$ ($138$ LBGs) and $z\sim6$ ($82$ LBGs) separately (Table~\ref{tab:lya_ew_dist}). 
At $z\sim5$, we find that $35^{+7}_{-7}$~per~cent of our LBGs show strong Ly$\alpha$ emission with EW $>25$~\AA. 
Comparing to the Ly$\alpha$ fractions derived at $z\sim3-4$ ($\simeq15-25$~per~cent; e.g., \citealt{Cassata2015,ArrabalHaro2018,deLaVieuville2020,Kusakabe2020,Goovaerts2023}), this result suggests that strong Ly$\alpha$ emission becomes more common from $z\sim3$ to $z\sim5$. 
This is consistent with previous studies \citep[e.g.,][]{Stark2011,ArrabalHaro2018,Kusakabe2020} and also consistent with the implied evolution of the Ly$\alpha$ escape fraction discussed above. 
We expect the Ly$\alpha$ fraction to increase at higher redshifts due to a combination of the larger sSFRs \citep[e.g.,][]{Salmon2015} and lower dust obscuration in higher redshift galaxies \citep[e.g.,][]{Topping2024}. 
The former leads to higher intrinsic Ly$\alpha$ EWs and the latter boosts the transmission of Ly$\alpha$ inside galaxies, both of which will enhance the observed Ly$\alpha$ EWs as we approach the reionization era. 


\begin{figure*}
\begin{center}
\includegraphics[width=\linewidth]{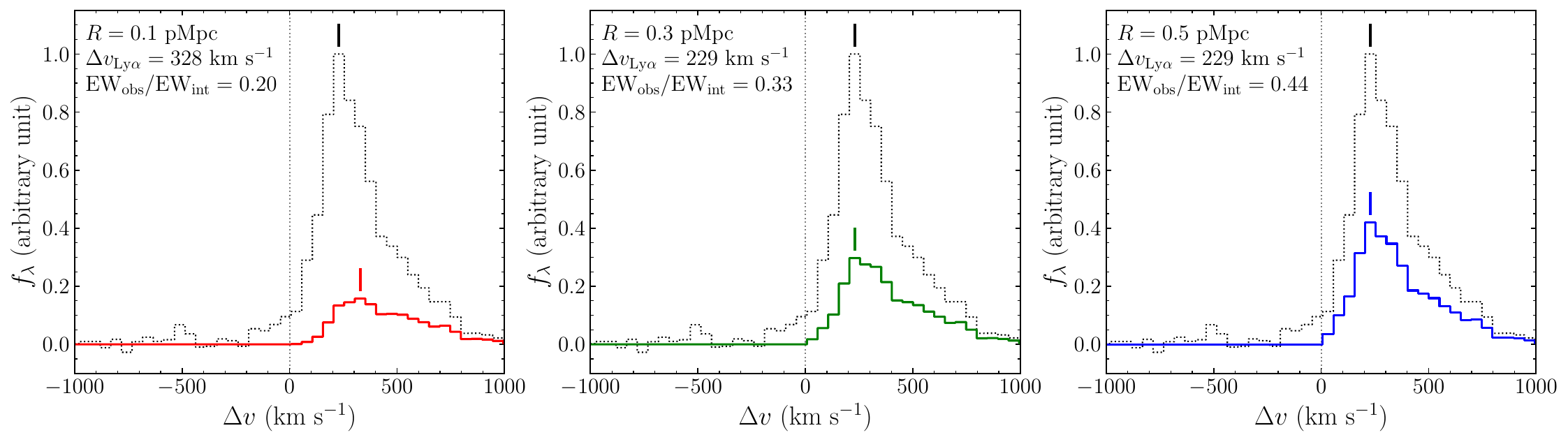}
\caption{Impact of IGM damping wing absorption to Ly$\alpha$ profile at $z=8.5$. The black dotted lines show the composite Ly$\alpha$ profile of moderately strong Ly$\alpha$ emitters (EW $=50-100$~\AA) at $z\simeq5-6$. We assume this profile is the intrinsic Ly$\alpha$ profile of Ly$\alpha$ emitters at $z\gtrsim7$. To estimate the IGM damping wing absorption, we consider a galaxy at $z=8.5$ which reside in a series sizes of ionized bubbles ($R=0.1$~pMpc, left; $R=0.3$~pMpc, middle; $R=0.5$~pMpc, right) and adopting the damping wing attenuation calculation in \citet{Miralda-Escude1998}. The resulting Ly$\alpha$ profiles after IGM absorption are shown by solid lines. We mark the Ly$\alpha$ peak velocities of the intrinsic profiles with black vertical lines and the IGM attenuated profiles with colored vertical lines. In a small bubble ($R=0.1$~pMpc), the Ly$\alpha$ peak velocity shifts from $\simeq230$~km~s$^{-1}$ to $\simeq330$~km~s$^{-1}$. In each panel, we list the Ly$\alpha$ velocity offset after IGM attenuation and the fraction of Ly$\alpha$ photons transmitted through the IGM comparing to the intrinsic value (EW$_{\rm{obs}}/\rm{EW}_{\rm{int}}$).}
\label{fig:lya_bubble}
\end{center}
\end{figure*}

We also consider the evolution in the Ly$\alpha$ fraction between $z\simeq5$ and $z\simeq6$. 
Whereas we find that the Ly$\alpha$ fraction increases between $z\simeq3$ and $z\simeq5$, we do not find clear evidence that this trend continues in the $250$~Myr between $z\simeq5$ and $z\simeq6$. 
For systems with $-20.25<{\rm M}_{\rm UV}<-18.75$, the Ly$\alpha$ fraction with EW$>$25~\AA\ is broadly consistent at $z\sim5$ ($35^{+7}_{-7}$~per~cent) and $z\sim6$ ($28^{+10}_{-10}$~per~cent). 
The less rapid evolution at $5<z<6$ (relative to $3<z<5$) is broadly consistent with the trends found in previous studies \citep[e.g.,][]{deLaVieuville2020,Kusakabe2020,Goovaerts2023}. 
We find that the Ly$\alpha$ escape fraction may decrease between $z\simeq5$ and $z\simeq6$. 
At $z\simeq5$, we find that $28^{+6}_{-6}$~per~cent of our galaxies have Ly$\alpha$ escape fractions with $f_{{\rm esc,Ly}\alpha}>0.2$. 
At $z\sim6$, this fraction decreases to $16^{+7}_{-7}$~per~cent. 
This trend may reflect the impact of the IGM on galaxy samples at $z\simeq6$.
Because the IGM at $z\sim6$ is not only denser but also slightly more neutral than that at $z\sim5$ \citep[e.g.,][]{Fan2023}, the Ly$\alpha$ photons emerging from $z\sim6$ galaxies are more likely to be scattered by the residual H~{\small I} in the IGM, decreasing the Ly$\alpha$ escape fractions.


\section{Discussion} \label{sec:discussion}

In this section, we use our Ly$\alpha$ distributions at $z\simeq5-6$ to investigate Ly$\alpha$ emission in galaxies at $z\gtrsim8$. 
We make predictions for $z\gtrsim8$ Ly$\alpha$ emission line profiles in Section~\ref{sec:lya_pred}, and discuss implications of a recently discovered Ly$\alpha$ emitter at $z\simeq8.5$ in Section~\ref{sec:z8p5_lae}.

\subsection{Expectations for Ly$\alpha$ Deep in the Reionization Era} \label{sec:lya_pred}

In Section~\ref{sec:lya_lbg}, we have derived the range of Ly$\alpha$ properties in Lyman break selected galaxies at $z\simeq5-6$, when the IGM is likely to be mostly ionized. 
Typical UV-faint galaxies at these redshifts have moderate EW Ly$\alpha$ lines, with median Ly$\alpha$ EW $=15$~\AA\ and Ly$\alpha$ escape fraction $=0.13$. 
The strongest Ly$\alpha$ lines are found more rarely, with only $11$~per~cent of the population seen with Ly$\alpha$ EW $>100$~\AA. 
In spite of the large ionized fraction, the IGM is already leaving its imprint on the Ly$\alpha$ distributions at $z\simeq5-6$, with the residual H~{\small I} significantly attenuating those strong Ly$\alpha$ emitters with line flux emerging near the systemic redshift.

At yet higher redshifts, the Ly$\alpha$ will be further weakened by the damping wing from the neutral H~{\small I} outside of ionized bubbles, leading to the well-established drop in the Ly$\alpha$ fraction at $z\gtrsim7$ \citep[e.g.,][]{Caruana2012,Caruana2014,Schenker2014,Pentericci2018,Jones2024}. 
The current frontier of these investigations is at $z\gtrsim8$, where the neutral fraction is expected to be very large ($x_{\rm HI}\gtrsim0.8$; e.g., \citealt{Mason2019,Naidu2020,Umeda2023,Nakane2024}) and ionized bubbles are expected to be small. 
Around faint galaxies (M$_{\rm UV}>-19$), the median bubble size is predicted to be $\simeq0.1$~pMpc at $z\simeq8$ for standard models where reionization is driven by low mass galaxies \citep[e.g.,][]{Mason2020,Lu2024}. 
In contrast, only $\simeq10$~per~cent of faint galaxies (M$_{\rm UV}>-19$) are predicted to lie in bubbles with $R\gtrsim0.5$~pMpc in these models \citep{Lu2024}. 
If the bubbles are indeed this small at $z\gtrsim8$, we expect Ly$\alpha$ to be significantly weakened relative to our baseline $z\simeq5-6$ model. 
However these predictions are very sensitive to the nature of the ionizing sources. 
If reionization is driven by more massive galaxies \citep[e.g.,][]{Naidu2020}, we would expect the ionized volume in the IGM to be dominated by larger structures, with bubbles in excess of $1-2$~pMpc potentially present at $z\gtrsim8$.

Observationally we still have very few Ly$\alpha$ detections at $z\gtrsim8$, with results mostly revealing upper limits \citep[e.g.,][]{Bunker2023b,Curtis-Lake2023,Fujimoto2023,Nakajima2023,Harikane2024}. 
If large bubbles are present, we should begin to find signatures of them in Ly$\alpha$ datasets with {\it JWST}. 
We can use our $z\simeq5-6$ distributions to investigate the range of Ly$\alpha$ lines that are likely to be found at $z\gtrsim8$. 
We consider a range of bubble sizes likely to be common at such high redshifts ($0.1$, $0.3$, and $0.5$~pMpc), assuming the neutral fraction is very large (i.e., $x_{\rm{HI}}>0.8$). 
The IGM will strongly attenuate Ly$\alpha$ at $z\gtrsim8$, so the galaxies that do present Ly$\alpha$ are likely those that would appear as the very strongest line emitters (EW $>50$~\AA) at $z\simeq5-6$. 
We thus consider Ly$\alpha$ profiles from our $z\simeq5-6$ composites as our input ``intrinsic'' models prior to attenuation from the IGM H~{\small I} damping wing. 
We use both a moderate EW Ly$\alpha$ composite (EW $=50-100$~\AA) and one that includes extremely strong Ly$\alpha$ (EW $=100-500$~\AA). 
The composites are derived following the procedures described in Section~\ref{sec:lya_offset}. 
Both stacks reveal roughly similar profiles, with Ly$\alpha$ peaked at redshifted velocity of $\simeq210$~km~s$^{-1}$ (black dotted lines in Fig.~\ref{fig:lya_bubble}). 
We then apply the H~{\small I} damping wing opacity to these profiles, assuming our three different ionized bubble sizes. 
In this simple model, we assume that H~{\small I} transitions from fully ionized to fully neutral at the bubble radius. 


\begin{figure*}
\begin{center}
\includegraphics[width=\linewidth]{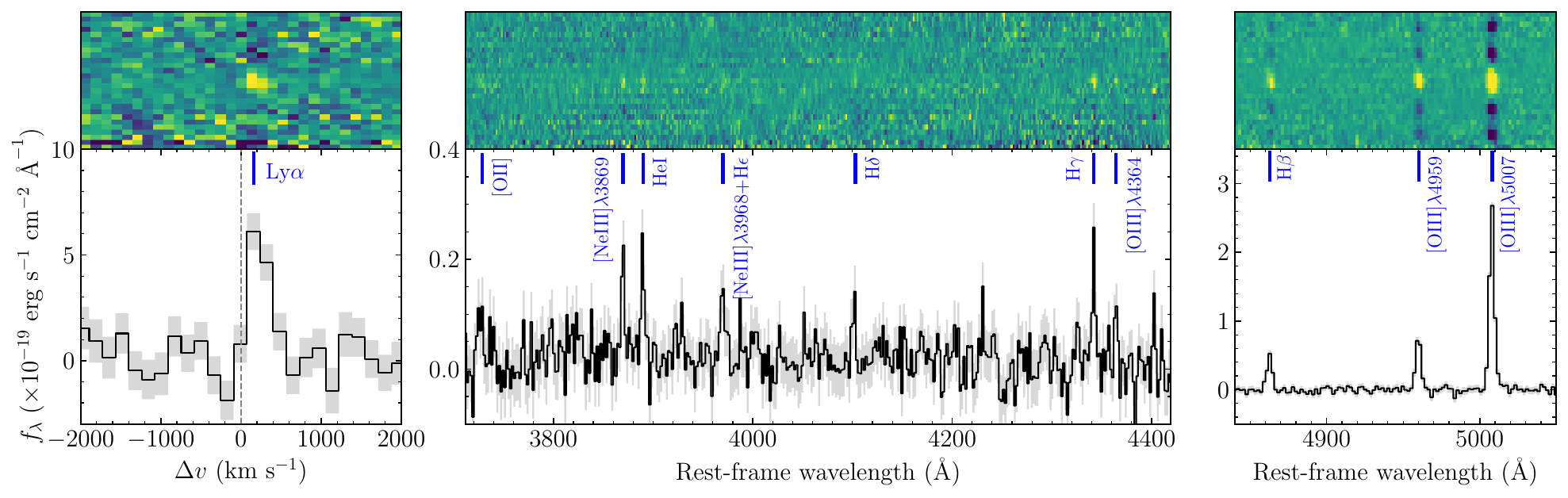}
\caption{JADES 2D (top) and 1D (bottom) NIRSpec medium resolution grating spectra of the $z=8.49$ Ly$\alpha$ emitter JADES-GS-z8.5-LAE. The error spectra are shown as the grey-shaded regions. Left panel: G140M/F070LP spectrum showing the region near the Ly$\alpha$ emission line. The detected Ly$\alpha$ line has EW $=17$~\AA\ and velocity offset $\Delta v_{{\rm Ly}\alpha}=156$~km~s$^{-1}$. Middle and right panels: G395M/F290LP spectrum showing rest-frame optical emission lines.}
\label{fig:z8p5_lae_spec}
\end{center}
\end{figure*}

The results are shown in Fig.~\ref{fig:lya_bubble}. 
Here we focus on the moderate EW Ly$\alpha$ composite (EW $=50-100$~\AA), but the results do not differ if we were instead to adopt the stronger Ly$\alpha$ composite. 
Both have similar line profiles and thus will have comparable attenuation from the IGM damping wing. 
We see that the Ly$\alpha$ transmission tracks the bubble size, increasing from $\mathcal{T}_{{\rm Ly}\alpha}=0.20$ ($R=0.1$~pMpc) to $0.33$ ($R=0.3$~pMpc), and finally to $0.44$ ($R=0.5$~pMpc). 
The IGM damping wing removes preferentially more flux near the line center, which can shift the emergent Ly$\alpha$ profile to higher velocities. 
This is most prominent in galaxies situated in the smallest bubbles. 
In the mock Ly$\alpha$ spectrum, we see that in the $R=0.1$ pMpc bubble the peak velocity shifts from $210$~km~s$^{-1}$ to $328$~km~s$^{-1}$. 
Even if we consider an input Ly$\alpha$ profile with small velocity offset of $\simeq100$~km~s$^{-1}$ (i.e., consistent with the lowest Ly$\alpha$ velocity offsets seen in our $z\simeq5-6$ sample), we still find that in the $R=0.1$ pMpc bubble the peak velocity shifts to $\gtrsim300$~km~s$^{-1}$. 
In the $R=0.3$ and $0.5$~pMpc bubbles, this effect is less significant given the reduced IGM opacities associated with these bubble sizes. 

Given the IGM transmission factors derived above, we expect that the strongest Ly$\alpha$ emitters in our baseline $z\simeq5-6$ sample (EW $>100$~\AA) would have much weaker lines ($\gtrsim20-40$~\AA) at $z\gtrsim8$. 
Recalling that only $10$~per~cent of $z\simeq5-6$ galaxies have Ly$\alpha$ with EW $>100$~\AA, this suggests that we must observe of order 10 $z\gtrsim8$ galaxies to recover a Ly$\alpha$ line with EW $>20$~\AA\ if typical bubble sizes are in the range $0.1-0.5$~pMpc. 
The velocity offsets of these strong Ly$\alpha$ emitters should be close to the intrinsic values seen at $z\simeq5-6$ ($\simeq200$~km~s$^{-1}$), provided bubble sizes are $R=0.3~$pMpc and larger. 
If we were to discover stronger Ly$\alpha$ emitters at $z\gtrsim8$, it is likely to be a signpost of an unexpected population of yet larger bubbles. 
We must further account for the aperture corrections required to predict the Ly$\alpha$ flux in the NIRSpec microshutters. 
We have estimated (see Section~\ref{sec:lya_flux}) that we are likely to recover $\simeq80$~per~cent of the emission recovered in our ground-based surveys, where this estimate is based on the expected Ly$\alpha$ surface brightness profiles (assuming they do not evolve with redshift), suggesting these EWs may be somewhat lower ($\gtrsim16-32$~\AA). 
Galaxies seen with moderate EW Ly$\alpha$ emission at $z\simeq5-6$ (EW $=50$~\AA) would be even weaker ($8-16$~\AA) at $z\gtrsim8$.
We note that these values may change slightly if the surface brightness profiles are different at $z\gtrsim8$, or if galaxies are significantly off-centered in the microshutter. 
Regardless, these estimates give a blueprint for what the strongest Ly$\alpha$ emitters are likely to look like at $z\gtrsim8$, both in terms of their EW and velocity offset. 
We apply this blueprint to the existing {\it JWST}/NIRSpec public database in the next subsection.

\subsection{New detection of Ly$\alpha$ emission at $z\simeq8.5$ with low $\Delta v_{{\rm Ly}\alpha}$} \label{sec:z8p5_lae}

With spectroscopic samples rapidly growing at $z\gtrsim8$ \citep[e.g.,][]{Bunker2023b,Curtis-Lake2023,Fujimoto2023,Nakajima2023,Harikane2024}, it is now possible to extend the search for Ly$\alpha$ emitting galaxies to this early epoch. 
To date, there are only three galaxies in the literature at $z>8$ with robust Ly$\alpha$ detections \citep{Zitrin2015,Larson2022,Bunker2023a}. 
All these three galaxies have large velocity offsets ($\Delta v_{{\rm Ly}\alpha}\gtrsim400$~km~s$^{-1}$) and relatively weak Ly$\alpha$ (EW $\lesssim10-20$~\AA; \citealt{Zitrin2015,Larson2022,Larson2023,Bunker2023a,Tang2023}). 
Two of these three galaxies with Ly$\alpha$ detections were identified in the Extended Groth Strip (EGS; \citealt{Davis2007}) field at $z\simeq8.7$ \citep{Zitrin2015,Larson2022}. 
Given their close proximity ($\simeq4$~pMpc) and the potential location of a galaxy overdensity in their surroundings \citep{Leonova2022,Whitler2024}, it has been suggested that these two systems trace a large ionized bubble at $z\simeq8.7$. 
An additional Ly$\alpha$ emitter at $z=7.98$, JADES-GS+53.15682-27.76716, has been reported in the literature \citep{Jones2024,Saxena2024,Witstok2024a}. 
This galaxy has a relatively low Ly$\alpha$ velocity offset ($\Delta v_{{\rm Ly}\alpha}=167$~km~s$^{-1}$) and strong Ly$\alpha$ emission (EW $\simeq29$~\AA) among the existing $z\gtrsim8$ galaxies with Ly$\alpha$ measurements. 
Further progress will require identification of more Ly$\alpha$ emitters at $z\gtrsim8$, with particular attention to those with elevated EWs ($>15$~\AA) and small velocity offsets ($\lesssim 200$~km~s$^{-1}$) which may be signposts of moderate or larger size ($R\gtrsim0.3$~pMpc) ionized bubbles (Section~\ref{sec:lya_pred}).

As part of an ongoing effort to build a large database of Ly$\alpha$ measurements in the reionization era, we have identified a new Ly$\alpha$ detection at $z\simeq8.5$, JADES-GS+53.15891-27.76508 (hereafter JADES-GS-z8.5-LAE), in the JADES Cycle 2 program 3215 \citep{Eisenstein2023b,DEugenio2024}. 
The discovery of this Ly$\alpha$ emitting galaxy is also presented and discussed in \citet[][ID: JADES-GS-z8-0-LA therein]{Witstok2024b}. 
This object was initially selected as a F105W-dropout from {\it HST} imaging in the HUDF \citep{Bouwens2010a,Bunker2010,McLure2010}. 
\citet{Lehnert2010} obtained spectroscopic observations of this galaxy with VLT/SINFONI \citep{Eisenhauer2003}, presenting a $6\sigma$ detection of Ly$\alpha$ emission at $z=8.5549$ with a line flux of $6.1\times10^{-18}$~erg~s$^{-1}$~cm$^{-2}$ and a rest-frame EW $\simeq200$~\AA. 
Later \citet{Bunker2013} reported the spectroscopic observations on this source with VLT/XSHOOTER \citep{DOdorico2006} and Subaru/MOIRCS \citep{Ichikawa2006}. 
These observations were unable to reproduce the line reported in \citet{Lehnert2010}, placing a $2\sigma$ upper limit on the line flux of $\lesssim2\times10^{-18}$~erg~s$^{-1}$~cm$^{-2}$. 
Then NIRCam imaging observations of JADES-GS-z8.5-LAE were obtained from the JADES Cycle 1 program \citep{Eisenstein2023a}. 
It was also selected as a candidate high redshift galaxy with photometric redshift $z_{\rm phot}=8.5$ using the dropout technique in \citet{Hainline2024b}. 
NIRSpec observations of this source were performed as a part of the program 3215 using the MSA in October 2023 over five sub-pointings. 
Each pointing has an exposure time of $33263$~s for the low spectral resolution ($R\sim100$) PRISM/CLEAR setup, and $8316$~s and $33263$~s for the medium resolution (MR; $R\sim1000$) G140M/F070LP and G395M/F290LP grating/filter setups, respectively. 
We refer readers to \citet{DEugenio2024} for details of the target selection and the follow up NIRSpec observations. 

The NIRSpec spectra used here were reduced following the procedures described in \citet{Tang2023}. 
In Fig.~\ref{fig:z8p5_lae_spec}, we show the MR grating spectra of JADES-GS-z8.5-LAE. 
Strong rest-frame optical emission lines (H$\beta$, [O~{\small III}]) are clearly seen in both the prism and the G395M/F290LP spectra. 
We derive the systemic redshift $z_{\rm sys}=8.4858\pm0.0004$ by simultaneously fitting those strong rest-frame optical lines measured in G395M/F290LP spectrum with Gaussian profiles. 
The Ly$\alpha$ emission line of JADES-GS-z8.5-LAE is detected in the G140M/F070LP spectrum (Fig.~\ref{fig:z8p5_lae_spec}). 
We measure the Ly$\alpha$ flux by directly integrating the flux density within a $\pm1000$~km~s$^{-1}$ window, obtaining $F_{{\rm Ly}\alpha}=7.44\pm1.27\times10^{-19}$~erg~s$^{-1}$~cm$^{-2}$. 
The rest-frame UV continuum of this galaxy is below the noise fluctuation in the G140M/F070LP spectrum, but is well detected in the prism spectrum. 
We measure the continuum flux density near the Ly$\alpha$ emission from the prism spectrum following the methods in \citet{Chen2024}. 
We derive that the rest-frame Ly$\alpha$ EW is $21\pm4$~\AA. 
Using the peak of the Ly$\alpha$ emission line measured in the G140M/F070LP spectrum, we calculate the Ly$\alpha$ redshift $z_{{\rm Ly}\alpha}=8.4907\pm0.0005$. 
This results in a Ly$\alpha$ velocity offset $\Delta v_{{\rm Ly}\alpha}=156\pm20$~km~s$^{-1}$. 
We also constrain the Ly$\alpha$ escape fraction of JADES-GS-z8.5-LAE using the H$\beta$ emission line flux (Table~\ref{tab:z8p5_lines}). 
Assuming case B recombination with $T_{\rm e}=10^4$~K, $n_{\rm e}=10^2$~cm$^{-3}$, the intrinsic Ly$\alpha$/H$\beta$ luminosity ratio is $24.9$. 
By measuring of the H$\gamma$/H$\beta$ ratio, we derive the dust attenuation is relatively small (details in the next paragraph) and correct the observed H$\beta$ luminosity to the intrinsic value. 
Then we calculate the Ly$\alpha$ escape fraction $f_{{\rm esc,Ly}\alpha}=0.10\pm0.02$. 
The Ly$\alpha$ emission line of JADES-GS-z8.5-LAE detected in the NIRSpec spectra is $\gtrsim2000$~km~s$^{-1}$ bluewards the line shown in the VLT/SINFONI spectrum ($z=8.5549$; \citealt{Lehnert2010}), with much lower line flux and EW than those reported in \citet[][$F_{{\rm Ly}\alpha}=6.1\times10^{-18}$~erg~s$^{-1}$~cm$^{-2}$, EW$_{{\rm Ly}\alpha}\simeq200$~\AA]{Lehnert2010}. 
These results are in disagreement with \citet{Lehnert2010}, consistent with the upper limit ($F_{{\rm Ly}\alpha}<2\times10^{-18}$~erg~s$^{-1}$~cm$^{-2}$) placed in \citet{Bunker2013}. 

Before discussing the potential bubble size this $z\simeq8.5$ source might sit in, we first investigate the physical properties of the galaxy, characterizing the rest-frame optical emission line properties and the NIRCam SED. 
Our analysis follows that presented in \citet{Tang2023}. 
We present the detected emission lines of JADES-GS-z8.5-LAE in Table~\ref{tab:z8p5_lines}. 
We find strong [O~{\small III}] emission lines with [O~{\small III}]~$\lambda5007$ EW $=1250$~\AA, consistent with systems dominated by very young stellar populations \citep[e.g.,][]{Tang2019}. 
The Balmer emission line detections (H$\gamma$, H$\beta$) allow us to estimate the dust attenuation in nebular gas. 
We measure H$\gamma$/H$\beta=0.442^{+0.097}_{-0.087}$, close to the intrinsic H$\gamma$/H$\beta$ ratio expected in the case B recombination ($0.468$, assuming $T_{\rm e}=10^4$~K; \citealt{Osterbrock2006}). 
This implies a relatively small extinction with $E(B-V)=0.11$ assuming the \citet{Cardelli1989} extinction curve. 
With dust attenuation inferred from Balmer decrement, we quantify the ionization-sensitive [O~{\small III}]/[O~{\small II}] (O32; dust-corrected) and [Ne~{\small III}]/[O~{\small II}] (Ne3O2) ratios. 
These line ratios are large, with O32 $=18^{+8}_{-4}$ and Ne3O2 $=1.1^{+0.6}_{-0.4}$. 
These values are consistent with those found in galaxies at $z=5-9$ \citep[e.g.,][]{Cameron2023a,Sanders2023,Saxena2024,Tang2023}, well above the average ratios measured in galaxies at $z<5$ (O32 $\simeq1-5$ and Ne3O2 $\simeq0.1-0.4$; e.g., \citealt{Sanders2016,Steidel2016,Shapley2023}). 
This indicates extreme ionizing conditions in this galaxy. 
We detect a tentative [O~{\small III}]~$\lambda4363$ emission line with S/N $=3$, enabling us to constrain the gas-phase oxygen abundance with direct method. 
Following the procedures in \citet{Izotov2006}, we derive that the nebular gas of JADES-GS-z8.5-LAE is very metal-poor with $12+\log{({\rm O/H})}=7.37^{+0.21}_{-0.10}$ ($0.05^{+0.02}_{-0.01}\ Z_{\odot}$, where the solar metallicity corresponds to a gas-phase oxygen abundance $12+\log{({\rm O/H})}=8.71$; \citealt{Gutkin2016}). 
Using the strong-line ratio R23 $\equiv{\rm ([O~{\small III}]+[O~{\small II}])/H}\beta$ ($=7.0^{+0.9}_{-0.6}$) and applying the empirical metallicity calibration derived from $z=2-9$ galaxies \citep{Sanders2024}, we derive the gas-phase oxygen abundance $12+\log{({\rm O/H})}=7.31^{+0.11}_{-0.09}$ similar to that inferred from direct method.


\begin{table}
\centering
\begin{tabular}{ccc}
\hline
Line & $\lambda_{\rm rest,vacuum}$ & Flux \\
 & (\AA) & ($\times10^{-20}$ erg s$^{-1}$ cm$^{-2}$) \\
\hline
Ly$\alpha$ & $1215.67$ & $74.4\pm12.7$ \\
$[{\rm O}$~{\scriptsize II}$]$ & $3727.1,3729.9$ & $7.2\pm2.2$ \\
$[{\rm Ne}$~{\scriptsize III}$]$ & $3869.8$ & $7.6\pm2.5$ \\
He~{\scriptsize I} & $3889.8$ & $8.2\pm2.1$ \\
$[{\rm Ne}$~{\scriptsize III}$]$+H$\epsilon$ & $3968.6,3971.2$ & $10.3\pm2.1$ \\
H$\delta$ & $4102.9$ & $6.0\pm1.8$ \\
H$\gamma$ & $4341.7$ & $9.6\pm1.7$ \\
$[{\rm O}$~{\scriptsize III}$]$ & $4364.4$ & $4.2\pm1.4$ \\
H$\beta$ & $4862.7$ & $21.6\pm2.5$ \\
$[{\rm O}$~{\scriptsize III}$]$ & $4960.3$ & $34.3\pm2.9$ \\
$[{\rm O}$~{\scriptsize III}$]$ & $5008.2$ & $110.0\pm2.6$ \\
\hline
\end{tabular}
\caption{Emission lines detected in the NIRSpec spectrum of JADES-GS+53.15891-27.76508. Line fluxes are measured from the MR grating spectrum.}
\label{tab:z8p5_lines}
\end{table}

\begin{figure}
\begin{center}
\includegraphics[width=\linewidth]{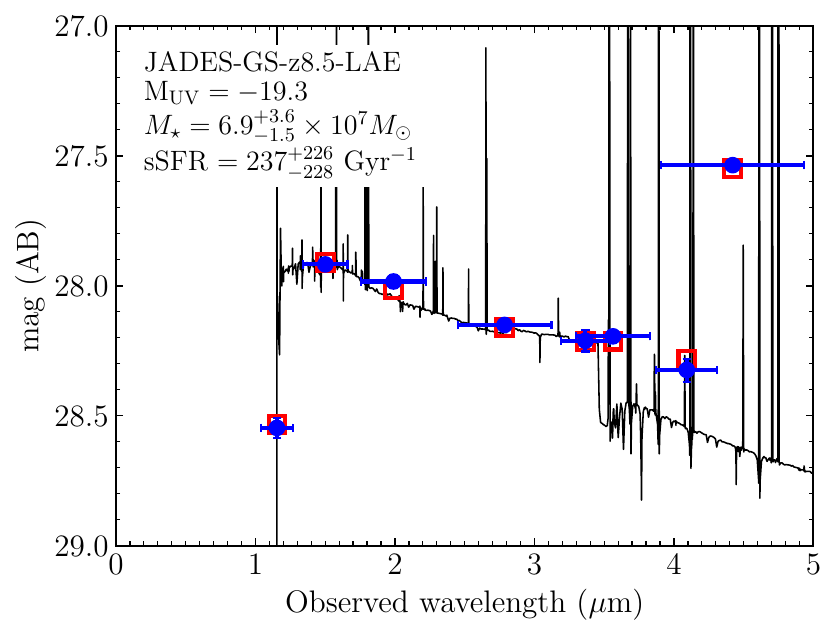}
\caption{JADES NIRCam SED of JADES-GS-z8.5-LAE. Observed NIRCam photometry is shown by blue circles. The \textsc{beagle} model spectrum is shown by the black line and the synthetic photometry is presented by open red squares.}
\label{fig:z8p5_lae_sed}
\end{center}
\end{figure}

The NIRCam photometry of JADES-GS-z8.5-LAE is computed following the procedures described in \citet{Rieke2023b} which will also be fully described in Robertson et al. (in prep.). 
The SED is shown in Fig.~\ref{fig:z8p5_lae_sed}, demonstrating that it has M$_{\rm UV}=-19.3$. 
The UV slope is blue ($\beta=-2.2$), consistent with the low dust attenuation inferred from Balmer decrement measurement. 
We fit the NIRCam SED of this galaxy with \textsc{beagle} models following the same methods described in Section~\ref{sec:sed}. 
The SED fitting results demonstrate that JADES-GS-z8.5-LAE has relatively low stellar mass with $M_{\star}=6.9^{+3.6}_{-1.5}\times10^7\ M_{\odot}$. 
The rest-frame UV to optical light of this object is dominated by very young stellar populations (luminosity-weighted age $=1.6^{+0.7}_{-0.4}$~Myr), as expected for galaxies that have undergone a recent upturn in star formation.


The discovery of Ly$\alpha$ emission at $z\simeq8.5$ with a small velocity offset from the line center is suggestive of a moderate bubble size. 
Based on our mock Ly$\alpha$ profiles presented in Section~\ref{sec:lya_pred}, we would expect extremely small bubbles ($\lesssim0.1$~pMpc) to have shifted line emission to larger velocities ($>300$~km~s$^{-1}$). 
We may therefore expect the galaxy to be situated in a bubble of order $0.3$~pMpc or larger, significantly above the median value expected if the IGM neutral fraction is $x_{\rm{HI}}\simeq0.9$. 
If the bubble is $0.3$ or $0.5$~pMpc, the Ly$\alpha$ EW would be attenuated by $2-3\times$ by the IGM damping wing. 
This would imply the galaxy would have been observed with Ly$\alpha$ EW $=40-50$~\AA\ at $z\simeq5-6$ (or $50-60$~\AA\ after accounting for the small aperture correction). 

More robust inferences of bubble sizes will require Ly$\alpha$ observations of other galaxies surrounding JADES-GS-z8.5-LAE. 
If the galaxy is situated in a moderate-sized bubble, we would expect to see numerous other ionizing sources surrounding the galaxy, many of which should also show Ly$\alpha$. 
We search for additional sources at $z\simeq8.5$ within $2$~arcmin (physical separation $\simeq0.6$~pMpc in projection at $z=8.5$) from JADES-GS-z8.5-LAE. 
We utilize the up-to-date version of the JADES photometric redshift catalog created by \citet{Hainline2024b}. 
The photometric redshifts in \citet{Hainline2024b} catalog were computed using the \textsc{eazy} code \citep{Brammer2008}, taking advantage of {\it JWST}/NIRCam and {\it HST}/ACS imaging data. 
We select sources with $8.3\le z_{\rm phot}\le 8.7$ and $<2$~arcmin away from JADES-GS-z8.5-LAE. 
To ensure the photometric redshifts are robust, we additionally require the $1\sigma$ confidence interval of the photometric redshift of each object is $z_{\rm l68}\ge8.1$ and $z_{\rm u68}\le8.9$, where $z_{\rm l68}$ and $z_{\rm u68}$ are the lower and upper bound of the $1\sigma$ confidence interval. 
Adopting the above criteria we identify $24$ sources at $z\simeq8.5$ within $2$~arcmin around JADES-GS-z8.5-LAE (Table~\ref{tab:z8p5_od}). 
Most of these $24$ galaxies are less luminous than JADES-GS-z8.5-LAE (M$_{\rm UV}=-19.3$), with M$_{\rm UV}$ spanning a range from $-19.7$ to $-16.8$ with a median of M$_{\rm UV}=-17.9$. 


\begin{table}
\centering
\begin{tabular}{ccc}
\hline
JADES ID & $z_{\rm phot}$ & M$_{\rm UV}$ \\
\hline
JADES-GS+53.13364-27.77895 & $8.30^{+0.14}_{-0.17}$ & $-18.2\pm0.1$ \\
JADES-GS+53.13939-27.78334 & $8.35^{+0.09}_{-0.24}$ & $-18.0\pm0.1$ \\
JADES-GS+53.15047-27.79684 & $8.37^{+0.05}_{-0.24}$ & $-17.4\pm0.1$ \\
JADES-GS+53.13621-27.77716 & $8.43^{+0.19}_{-0.26}$ & $-17.8\pm0.1$ \\
JADES-GS+53.18075-27.76077 & $8.43^{+0.21}_{-0.26}$ & $-17.4\pm0.1$ \\
JADES-GS+53.19495-27.78067 & $8.45^{+0.14}_{-0.22}$ & $-18.1\pm0.1$ \\
JADES-GS+53.15576-27.75095 & $8.45^{+0.03}_{-0.30}$ & $-17.5\pm0.1$ \\
JADES-GS+53.13356-27.77870 & $8.48^{+0.21}_{-0.22}$ & $-18.1\pm0.2$ \\
JADES-GS+53.17292-27.76656 & $8.48^{+0.34}_{-0.37}$ & $-17.1\pm0.1$ \\
JADES-GS+53.16326-27.76361 & $8.48^{+0.08}_{-0.07}$ & $-17.5\pm0.1$ \\
JADES-GS+53.16717-27.75152 & $8.49^{+0.22}_{-0.24}$ & $-17.9\pm0.1$ \\
JADES-GS+53.17861-27.74042 & $8.50^{+0.24}_{-0.34}$ & $-18.5\pm0.1$ \\
JADES-GS+53.16447-27.80218$^{\rm a}$ & $8.51^{+0.01}_{-0.02}$ & $-18.7\pm0.1$ \\
JADES-GS+53.17770-27.78477 & $8.52^{+0.12}_{-0.03}$ & $-18.2\pm0.1$ \\
JADES-GS+53.18354-27.77014$^{\rm b}$ & $8.54^{+0.20}_{-0.02}$ & $-19.7\pm0.1$ \\
JADES-GS+53.16337-27.77569 & $8.54^{+0.20}_{-0.18}$ & $-17.8\pm0.1$ \\
JADES-GS+53.16668-27.76126 & $8.54^{+0.31}_{-0.41}$ & $-17.3\pm0.2$ \\
JADES-GS+53.13377-27.78061 & $8.54^{+0.26}_{-0.39}$ & $-18.0\pm0.3$ \\
JADES-GS+53.14091-27.77279 & $8.56^{+0.20}_{-0.03}$ & $-17.7\pm0.1$ \\
JADES-GS+53.15106-27.75698 & $8.59^{+0.17}_{-0.40}$ & $-18.1\pm0.1$ \\
JADES-GS+53.14453-27.75413 & $8.60^{+0.19}_{-0.36}$ & $-17.7\pm0.1$ \\
JADES-GS+53.17257-27.79305 & $8.65^{+0.19}_{-0.31}$ & $-16.8\pm0.2$ \\
JADES-GS+53.17121-27.76998 & $8.68^{+0.21}_{-0.12}$ & $-16.9\pm0.2$ \\
JADES-GS+53.18467-27.79088 & $8.69^{+0.13}_{-0.07}$ & $-18.0\pm0.1$ \\
\hline
\end{tabular}
\caption{JADES sources at $z\simeq8.5$ within $2$~arcmin from JADES-GS-z8.5-LAE identified in \citet{Hainline2024b} photometric redshift catalog. a: JADES-GS+53.16447-27.80218 has spectroscopic redshift ($z_{\rm spec}=8.473$) measured from JADES NIRSpec spectrum \citep{Bunker2023b}. b: JADES-GS+53.18354-27.77014 has spectroscopic redshift ($z_{\rm spec}=8.385$) measured from FRESCO NIRCam grism spectrum (Sun et al. in prep.).}
\label{tab:z8p5_od}
\end{table}

Among the $24$ photometric redshift selected galaxies around JADES-GS-z8.5-LAE, two have already been spectroscopically confirmed. 
JADES-GS+53.18354-27.77014 has spectroscopic redshift measured from the FRESCO dataset \citep{Oesch2023}. 
The resolved [O~{\small III}]~$\lambda4959$ and [O~{\small III}]~$\lambda5007$ doublet is detected in the NIRCam F444W grism spectrum obtained by the FRESCO survey, and the spectroscopic redshift measured for this galaxy is $z_{\rm spec}=8.385$ (Sun et al. in prep.). 
This galaxy is the brightest source among the above $24$ objects around JADES-GS-z8.5-LAE, with $H_{150}=27.5$ and M$_{\rm UV}=-19.7$. 
In physical distance, JADES-GS+53.18354-27.77014 is $2.9$~pMpc away from JADES-GS-z8.5-LAE, likely outside of any bubble that exists around the Ly$\alpha$ emitter.
JADES-GS+53.16447-27.80218 is another galaxy with spectroscopic observation around JADES-GS-z8.5-LAE. 
It was observed with NIRSpec in the JADES program 1210, and the spectroscopic measurements were presented in \citet{Bunker2023b}. 
[O~{\small III}] and H$\beta$ emission lines of this galaxy are detected from the NIRSpec spectrum, revealing a spectroscopic redshift $z_{\rm spec}=8.473$. 
This object has $H_{150}=28.5$ and M$_{\rm UV}=-18.7$ and is only $0.7$~pMpc away from JADES-GS-z8.5-LAE, much closer than the other spectroscopically confirmed source. 
Deep spectroscopic follow-up of the other 22 photometric sources should better characterize the overdensity around JADES-GS-z8.5-LAE. 
If the bubble is large, we would expect additional sources to show Ly$\alpha$ with small velocity offsets. 
A separate analysis of the overdensity and the ionized bubble around JADES-GS-z8.5-LAE is presented in \citet{Witstok2024b}.


\section{Summary} \label{sec:summary}

{\it JWST} has recently sparked a new era of Ly$\alpha$ spectroscopy at $z\gtrsim7$. 
To fully realize the potential of these observations to provide a new insight into reionization, we need a much-improved understanding of Ly$\alpha$ emission in galaxies at $z\simeq5-6$ when the IGM is mostly ionized. 
Using Ly$\alpha$ emission lines measured from ground-based Keck and VLT spectroscopic surveys and {\it JWST} observations from the JADES and FRESCO surveys, we characterize the Ly$\alpha$ EWs, escape fractions, and velocity offsets of $z\simeq5-6$ galaxies. 
These measurements are meant to provide an ``intrinsic'' model to interpret the impact of the IGM H~{\small I} damping wing absorption on Ly$\alpha$ emission at $z\gtrsim7-12$.
We summarize our key results below. 

1. We identify $79$ galaxies with Ly$\alpha$ and H$\alpha$ emission line detections at $z\simeq5-6$, and we measure the Ly$\alpha$ EWs and Ly$\alpha$ escape fractions. 
We investigate the nature of Ly$\alpha$ emitters with EW $>100$~\AA, a population that is becoming very important in efforts to study the IGM at $z\gtrsim7$ \citep{Saxena2023,Chen2024}. 
We find that the most significant difference between galaxies with moderate-EW Ly$\alpha$ (EW $=10-50$~\AA) and extremely strong Ly$\alpha$ ($>100-300$~\AA) is the transmission of Ly$\alpha$ through the ISM and CGM, with the most intense Ly$\alpha$ emitters tending to be those that leak over $60$~per~cent of their Ly$\alpha$ emission (compared to $10-20$~per~cent for the weaker line emitters). 
While Ly$\alpha$ EW also increases with ionizing photon production efficiency, we find that this quantity is not significantly different between moderate-EW and strong Ly$\alpha$ emitters.

2. We characterize the Ly$\alpha$ velocity offsets of the $79$ Ly$\alpha$ emitters with H$\alpha$ detections. 
We find significant evolution in the velocity profiles of the strongest $z\simeq5-6$ Ly$\alpha$ emitters with respect to those at $z\simeq2-3$. 
At the lower redshifts, very intense Ly$\alpha$ emitters ($>100$~\AA) tend to have profiles that peak near the line center, likely reflecting direct escape of Ly$\alpha$ through very low H~{\small I} density channels in the ISM and CGM. 
However at $z\simeq5-6$, we find that the strongest Ly$\alpha$ emitters are significantly redshifted from the line center (median $\simeq200$~km~s$^{-1}$) with negligible Ly$\alpha$ central escape fractions. 
Galaxies with low velocity offsets ($<100$~km~s$^{-1}$) are extremely rare in our $z\simeq5-6$ sample. 
The disappearance of Ly$\alpha$ emitters with very low velocity offsets and large Ly$\alpha$ central escape fractions at $z\simeq5-6$ is likely driven by the IGM, with the residual H~{\small I} fraction ($x_{\rm HI}\gtrsim10^{-4}$; e.g., \citealt{Yang2020b,Bosman2022}) large enough to resonantly scatter line photons near the line center given the high IGM density at $z\simeq5-6$. 
Given the link between low velocity offsets and LyC leakage \citep[e.g.,][]{Verhamme2015,Choustikov2024b}, these results suggest that strong Ly$\alpha$ emitters may not provide the best indicator of ionizing photon escape at $z\simeq5-6$.

3. The blue side of Ly$\alpha$ is also strongly attenuated by the mostly-ionized IGM at $z\simeq5-6$. 
Several recent detections of blue peaks at these redshifts have been challenging to explain, requiring either inflows or locally intense radiation fields. 
We constrain the strength of blue peak Ly$\alpha$ emission for the $79$ galaxies with Ly$\alpha$ and H$\alpha$ detections in our $z\simeq5-6$ sample. 
In $77$ of the $79$ galaxies we do not find blue peak Ly$\alpha$ emission. 
The average blue-to-red flux ratio ($<0.04$ at $5\sigma$) is much smaller than that of Ly$\alpha$ emitters at $z\simeq0-2$ ($\simeq0.3$; e.g., \citealt{Hayes2021,Matthee2021}), consistent with expectations given the increasing IGM opacity at $z\simeq5-6$. 
We identify blue peak Ly$\alpha$ emission in two galaxies in our sample, with blue-to-red flux ratio $=0.06$ and $0.07$. 
This is $5-10\times$ below the ratios found in several cases in the literature at $z\simeq5-6$. 
These results suggest that very prominent blue peaks are rare at $z\simeq5-6$.

4. We derive statistical distributions of Ly$\alpha$ properties in a Lyman break selected sample at $z\simeq5-6$, with the goal of providing baseline models for comparison against $z\gtrsim7$ studies. 
We find that galaxies with large Ly$\alpha$ escape fractions ($f_{{\rm esc,Ly}\alpha}>0.2$) or Ly$\alpha$ EWs ($>25$~\AA) are common at $z\simeq5-6$, comprising $\simeq30-40$~per~cent of the Lyman break selected population. 
Our results suggest that strong Ly$\alpha$ emission with EW $>25$~\AA\ becomes more common from $z\sim3$ to $z\sim5$, consistent with previous findings. 
We find that the evolution of the Ly$\alpha$ fraction begins to plateau between $z\sim5$ ($35^{+7}_{-7}$~per~cent) and $z\sim6$ ($28^{+10}_{-10}$~per~cent), likely reflecting the attenuation provided by the IGM at higher redshifts. 
We investigate the dependence of Ly$\alpha$ escape fraction and Ly$\alpha$ EW on galaxy properties, quantifying trends with UV luminosity, UV slope, and [O~{\small III}]+H$\beta$ EW. 

5. Using the statistical distributions at $z\simeq5-6$, we investigate the likely impact of the IGM damping wing on Ly$\alpha$ at $z\gtrsim8$. 
At these very high redshifts, little is still known about Ly$\alpha$. 
We demonstrate that typical lines are likely to be attenuated by $3-5\times$ owing to the strong damping wing associated with small ionized bubbles ($\lesssim 0.3$~pMpc). 
We show that small velocity offsets ($\Delta v_{{\rm Ly}\alpha}\lesssim250$~km~s$^{-1}$) are mostly likely to be observed in moderate-size bubbles ($\gtrsim0.3$~pMpc) at $z\gtrsim8$.

6. We present a recently-discovered Ly$\alpha$ emitter at $z=8.5$ from the JADES Cycle 2 program 3215 \citep{Eisenstein2023b}. 
This discovery is also described in \citet{Witstok2024b}. 
The systemic redshift is confidently determined from very strong rest-frame optical emission lines. 
We measure Ly$\alpha$ emission with EW $=17$~\AA\ and a relatively low Ly$\alpha$ velocity offset $\Delta v_{{\rm Ly}\alpha}=156$~km~s$^{-1}$. 
This is one of just five robustly confirmed Ly$\alpha$ emitters at $z\gtrsim8$. 
The small velocity offset may provide a signpost of a moderate-size bubble ($\gtrsim0.3$~pMpc) for the $z\simeq8.5$ Universe. 
In this case, we may expect numerous ionizing sources in the vicinity of the Ly$\alpha$ emitter. 
We identify $24$ photometric sources at $z\simeq8.5$ within $2$~arcmin from this galaxy, with two of them currently spectroscopically confirmed. 
Deep {\it JWST} spectroscopic follow-up of the neighboring sources will better characterize the overdensity and constrain the ionized bubble around the Ly$\alpha$ emitter.


\section*{Acknowledgements}

The authors thank the anonymous referee for insightful comments which improved the manuscript. 
We would like to thank the entire FRESCO team for their effort designing and executing this program and developing their observing program with a zero-exclusive-access period. 
We also thank Jorryt Matthee for kindly sharing data from the X-SHOOTER Lyman $\alpha$ survey at $z=2$ (XLS-$z2$; \citealt{Matthee2021}). 
MT acknowledges funding from the {\it JWST} Arizona/Steward Postdoc in Early galaxies and Reionization (JASPER) Scholar contract at the University of Arizona. 
DPS acknowledges support from the National Science Foundation through the grant AST-2109066. 
RSE acknowledges funding from the European Research Council (ERC) under the European Union's Horizon 2020 research and innovation program (grant agreement No. 669253).
FS acknowledges {\it JWST}/NIRCam contract to the University of Arizona NAS5-02015. 
BER acknowledges support from the NIRCam Science Team contract to the University of Arizona, NAS5-02015, and {\it JWST} Program 3215. 
SA acknowledges support from Grant PID2021-127718NB-I00 funded by the Spanish Ministry of Science and Innovation/State Agency of Research (MICIN/AEI/10.13039/501100011033). 
WB, RM, and JW acknowledge support from the Science and Technology Facilities Council (STFC), by the ERC through Advanced Grant 695671 ``QUENCH'', by the UKRI Frontier Research grant RISEandFALL.
RM also acknowledges funding from a research professorship from the Royal Society.
KB is supported by the Australian Research Council Centre of Excellence for All Sky Astrophysics in 3 Dimensions (ASTRO 3D), through project number CE170100013. 
AJB, JC, GCJ, and AS acknowledge funding from the “FirstGalaxies” Advanced Grant from the ERC under the European Union's Horizon 2020 research and innovation program (grant agreement No. 789056).
LW acknowledges support from the National Science Foundation Graduate Research Fellowship under Grant No. DGE-2137419. 
The research of CCW is supported by NOIRLab, which is managed by the Association of Universities for Research in Astronomy (AURA) under a cooperative agreement with the National Science Foundation.

This work is based in part on observations taken by the MUSE-Wide Survey and the MUSE {\it Hubble} Ultra Deep Field Survey as part of the MUSE Consortium. 
Part of the data presented in this work were obtained at Keck Observatory. 
The authors wish to recognize and acknowledge the very significant cultural role and reverence that the summit of Maunakea has always had within the Native Hawaiian community. 
We are most fortunate to have the opportunity to conduct observations from this mountain.
This research is based in part on observations made with the NASA/ESA/CSA {\it James Webb Space Telescope} and the NASA/ESA {\it Hubble Space Telescope} from the Space Telescope Science Institute, which are operated by the Association of Universities for Research in Astronomy, Inc., under NASA contract NAS 5-03127 for {\it JWST} and NAS 5-26555 for {\it HST}. 
These observations are associated with programs \# 1180, 1181, 3215, and 1895. 
The {\it JWST} and the {\it HST} data were obtained from the Mikulski Archive for Space Telescopes at the Space Telescope Science Institute. 
The authors acknowledge use of the lux supercomputer at UC Santa Cruz, funded by NSF MRI grant AST 1828315. 
This work is based in part upon High Performance Computing (HPC) resources supported by the University of Arizona TRIF, UITS, and Research, Innovation, and Impact (RII) and maintained by the UArizona Research Technologies department.

This research made use of the following software: 
\textsc{numpy} \citep{Harris2020}, \textsc{matplotlib} \citep{Hunter2007}, \textsc{scipy} \citep{Virtanen2020}, \textsc{astropy}, a community-developed core Python package for Astronomy \citep{AstropyCollaboration2013}, and \textsc{beagle} \citep{Chevallard2016}.


\section*{Data Availability}

The VLT/MUSE data used in this work are available from MUSE-Wide (\url{https://musewide.aip.de/project/}) and AMUSED (\url{https://amused.univ-lyon1.fr/}). 
The {\it HST} data utilized in this work are available from the {\it Hubble} Legacy Field archive (\url{https://archive.stsci.edu/prepds/hlf/}). 
The {\it JWST} data used here are available on the Mikulski Archive for Space Telescopes (\url{https://mast.stsci.edu/}). 
Other data underlying this article will be shared on reasonable request to the corresponding author.



\bibliographystyle{mnras}
\bibliography{z56_Lya} 

\begin{thebibliography}{}
\makeatletter
\relax
\def\mn@urlcharsother{\let\do\@makeother \do\$\do\&\do\#\do\^\do\_\do\%\do\~}
\def\mn@doi{\begingroup\mn@urlcharsother \@ifnextchar [ {\mn@doi@}
  {\mn@doi@[]}}
\def\mn@doi@[#1]#2{\def\@tempa{#1}\ifx\@tempa\@empty \href
  {http://dx.doi.org/#2} {doi:#2}\else \href {http://dx.doi.org/#2} {#1}\fi
  \endgroup}
\def\mn@eprint#1#2{\mn@eprint@#1:#2::\@nil}
\def\mn@eprint@arXiv#1{\href {http://arxiv.org/abs/#1} {{\tt arXiv:#1}}}
\def\mn@eprint@dblp#1{\href {http://dblp.uni-trier.de/rec/bibtex/#1.xml}
  {dblp:#1}}
\def\mn@eprint@#1:#2:#3:#4\@nil{\def\@tempa {#1}\def\@tempb {#2}\def\@tempc
  {#3}\ifx \@tempc \@empty \let \@tempc \@tempb \let \@tempb \@tempa \fi \ifx
  \@tempb \@empty \def\@tempb {arXiv}\fi \@ifundefined
  {mn@eprint@\@tempb}{\@tempb:\@tempc}{\expandafter \expandafter \csname
  mn@eprint@\@tempb\endcsname \expandafter{\@tempc}}}

\bibitem[\protect\citeauthoryear{{Almada Monter} \& {Gronke}}{{Almada Monter}
  \& {Gronke}}{2024}]{Almada-Monter2024}
{Almada Monter} S.,  {Gronke} M.,  2024, \mn@doi [arXiv e-prints]
  {10.48550/arXiv.2404.07169}, \href
  {https://ui.adsabs.harvard.edu/abs/2024arXiv240407169A} {p. arXiv:2404.07169}

\bibitem[\protect\citeauthoryear{{Ando}, {Ohta}, {Iwata}, {Akiyama}, {Aoki}  \&
  {Tamura}}{{Ando} et~al.}{2006}]{Ando2006}
{Ando} M.,  {Ohta} K.,  {Iwata} I.,  {Akiyama} M.,  {Aoki} K.,   {Tamura} N.,
  2006, \mn@doi [\apjl] {10.1086/505652}, \href
  {https://ui.adsabs.harvard.edu/abs/2006ApJ...645L...9A} {645, L9}

\bibitem[\protect\citeauthoryear{{Ao} et~al.,}{{Ao} et~al.}{2020}]{Ao2020}
{Ao} Y.,  et~al., 2020, \mn@doi [Nature Astronomy] {10.1038/s41550-020-1033-3},
  \href {https://ui.adsabs.harvard.edu/abs/2020NatAs...4..670A} {4, 670}

\bibitem[\protect\citeauthoryear{{Arrabal Haro} et~al.,}{{Arrabal Haro}
  et~al.}{2018}]{ArrabalHaro2018}
{Arrabal Haro} P.,  et~al., 2018, \mn@doi [\mnras] {10.1093/mnras/sty1106},
  \href {https://ui.adsabs.harvard.edu/abs/2018MNRAS.478.3740A} {478, 3740}

\bibitem[\protect\citeauthoryear{{Astropy Collaboration} et~al.,}{{Astropy
  Collaboration} et~al.}{2013}]{AstropyCollaboration2013}
{Astropy Collaboration} et~al., 2013, \mn@doi [\aap]
  {10.1051/0004-6361/201322068}, \href
  {https://ui.adsabs.harvard.edu/abs/2013A&A...558A..33A} {558, A33}

\bibitem[\protect\citeauthoryear{{Ba{\~n}ados} et~al.,}{{Ba{\~n}ados}
  et~al.}{2018}]{Banados2018}
{Ba{\~n}ados} E.,  et~al., 2018, \mn@doi [\nat] {10.1038/nature25180}, \href
  {https://ui.adsabs.harvard.edu/abs/2018Natur.553..473B} {553, 473}

\bibitem[\protect\citeauthoryear{{Bacon} et~al.,}{{Bacon}
  et~al.}{2010}]{Bacon2010}
{Bacon} R.,  et~al., 2010, in {McLean} I.~S.,  {Ramsay} S.~K.,   {Takami} H.,
  eds,  Society of Photo-Optical Instrumentation Engineers (SPIE) Conference
  Series Vol. 7735, Ground-based and Airborne Instrumentation for Astronomy
  III. p. 773508 (\mn@eprint {arXiv} {2211.16795}), \mn@doi{10.1117/12.856027}

\bibitem[\protect\citeauthoryear{{Bacon} et~al.,}{{Bacon}
  et~al.}{2017}]{Bacon2017}
{Bacon} R.,  et~al., 2017, \mn@doi [\aap] {10.1051/0004-6361/201730833}, \href
  {https://ui.adsabs.harvard.edu/abs/2017A&A...608A...1B} {608, A1}

\bibitem[\protect\citeauthoryear{{Bacon} et~al.,}{{Bacon}
  et~al.}{2023}]{Bacon2023}
{Bacon} R.,  et~al., 2023, \mn@doi [\aap] {10.1051/0004-6361/202244187}, \href
  {https://ui.adsabs.harvard.edu/abs/2023A&A...670A...4B} {670, A4}

\bibitem[\protect\citeauthoryear{{Becker}, {Bolton}, {Madau}, {Pettini},
  {Ryan-Weber}  \& {Venemans}}{{Becker} et~al.}{2015}]{Becker2015}
{Becker} G.~D.,  {Bolton} J.~S.,  {Madau} P.,  {Pettini} M.,  {Ryan-Weber}
  E.~V.,   {Venemans} B.~P.,  2015, \mn@doi [\mnras] {10.1093/mnras/stu2646},
  \href {https://ui.adsabs.harvard.edu/abs/2015MNRAS.447.3402B} {447, 3402}

\bibitem[\protect\citeauthoryear{{Becker}, {D'Aloisio}, {Christenson}, {Zhu},
  {Worseck}  \& {Bolton}}{{Becker} et~al.}{2021}]{Becker2021}
{Becker} G.~D.,  {D'Aloisio} A.,  {Christenson} H.~M.,  {Zhu} Y.,  {Worseck}
  G.,   {Bolton} J.~S.,  2021, \mn@doi [\mnras] {10.1093/mnras/stab2696}, \href
  {https://ui.adsabs.harvard.edu/abs/2021MNRAS.508.1853B} {508, 1853}

\bibitem[\protect\citeauthoryear{{Beckwith} et~al.,}{{Beckwith}
  et~al.}{2006}]{Beckwith2006}
{Beckwith} S. V.~W.,  et~al., 2006, \mn@doi [\aj] {10.1086/507302}, \href
  {https://ui.adsabs.harvard.edu/abs/2006AJ....132.1729B} {132, 1729}

\bibitem[\protect\citeauthoryear{{Behrens}, {Dijkstra}  \&
  {Niemeyer}}{{Behrens} et~al.}{2014}]{Behrens2014}
{Behrens} C.,  {Dijkstra} M.,   {Niemeyer} J.~C.,  2014, \mn@doi [\aap]
  {10.1051/0004-6361/201322949}, \href
  {https://ui.adsabs.harvard.edu/abs/2014A&A...563A..77B} {563, A77}

\bibitem[\protect\citeauthoryear{{Blaizot} et~al.,}{{Blaizot}
  et~al.}{2023}]{Blaizot2023}
{Blaizot} J.,  et~al., 2023, \mn@doi [\mnras] {10.1093/mnras/stad1523}, \href
  {https://ui.adsabs.harvard.edu/abs/2023MNRAS.523.3749B} {523, 3749}

\bibitem[\protect\citeauthoryear{{Bolan} et~al.,}{{Bolan}
  et~al.}{2022}]{Bolan2022}
{Bolan} P.,  et~al., 2022, \mn@doi [\mnras] {10.1093/mnras/stac1963}, \href
  {https://ui.adsabs.harvard.edu/abs/2022MNRAS.517.3263B} {517, 3263}

\bibitem[\protect\citeauthoryear{{Bosman}, {Fan}, {Jiang}, {Reed}, {Matsuoka},
  {Becker}  \& {Haehnelt}}{{Bosman} et~al.}{2018}]{Bosman2018}
{Bosman} S. E.~I.,  {Fan} X.,  {Jiang} L.,  {Reed} S.,  {Matsuoka} Y.,
  {Becker} G.,   {Haehnelt} M.,  2018, \mn@doi [\mnras]
  {10.1093/mnras/sty1344}, \href
  {https://ui.adsabs.harvard.edu/abs/2018MNRAS.479.1055B} {479, 1055}

\bibitem[\protect\citeauthoryear{{Bosman}, {Kakiichi}, {Meyer}, {Gronke},
  {Laporte}  \& {Ellis}}{{Bosman} et~al.}{2020}]{Bosman2020}
{Bosman} S. E.~I.,  {Kakiichi} K.,  {Meyer} R.~A.,  {Gronke} M.,  {Laporte} N.,
    {Ellis} R.~S.,  2020, \mn@doi [\apj] {10.3847/1538-4357/ab85cd}, \href
  {https://ui.adsabs.harvard.edu/abs/2020ApJ...896...49B} {896, 49}

\bibitem[\protect\citeauthoryear{{Bosman} et~al.,}{{Bosman}
  et~al.}{2022}]{Bosman2022}
{Bosman} S. E.~I.,  et~al., 2022, \mn@doi [\mnras] {10.1093/mnras/stac1046},
  \href {https://ui.adsabs.harvard.edu/abs/2022MNRAS.514...55B} {514, 55}

\bibitem[\protect\citeauthoryear{{Bouwens} et~al.,}{{Bouwens}
  et~al.}{2009}]{Bouwens2009}
{Bouwens} R.~J.,  et~al., 2009, \mn@doi [\apj] {10.1088/0004-637X/705/1/936},
  \href {https://ui.adsabs.harvard.edu/abs/2009ApJ...705..936B} {705, 936}

\bibitem[\protect\citeauthoryear{{Bouwens} et~al.,}{{Bouwens}
  et~al.}{2010a}]{Bouwens2010b}
{Bouwens} R.~J.,  et~al., 2010a, \mn@doi [\apjl] {10.1088/2041-8205/708/2/L69},
  \href {https://ui.adsabs.harvard.edu/abs/2010ApJ...708L..69B} {708, L69}

\bibitem[\protect\citeauthoryear{{Bouwens} et~al.,}{{Bouwens}
  et~al.}{2010b}]{Bouwens2010a}
{Bouwens} R.~J.,  et~al., 2010b, \mn@doi [\apjl]
  {10.1088/2041-8205/709/2/L133}, \href
  {https://ui.adsabs.harvard.edu/abs/2010ApJ...709L.133B} {709, L133}

\bibitem[\protect\citeauthoryear{{Bouwens} et~al.,}{{Bouwens}
  et~al.}{2012}]{Bouwens2012}
{Bouwens} R.~J.,  et~al., 2012, \mn@doi [\apj] {10.1088/0004-637X/754/2/83},
  \href {https://ui.adsabs.harvard.edu/abs/2012ApJ...754...83B} {754, 83}

\bibitem[\protect\citeauthoryear{{Bouwens} et~al.,}{{Bouwens}
  et~al.}{2015}]{Bouwens2015}
{Bouwens} R.~J.,  et~al., 2015, \mn@doi [\apj] {10.1088/0004-637X/803/1/34},
  \href {https://ui.adsabs.harvard.edu/abs/2015ApJ...803...34B} {803, 34}

\bibitem[\protect\citeauthoryear{{Bouwens} et~al.,}{{Bouwens}
  et~al.}{2021}]{Bouwens2021}
{Bouwens} R.~J.,  et~al., 2021, \mn@doi [\aj] {10.3847/1538-3881/abf83e}, \href
  {https://ui.adsabs.harvard.edu/abs/2021AJ....162...47B} {162, 47}

\bibitem[\protect\citeauthoryear{{Boyett}, {Stark}, {Bunker}, {Tang}  \&
  {Maseda}}{{Boyett} et~al.}{2022}]{Boyett2022}
{Boyett} K. N.~K.,  {Stark} D.~P.,  {Bunker} A.~J.,  {Tang} M.,   {Maseda}
  M.~V.,  2022, \mn@doi [\mnras] {10.1093/mnras/stac1109}, \href
  {https://ui.adsabs.harvard.edu/abs/2022MNRAS.513.4451B} {513, 4451}

\bibitem[\protect\citeauthoryear{{Boyett} et~al.,}{{Boyett}
  et~al.}{2024}]{Boyett2024}
{Boyett} K.,  et~al., 2024, \mn@doi [arXiv e-prints]
  {10.48550/arXiv.2401.16934}, \href
  {https://ui.adsabs.harvard.edu/abs/2024arXiv240116934B} {p. arXiv:2401.16934}

\bibitem[\protect\citeauthoryear{{Brada{\v{c}}} et~al.,}{{Brada{\v{c}}}
  et~al.}{2017}]{Bradac2017}
{Brada{\v{c}}} M.,  et~al., 2017, \mn@doi [\apjl] {10.3847/2041-8213/836/1/L2},
  \href {https://ui.adsabs.harvard.edu/abs/2017ApJ...836L...2B} {836, L2}

\bibitem[\protect\citeauthoryear{{Brammer}, {van Dokkum}  \& {Coppi}}{{Brammer}
  et~al.}{2008}]{Brammer2008}
{Brammer} G.~B.,  {van Dokkum} P.~G.,   {Coppi} P.,  2008, \mn@doi [\apj]
  {10.1086/591786}, \href
  {https://ui.adsabs.harvard.edu/abs/2008ApJ...686.1503B} {686, 1503}

\bibitem[\protect\citeauthoryear{{Bruzual} \& {Charlot}}{{Bruzual} \&
  {Charlot}}{2003}]{Bruzual2003}
{Bruzual} G.,  {Charlot} S.,  2003, \mn@doi [\mnras]
  {10.1046/j.1365-8711.2003.06897.x}, \href
  {https://ui.adsabs.harvard.edu/abs/2003MNRAS.344.1000B} {344, 1000}

\bibitem[\protect\citeauthoryear{{Bunker}, {Stanway}, {Ellis}  \&
  {McMahon}}{{Bunker} et~al.}{2004}]{Bunker2004}
{Bunker} A.~J.,  {Stanway} E.~R.,  {Ellis} R.~S.,   {McMahon} R.~G.,  2004,
  \mn@doi [\mnras] {10.1111/j.1365-2966.2004.08326.x}, \href
  {https://ui.adsabs.harvard.edu/abs/2004MNRAS.355..374B} {355, 374}

\bibitem[\protect\citeauthoryear{{Bunker} et~al.,}{{Bunker}
  et~al.}{2010}]{Bunker2010}
{Bunker} A.~J.,  et~al., 2010, \mn@doi [\mnras]
  {10.1111/j.1365-2966.2010.17350.x}, \href
  {https://ui.adsabs.harvard.edu/abs/2010MNRAS.409..855B} {409, 855}

\bibitem[\protect\citeauthoryear{{Bunker}, {Caruana}, {Wilkins}, {Stanway},
  {Lorenzoni}, {Lacy}, {Jarvis}  \& {Hickey}}{{Bunker}
  et~al.}{2013}]{Bunker2013}
{Bunker} A.~J.,  {Caruana} J.,  {Wilkins} S.~M.,  {Stanway} E.~R.,  {Lorenzoni}
  S.,  {Lacy} M.,  {Jarvis} M.~J.,   {Hickey} S.,  2013, \mn@doi [\mnras]
  {10.1093/mnras/stt132}, \href
  {https://ui.adsabs.harvard.edu/abs/2013MNRAS.430.3314B} {430, 3314}

\bibitem[\protect\citeauthoryear{{Bunker} et~al.,}{{Bunker}
  et~al.}{2023a}]{Bunker2023b}
{Bunker} A.~J.,  et~al., 2023a, \mn@doi [arXiv e-prints]
  {10.48550/arXiv.2306.02467}, \href
  {https://ui.adsabs.harvard.edu/abs/2023arXiv230602467B} {p. arXiv:2306.02467}

\bibitem[\protect\citeauthoryear{{Bunker} et~al.,}{{Bunker}
  et~al.}{2023b}]{Bunker2023a}
{Bunker} A.~J.,  et~al., 2023b, \mn@doi [\aap] {10.1051/0004-6361/202346159},
  \href {https://ui.adsabs.harvard.edu/abs/2023A&A...677A..88B} {677, A88}

\bibitem[\protect\citeauthoryear{{Bushouse} et~al.,}{{Bushouse}
  et~al.}{2024}]{Bushouse2024}
{Bushouse} H.,  et~al., 2024, {JWST Calibration Pipeline},
  \mn@doi{10.5281/zenodo.6984365}

\bibitem[\protect\citeauthoryear{{Caffau}, {Ludwig}, {Steffen}, {Freytag}  \&
  {Bonifacio}}{{Caffau} et~al.}{2011}]{Caffau2011}
{Caffau} E.,  {Ludwig} H.~G.,  {Steffen} M.,  {Freytag} B.,   {Bonifacio} P.,
  2011, \mn@doi [\solphys] {10.1007/s11207-010-9541-4}, \href
  {https://ui.adsabs.harvard.edu/abs/2011SoPh..268..255C} {268, 255}

\bibitem[\protect\citeauthoryear{{Calzetti}, {Kinney}  \&
  {Storchi-Bergmann}}{{Calzetti} et~al.}{1994}]{Calzetti1994}
{Calzetti} D.,  {Kinney} A.~L.,   {Storchi-Bergmann} T.,  1994, \mn@doi [\apj]
  {10.1086/174346}, \href
  {https://ui.adsabs.harvard.edu/abs/1994ApJ...429..582C} {429, 582}

\bibitem[\protect\citeauthoryear{{Cameron}, {Katz}, {Witten}, {Saxena},
  {Laporte}  \& {Bunker}}{{Cameron} et~al.}{2023a}]{Cameron2023b}
{Cameron} A.~J.,  {Katz} H.,  {Witten} C.,  {Saxena} A.,  {Laporte} N.,
  {Bunker} A.~J.,  2023a, \mn@doi [arXiv e-prints] {10.48550/arXiv.2311.02051},
  \href {https://ui.adsabs.harvard.edu/abs/2023arXiv231102051C} {p.
  arXiv:2311.02051}

\bibitem[\protect\citeauthoryear{{Cameron} et~al.,}{{Cameron}
  et~al.}{2023b}]{Cameron2023a}
{Cameron} A.~J.,  et~al., 2023b, \mn@doi [\aap] {10.1051/0004-6361/202346107},
  \href {https://ui.adsabs.harvard.edu/abs/2023A&A...677A.115C} {677, A115}

\bibitem[\protect\citeauthoryear{{Cardelli}, {Clayton}  \& {Mathis}}{{Cardelli}
  et~al.}{1989}]{Cardelli1989}
{Cardelli} J.~A.,  {Clayton} G.~C.,   {Mathis} J.~S.,  1989, \mn@doi [\apj]
  {10.1086/167900}, \href
  {https://ui.adsabs.harvard.edu/abs/1989ApJ...345..245C} {345, 245}

\bibitem[\protect\citeauthoryear{{Caruana}, {Bunker}, {Wilkins}, {Stanway},
  {Lacy}, {Jarvis}, {Lorenzoni}  \& {Hickey}}{{Caruana}
  et~al.}{2012}]{Caruana2012}
{Caruana} J.,  {Bunker} A.~J.,  {Wilkins} S.~M.,  {Stanway} E.~R.,  {Lacy} M.,
  {Jarvis} M.~J.,  {Lorenzoni} S.,   {Hickey} S.,  2012, \mn@doi [\mnras]
  {10.1111/j.1365-2966.2012.21996.x}, \href
  {https://ui.adsabs.harvard.edu/abs/2012MNRAS.427.3055C} {427, 3055}

\bibitem[\protect\citeauthoryear{{Caruana}, {Bunker}, {Wilkins}, {Stanway},
  {Lorenzoni}, {Jarvis}  \& {Ebert}}{{Caruana} et~al.}{2014}]{Caruana2014}
{Caruana} J.,  {Bunker} A.~J.,  {Wilkins} S.~M.,  {Stanway} E.~R.,  {Lorenzoni}
  S.,  {Jarvis} M.~J.,   {Ebert} H.,  2014, \mn@doi [\mnras]
  {10.1093/mnras/stu1341}, \href
  {https://ui.adsabs.harvard.edu/abs/2014MNRAS.443.2831C} {443, 2831}

\bibitem[\protect\citeauthoryear{{Cassata} et~al.,}{{Cassata}
  et~al.}{2015}]{Cassata2015}
{Cassata} P.,  et~al., 2015, \mn@doi [\aap] {10.1051/0004-6361/201423824},
  \href {https://ui.adsabs.harvard.edu/abs/2015A&A...573A..24C} {573, A24}

\bibitem[\protect\citeauthoryear{{Cassata} et~al.,}{{Cassata}
  et~al.}{2020}]{Cassata2020}
{Cassata} P.,  et~al., 2020, \mn@doi [\aap] {10.1051/0004-6361/202037517},
  \href {https://ui.adsabs.harvard.edu/abs/2020A&A...643A...6C} {643, A6}

\bibitem[\protect\citeauthoryear{{Chabrier}}{{Chabrier}}{2003}]{Chabrier2003}
{Chabrier} G.,  2003, \mn@doi [\pasp] {10.1086/376392}, \href
  {https://ui.adsabs.harvard.edu/abs/2003PASP..115..763C} {115, 763}

\bibitem[\protect\citeauthoryear{{Chen}, {Stark}, {Mason}, {Topping},
  {Whitler}, {Tang}, {Endsley}  \& {Charlot}}{{Chen} et~al.}{2024}]{Chen2024}
{Chen} Z.,  {Stark} D.~P.,  {Mason} C.,  {Topping} M.~W.,  {Whitler} L.,
  {Tang} M.,  {Endsley} R.,   {Charlot} S.,  2024, \mn@doi [\mnras]
  {10.1093/mnras/stae455}, \href
  {https://ui.adsabs.harvard.edu/abs/2024MNRAS.528.7052C} {528, 7052}

\bibitem[\protect\citeauthoryear{{Chevallard} \& {Charlot}}{{Chevallard} \&
  {Charlot}}{2016}]{Chevallard2016}
{Chevallard} J.,  {Charlot} S.,  2016, \mn@doi [\mnras]
  {10.1093/mnras/stw1756}, \href
  {https://ui.adsabs.harvard.edu/abs/2016MNRAS.462.1415C} {462, 1415}

\bibitem[\protect\citeauthoryear{{Chevallard} et~al.,}{{Chevallard}
  et~al.}{2018}]{Chevallard2018}
{Chevallard} J.,  et~al., 2018, \mn@doi [\mnras] {10.1093/mnras/sty1461}, \href
  {https://ui.adsabs.harvard.edu/abs/2018MNRAS.479.3264C} {479, 3264}

\bibitem[\protect\citeauthoryear{{Chisholm} et~al.,}{{Chisholm}
  et~al.}{2018}]{Chisholm2018}
{Chisholm} J.,  et~al., 2018, \mn@doi [\aap] {10.1051/0004-6361/201832758},
  \href {https://ui.adsabs.harvard.edu/abs/2018A&A...616A..30C} {616, A30}

\bibitem[\protect\citeauthoryear{{Chisholm}, {Prochaska}, {Schaerer},
  {Gazagnes}  \& {Henry}}{{Chisholm} et~al.}{2020}]{Chisholm2020}
{Chisholm} J.,  {Prochaska} J.~X.,  {Schaerer} D.,  {Gazagnes} S.,   {Henry}
  A.,  2020, \mn@doi [\mnras] {10.1093/mnras/staa2470}, \href
  {https://ui.adsabs.harvard.edu/abs/2020MNRAS.498.2554C} {498, 2554}

\bibitem[\protect\citeauthoryear{{Chisholm} et~al.,}{{Chisholm}
  et~al.}{2022}]{Chisholm2022}
{Chisholm} J.,  et~al., 2022, \mn@doi [\mnras] {10.1093/mnras/stac2874}, \href
  {https://ui.adsabs.harvard.edu/abs/2022MNRAS.517.5104C} {517, 5104}

\bibitem[\protect\citeauthoryear{{Choustikov} et~al.,}{{Choustikov}
  et~al.}{2024a}]{Choustikov2024b}
{Choustikov} N.,  et~al., 2024a, \mn@doi [arXiv e-prints]
  {10.48550/arXiv.2401.09557}, \href
  {https://ui.adsabs.harvard.edu/abs/2024arXiv240109557C} {p. arXiv:2401.09557}

\bibitem[\protect\citeauthoryear{{Choustikov} et~al.,}{{Choustikov}
  et~al.}{2024b}]{Choustikov2024a}
{Choustikov} N.,  et~al., 2024b, \mn@doi [\mnras] {10.1093/mnras/stae776},
  \href {https://ui.adsabs.harvard.edu/abs/2024MNRAS.529.3751C} {529, 3751}

\bibitem[\protect\citeauthoryear{{Cooper} et~al.,}{{Cooper}
  et~al.}{2023}]{Cooper2023}
{Cooper} O.~R.,  et~al., 2023, \mn@doi [arXiv e-prints]
  {10.48550/arXiv.2309.06656}, \href
  {https://ui.adsabs.harvard.edu/abs/2023arXiv230906656C} {p. arXiv:2309.06656}

\bibitem[\protect\citeauthoryear{{Curtis-Lake} et~al.,}{{Curtis-Lake}
  et~al.}{2023}]{Curtis-Lake2023}
{Curtis-Lake} E.,  et~al., 2023, \mn@doi [Nature Astronomy]
  {10.1038/s41550-023-01918-w}, \href
  {https://ui.adsabs.harvard.edu/abs/2023NatAs...7..622C} {7, 622}

\bibitem[\protect\citeauthoryear{{D'Eugenio} et~al.,}{{D'Eugenio}
  et~al.}{2024}]{DEugenio2024}
{D'Eugenio} F.,  et~al., 2024, \mn@doi [arXiv e-prints]
  {10.48550/arXiv.2404.06531}, \href
  {https://ui.adsabs.harvard.edu/abs/2024arXiv240406531D} {p. arXiv:2404.06531}

\bibitem[\protect\citeauthoryear{{D'Odorico} et~al.,}{{D'Odorico}
  et~al.}{2006}]{DOdorico2006}
{D'Odorico} S.,  et~al., 2006, in {McLean} I.~S.,  {Iye} M.,  eds,  Society of
  Photo-Optical Instrumentation Engineers (SPIE) Conference Series Vol. 6269,
  Ground-based and Airborne Instrumentation for Astronomy. p. 626933,
  \mn@doi{10.1117/12.672969}

\bibitem[\protect\citeauthoryear{{Davies} et~al.,}{{Davies}
  et~al.}{2018}]{Davies2018}
{Davies} F.~B.,  et~al., 2018, \mn@doi [\apj] {10.3847/1538-4357/aad6dc}, \href
  {https://ui.adsabs.harvard.edu/abs/2018ApJ...864..142D} {864, 142}

\bibitem[\protect\citeauthoryear{{Davis} et~al.,}{{Davis}
  et~al.}{2007}]{Davis2007}
{Davis} M.,  et~al., 2007, \mn@doi [\apjl] {10.1086/517931}, \href
  {https://ui.adsabs.harvard.edu/abs/2007ApJ...660L...1D} {660, L1}

\bibitem[\protect\citeauthoryear{{De Barros} et~al.,}{{De Barros}
  et~al.}{2017}]{DeBarros2017}
{De Barros} S.,  et~al., 2017, \mn@doi [\aap] {10.1051/0004-6361/201731476},
  \href {https://ui.adsabs.harvard.edu/abs/2017A&A...608A.123D} {608, A123}

\bibitem[\protect\citeauthoryear{{De Barros}, {Oesch}, {Labb{\'e}}, {Stefanon},
  {Gonz{\'a}lez}, {Smit}, {Bouwens}  \& {Illingworth}}{{De Barros}
  et~al.}{2019}]{DeBarros2019}
{De Barros} S.,  {Oesch} P.~A.,  {Labb{\'e}} I.,  {Stefanon} M.,
  {Gonz{\'a}lez} V.,  {Smit} R.,  {Bouwens} R.~J.,   {Illingworth} G.~D.,
  2019, \mn@doi [\mnras] {10.1093/mnras/stz940}, \href
  {https://ui.adsabs.harvard.edu/abs/2019MNRAS.489.2355D} {489, 2355}

\bibitem[\protect\citeauthoryear{{Dijkstra}}{{Dijkstra}}{2014}]{Dijkstra2014}
{Dijkstra} M.,  2014, \mn@doi [\pasa] {10.1017/pasa.2014.33}, \href
  {https://ui.adsabs.harvard.edu/abs/2014PASA...31...40D} {31, e040}

\bibitem[\protect\citeauthoryear{{Dijkstra}}{{Dijkstra}}{2017}]{Dijkstra2017}
{Dijkstra} M.,  2017, \mn@doi [arXiv e-prints] {10.48550/arXiv.1704.03416},
  \href {https://ui.adsabs.harvard.edu/abs/2017arXiv170403416D} {p.
  arXiv:1704.03416}

\bibitem[\protect\citeauthoryear{{Dijkstra}, {Lidz}  \& {Wyithe}}{{Dijkstra}
  et~al.}{2007}]{Dijkstra2007}
{Dijkstra} M.,  {Lidz} A.,   {Wyithe} J. S.~B.,  2007, \mn@doi [\mnras]
  {10.1111/j.1365-2966.2007.11666.x}, \href
  {https://ui.adsabs.harvard.edu/abs/2007MNRAS.377.1175D} {377, 1175}

\bibitem[\protect\citeauthoryear{{Dijkstra}, {Mesinger}  \&
  {Wyithe}}{{Dijkstra} et~al.}{2011}]{Dijkstra2011}
{Dijkstra} M.,  {Mesinger} A.,   {Wyithe} J. S.~B.,  2011, \mn@doi [\mnras]
  {10.1111/j.1365-2966.2011.18530.x}, \href
  {https://ui.adsabs.harvard.edu/abs/2011MNRAS.414.2139D} {414, 2139}

\bibitem[\protect\citeauthoryear{{Dijkstra}, {Gronke}  \&
  {Venkatesan}}{{Dijkstra} et~al.}{2016}]{Dijkstra2016}
{Dijkstra} M.,  {Gronke} M.,   {Venkatesan} A.,  2016, \mn@doi [\apj]
  {10.3847/0004-637X/828/2/71}, \href
  {https://ui.adsabs.harvard.edu/abs/2016ApJ...828...71D} {828, 71}

\bibitem[\protect\citeauthoryear{{Du}, {Shapley}, {Tang}, {Stark}, {Martin},
  {Mobasher}, {Topping}  \& {Chevallard}}{{Du} et~al.}{2020}]{Du2020}
{Du} X.,  {Shapley} A.~E.,  {Tang} M.,  {Stark} D.~P.,  {Martin} C.~L.,
  {Mobasher} B.,  {Topping} M.~W.,   {Chevallard} J.,  2020, \mn@doi [\apj]
  {10.3847/1538-4357/ab67b8}, \href
  {https://ui.adsabs.harvard.edu/abs/2020ApJ...890...65D} {890, 65}

\bibitem[\protect\citeauthoryear{{Eilers}, {Davies}  \& {Hennawi}}{{Eilers}
  et~al.}{2018}]{Eilers2018}
{Eilers} A.-C.,  {Davies} F.~B.,   {Hennawi} J.~F.,  2018, \mn@doi [\apj]
  {10.3847/1538-4357/aad4fd}, \href
  {https://ui.adsabs.harvard.edu/abs/2018ApJ...864...53E} {864, 53}

\bibitem[\protect\citeauthoryear{{Eisenhauer} et~al.,}{{Eisenhauer}
  et~al.}{2003}]{Eisenhauer2003}
{Eisenhauer} F.,  et~al., 2003, in {Iye} M.,  {Moorwood} A. F.~M.,  eds,
  Society of Photo-Optical Instrumentation Engineers (SPIE) Conference Series
  Vol. 4841, Instrument Design and Performance for Optical/Infrared
  Ground-based Telescopes. pp 1548--1561 (\mn@eprint {arXiv}
  {astro-ph/0306191}), \mn@doi{10.1117/12.459468}

\bibitem[\protect\citeauthoryear{{Eisenstein} et~al.,}{{Eisenstein}
  et~al.}{2023a}]{Eisenstein2023a}
{Eisenstein} D.~J.,  et~al., 2023a, \mn@doi [arXiv e-prints]
  {10.48550/arXiv.2306.02465}, \href
  {https://ui.adsabs.harvard.edu/abs/2023arXiv230602465E} {p. arXiv:2306.02465}

\bibitem[\protect\citeauthoryear{{Eisenstein} et~al.,}{{Eisenstein}
  et~al.}{2023b}]{Eisenstein2023b}
{Eisenstein} D.~J.,  et~al., 2023b, \mn@doi [arXiv e-prints]
  {10.48550/arXiv.2310.12340}, \href
  {https://ui.adsabs.harvard.edu/abs/2023arXiv231012340E} {p. arXiv:2310.12340}

\bibitem[\protect\citeauthoryear{{Ellis} et~al.,}{{Ellis}
  et~al.}{2013}]{Ellis2013}
{Ellis} R.~S.,  et~al., 2013, \mn@doi [\apjl] {10.1088/2041-8205/763/1/L7},
  \href {https://ui.adsabs.harvard.edu/abs/2013ApJ...763L...7E} {763, L7}

\bibitem[\protect\citeauthoryear{{Endsley} \& {Stark}}{{Endsley} \&
  {Stark}}{2022}]{Endsley2022a}
{Endsley} R.,  {Stark} D.~P.,  2022, \mn@doi [\mnras] {10.1093/mnras/stac524},
  \href {https://ui.adsabs.harvard.edu/abs/2022MNRAS.511.6042E} {511, 6042}

\bibitem[\protect\citeauthoryear{{Endsley}, {Stark}, {Charlot}, {Chevallard},
  {Robertson}, {Bouwens}  \& {Stefanon}}{{Endsley} et~al.}{2021}]{Endsley2021b}
{Endsley} R.,  {Stark} D.~P.,  {Charlot} S.,  {Chevallard} J.,  {Robertson} B.,
   {Bouwens} R.~J.,   {Stefanon} M.,  2021, \mn@doi [\mnras]
  {10.1093/mnras/stab432}, \href
  {https://ui.adsabs.harvard.edu/abs/2021MNRAS.502.6044E} {502, 6044}

\bibitem[\protect\citeauthoryear{{Endsley} et~al.,}{{Endsley}
  et~al.}{2022}]{Endsley2022b}
{Endsley} R.,  et~al., 2022, \mn@doi [\mnras] {10.1093/mnras/stac3064}, \href
  {https://ui.adsabs.harvard.edu/abs/2022MNRAS.517.5642E} {517, 5642}

\bibitem[\protect\citeauthoryear{{Endsley} et~al.,}{{Endsley}
  et~al.}{2023a}]{Endsley2023b}
{Endsley} R.,  et~al., 2023a, \mn@doi [arXiv e-prints]
  {10.48550/arXiv.2306.05295}, \href
  {https://ui.adsabs.harvard.edu/abs/2023arXiv230605295E} {p. arXiv:2306.05295}

\bibitem[\protect\citeauthoryear{{Endsley}, {Stark}, {Whitler}, {Topping},
  {Chen}, {Plat}, {Chisholm}  \& {Charlot}}{{Endsley}
  et~al.}{2023b}]{Endsley2023a}
{Endsley} R.,  {Stark} D.~P.,  {Whitler} L.,  {Topping} M.~W.,  {Chen} Z.,
  {Plat} A.,  {Chisholm} J.,   {Charlot} S.,  2023b, \mn@doi [\mnras]
  {10.1093/mnras/stad1919}, \href
  {https://ui.adsabs.harvard.edu/abs/2023MNRAS.524.2312E} {524, 2312}

\bibitem[\protect\citeauthoryear{{Erb}}{{Erb}}{2015}]{Erb2015}
{Erb} D.~K.,  2015, \mn@doi [\nat] {10.1038/nature14454}, \href
  {https://ui.adsabs.harvard.edu/abs/2015Natur.523..169E} {523, 169}

\bibitem[\protect\citeauthoryear{{Erb} et~al.,}{{Erb} et~al.}{2014}]{Erb2014}
{Erb} D.~K.,  et~al., 2014, \mn@doi [\apj] {10.1088/0004-637X/795/1/33}, \href
  {https://ui.adsabs.harvard.edu/abs/2014ApJ...795...33E} {795, 33}

\bibitem[\protect\citeauthoryear{{Faber} et~al.,}{{Faber}
  et~al.}{2003}]{Faber2003}
{Faber} S.~M.,  et~al., 2003, in {Iye} M.,  {Moorwood} A. F.~M.,  eds,  Society
  of Photo-Optical Instrumentation Engineers (SPIE) Conference Series Vol.
  4841, Instrument Design and Performance for Optical/Infrared Ground-based
  Telescopes. pp 1657--1669, \mn@doi{10.1117/12.460346}

\bibitem[\protect\citeauthoryear{{Fabricant} et~al.,}{{Fabricant}
  et~al.}{2019}]{Fabricant2019}
{Fabricant} D.,  et~al., 2019, \mn@doi [\pasp] {10.1088/1538-3873/ab1d78},
  \href {https://ui.adsabs.harvard.edu/abs/2019PASP..131g5004F} {131, 075004}

\bibitem[\protect\citeauthoryear{{Fan}, {Ba{\~n}ados}  \& {Simcoe}}{{Fan}
  et~al.}{2023}]{Fan2023}
{Fan} X.,  {Ba{\~n}ados} E.,   {Simcoe} R.~A.,  2023, \mn@doi [\araa]
  {10.1146/annurev-astro-052920-102455}, \href
  {https://ui.adsabs.harvard.edu/abs/2023ARA&A..61..373F} {61, 373}

\bibitem[\protect\citeauthoryear{{Ferland} et~al.,}{{Ferland}
  et~al.}{2013}]{Ferland2013}
{Ferland} G.~J.,  et~al., 2013, \mn@doi [\rmxaa] {10.48550/arXiv.1302.4485},
  \href {https://ui.adsabs.harvard.edu/abs/2013RMxAA..49..137F} {49, 137}

\bibitem[\protect\citeauthoryear{{Ferruit} et~al.,}{{Ferruit}
  et~al.}{2022}]{Ferruit2022}
{Ferruit} P.,  et~al., 2022, \mn@doi [\aap] {10.1051/0004-6361/202142673},
  \href {https://ui.adsabs.harvard.edu/abs/2022A&A...661A..81F} {661, A81}

\bibitem[\protect\citeauthoryear{{Finkelstein} et~al.,}{{Finkelstein}
  et~al.}{2011}]{Finkelstein2011}
{Finkelstein} S.~L.,  et~al., 2011, \mn@doi [\apj]
  {10.1088/0004-637X/729/2/140}, \href
  {https://ui.adsabs.harvard.edu/abs/2011ApJ...729..140F} {729, 140}

\bibitem[\protect\citeauthoryear{{Fletcher}, {Tang}, {Robertson}, {Nakajima},
  {Ellis}, {Stark}  \& {Inoue}}{{Fletcher} et~al.}{2019}]{Fletcher2019}
{Fletcher} T.~J.,  {Tang} M.,  {Robertson} B.~E.,  {Nakajima} K.,  {Ellis}
  R.~S.,  {Stark} D.~P.,   {Inoue} A.,  2019, \mn@doi [\apj]
  {10.3847/1538-4357/ab2045}, \href
  {https://ui.adsabs.harvard.edu/abs/2019ApJ...878...87F} {878, 87}

\bibitem[\protect\citeauthoryear{{Flury} et~al.,}{{Flury}
  et~al.}{2022}]{Flury2022}
{Flury} S.~R.,  et~al., 2022, \mn@doi [\apj] {10.3847/1538-4357/ac61e4}, \href
  {https://ui.adsabs.harvard.edu/abs/2022ApJ...930..126F} {930, 126}

\bibitem[\protect\citeauthoryear{{Foreman-Mackey}, {Hogg}, {Lang}  \&
  {Goodman}}{{Foreman-Mackey} et~al.}{2013}]{Foreman-Mackey2013}
{Foreman-Mackey} D.,  {Hogg} D.~W.,  {Lang} D.,   {Goodman} J.,  2013, \mn@doi
  [\pasp] {10.1086/670067}, \href
  {https://ui.adsabs.harvard.edu/abs/2013PASP..125..306F} {125, 306}

\bibitem[\protect\citeauthoryear{{Fujimoto} et~al.,}{{Fujimoto}
  et~al.}{2023}]{Fujimoto2023}
{Fujimoto} S.,  et~al., 2023, \mn@doi [arXiv e-prints]
  {10.48550/arXiv.2308.11609}, \href
  {https://ui.adsabs.harvard.edu/abs/2023arXiv230811609F} {p. arXiv:2308.11609}

\bibitem[\protect\citeauthoryear{{Furlanetto}, {Zaldarriaga}  \&
  {Hernquist}}{{Furlanetto} et~al.}{2006}]{Furlanetto2006}
{Furlanetto} S.~R.,  {Zaldarriaga} M.,   {Hernquist} L.,  2006, \mn@doi
  [\mnras] {10.1111/j.1365-2966.2005.09785.x}, \href
  {https://ui.adsabs.harvard.edu/abs/2006MNRAS.365.1012F} {365, 1012}

\bibitem[\protect\citeauthoryear{{Gardner} et~al.,}{{Gardner}
  et~al.}{2023}]{Gardner2023}
{Gardner} J.~P.,  et~al., 2023, \mn@doi [\pasp] {10.1088/1538-3873/acd1b5},
  \href {https://ui.adsabs.harvard.edu/abs/2023PASP..135f8001G} {135, 068001}

\bibitem[\protect\citeauthoryear{{Garel}, {Blaizot}, {Rosdahl},
  {Michel-Dansac}, {Haehnelt}, {Katz}, {Kimm}  \& {Verhamme}}{{Garel}
  et~al.}{2021}]{Garel2021}
{Garel} T.,  {Blaizot} J.,  {Rosdahl} J.,  {Michel-Dansac} L.,  {Haehnelt}
  M.~G.,  {Katz} H.,  {Kimm} T.,   {Verhamme} A.,  2021, \mn@doi [\mnras]
  {10.1093/mnras/stab990}, \href
  {https://ui.adsabs.harvard.edu/abs/2021MNRAS.504.1902G} {504, 1902}

\bibitem[\protect\citeauthoryear{{Gazagnes}, {Chisholm}, {Schaerer}, {Verhamme}
   \& {Izotov}}{{Gazagnes} et~al.}{2020}]{Gazagnes2020}
{Gazagnes} S.,  {Chisholm} J.,  {Schaerer} D.,  {Verhamme} A.,   {Izotov} Y.,
  2020, \mn@doi [\aap] {10.1051/0004-6361/202038096}, \href
  {https://ui.adsabs.harvard.edu/abs/2020A&A...639A..85G} {639, A85}

\bibitem[\protect\citeauthoryear{{Giavalisco} et~al.,}{{Giavalisco}
  et~al.}{2004}]{Giavalisco2004}
{Giavalisco} M.,  et~al., 2004, \mn@doi [\apjl] {10.1086/379232}, \href
  {https://ui.adsabs.harvard.edu/abs/2004ApJ...600L..93G} {600, L93}

\bibitem[\protect\citeauthoryear{{Goovaerts} et~al.,}{{Goovaerts}
  et~al.}{2023}]{Goovaerts2023}
{Goovaerts} I.,  et~al., 2023, \mn@doi [\aap] {10.1051/0004-6361/202347110},
  \href {https://ui.adsabs.harvard.edu/abs/2023A&A...678A.174G} {678, A174}

\bibitem[\protect\citeauthoryear{{Goto} et~al.,}{{Goto}
  et~al.}{2021}]{Goto2021}
{Goto} H.,  et~al., 2021, \mn@doi [\apj] {10.3847/1538-4357/ac308b}, \href
  {https://ui.adsabs.harvard.edu/abs/2021ApJ...923..229G} {923, 229}

\bibitem[\protect\citeauthoryear{{Greene} et~al.,}{{Greene}
  et~al.}{2017}]{Greene2017}
{Greene} T.~P.,  et~al., 2017, \mn@doi [Journal of Astronomical Telescopes,
  Instruments, and Systems] {10.1117/1.JATIS.3.3.035001}, \href
  {https://ui.adsabs.harvard.edu/abs/2017JATIS...3c5001G} {3, 035001}

\bibitem[\protect\citeauthoryear{{Greig}, {Mesinger}, {Davies}, {Wang}, {Yang}
  \& {Hennawi}}{{Greig} et~al.}{2022}]{Greig2022}
{Greig} B.,  {Mesinger} A.,  {Davies} F.~B.,  {Wang} F.,  {Yang} J.,
  {Hennawi} J.~F.,  2022, \mn@doi [\mnras] {10.1093/mnras/stac825}, \href
  {https://ui.adsabs.harvard.edu/abs/2022MNRAS.512.5390G} {512, 5390}

\bibitem[\protect\citeauthoryear{{Gronke} \& {Dijkstra}}{{Gronke} \&
  {Dijkstra}}{2016}]{Gronke2016}
{Gronke} M.,  {Dijkstra} M.,  2016, \mn@doi [\apj]
  {10.3847/0004-637X/826/1/14}, \href
  {https://ui.adsabs.harvard.edu/abs/2016ApJ...826...14G} {826, 14}

\bibitem[\protect\citeauthoryear{{Gunn} \& {Peterson}}{{Gunn} \&
  {Peterson}}{1965}]{Gunn1965}
{Gunn} J.~E.,  {Peterson} B.~A.,  1965, \mn@doi [\apj] {10.1086/148444}, \href
  {https://ui.adsabs.harvard.edu/abs/1965ApJ...142.1633G} {142, 1633}

\bibitem[\protect\citeauthoryear{{Guo} et~al.,}{{Guo} et~al.}{2023}]{Guo2023}
{Guo} Y.,  et~al., 2023, \mn@doi [arXiv e-prints] {10.48550/arXiv.2309.05513},
  \href {https://ui.adsabs.harvard.edu/abs/2023arXiv230905513G} {p.
  arXiv:2309.05513}

\bibitem[\protect\citeauthoryear{{Gutkin}, {Charlot}  \& {Bruzual}}{{Gutkin}
  et~al.}{2016}]{Gutkin2016}
{Gutkin} J.,  {Charlot} S.,   {Bruzual} G.,  2016, \mn@doi [\mnras]
  {10.1093/mnras/stw1716}, \href
  {https://ui.adsabs.harvard.edu/abs/2016MNRAS.462.1757G} {462, 1757}

\bibitem[\protect\citeauthoryear{{Hainline} et~al.,}{{Hainline}
  et~al.}{2024a}]{Hainline2024a}
{Hainline} K.~N.,  et~al., 2024a, \mn@doi [\apj] {10.3847/1538-4357/ad20d1},
  \href {https://ui.adsabs.harvard.edu/abs/2024ApJ...964...66H} {964, 66}

\bibitem[\protect\citeauthoryear{{Hainline} et~al.,}{{Hainline}
  et~al.}{2024b}]{Hainline2024b}
{Hainline} K.~N.,  et~al., 2024b, \mn@doi [\apj] {10.3847/1538-4357/ad1ee4},
  \href {https://ui.adsabs.harvard.edu/abs/2024ApJ...964...71H} {964, 71}

\bibitem[\protect\citeauthoryear{{Harikane}, {Nakajima}, {Ouchi}, {Umeda},
  {Isobe}, {Ono}, {Xu}  \& {Zhang}}{{Harikane} et~al.}{2024}]{Harikane2024}
{Harikane} Y.,  {Nakajima} K.,  {Ouchi} M.,  {Umeda} H.,  {Isobe} Y.,  {Ono}
  Y.,  {Xu} Y.,   {Zhang} Y.,  2024, \mn@doi [\apj] {10.3847/1538-4357/ad0b7e},
  \href {https://ui.adsabs.harvard.edu/abs/2024ApJ...960...56H} {960, 56}

\bibitem[\protect\citeauthoryear{{Harris} et~al.,}{{Harris}
  et~al.}{2020}]{Harris2020}
{Harris} C.~R.,  et~al., 2020, \mn@doi [\nat] {10.1038/s41586-020-2649-2},
  \href {https://ui.adsabs.harvard.edu/abs/2020Natur.585..357H} {585, 357}

\bibitem[\protect\citeauthoryear{{Hashimoto}, {Ouchi}, {Shimasaku}, {Ono},
  {Nakajima}, {Rauch}, {Lee}  \& {Okamura}}{{Hashimoto}
  et~al.}{2013}]{Hashimoto2013}
{Hashimoto} T.,  {Ouchi} M.,  {Shimasaku} K.,  {Ono} Y.,  {Nakajima} K.,
  {Rauch} M.,  {Lee} J.,   {Okamura} S.,  2013, \mn@doi [\apj]
  {10.1088/0004-637X/765/1/70}, \href
  {https://ui.adsabs.harvard.edu/abs/2013ApJ...765...70H} {765, 70}

\bibitem[\protect\citeauthoryear{{Hashimoto} et~al.,}{{Hashimoto}
  et~al.}{2015}]{Hashimoto2015}
{Hashimoto} T.,  et~al., 2015, \mn@doi [\apj] {10.1088/0004-637X/812/2/157},
  \href {https://ui.adsabs.harvard.edu/abs/2015ApJ...812..157H} {812, 157}

\bibitem[\protect\citeauthoryear{{Hashimoto} et~al.,}{{Hashimoto}
  et~al.}{2019}]{Hashimoto2019}
{Hashimoto} T.,  et~al., 2019, \mn@doi [\pasj] {10.1093/pasj/psz049}, \href
  {https://ui.adsabs.harvard.edu/abs/2019PASJ...71...71H} {71, 71}

\bibitem[\protect\citeauthoryear{{Hayes}}{{Hayes}}{2015}]{Hayes2015}
{Hayes} M.,  2015, \mn@doi [\pasa] {10.1017/pasa.2015.25}, \href
  {https://ui.adsabs.harvard.edu/abs/2015PASA...32...27H} {32, e027}

\bibitem[\protect\citeauthoryear{{Hayes} et~al.,}{{Hayes}
  et~al.}{2010}]{Hayes2010}
{Hayes} M.,  et~al., 2010, \mn@doi [\nat] {10.1038/nature08881}, \href
  {https://ui.adsabs.harvard.edu/abs/2010Natur.464..562H} {464, 562}

\bibitem[\protect\citeauthoryear{{Hayes}, {Schaerer}, {{\"O}stlin},
  {Mas-Hesse}, {Atek}  \& {Kunth}}{{Hayes} et~al.}{2011}]{Hayes2011}
{Hayes} M.,  {Schaerer} D.,  {{\"O}stlin} G.,  {Mas-Hesse} J.~M.,  {Atek} H.,
  {Kunth} D.,  2011, \mn@doi [\apj] {10.1088/0004-637X/730/1/8}, \href
  {https://ui.adsabs.harvard.edu/abs/2011ApJ...730....8H} {730, 8}

\bibitem[\protect\citeauthoryear{{Hayes} et~al.,}{{Hayes}
  et~al.}{2013}]{Hayes2013}
{Hayes} M.,  et~al., 2013, \mn@doi [\apjl] {10.1088/2041-8205/765/2/L27}, \href
  {https://ui.adsabs.harvard.edu/abs/2013ApJ...765L..27H} {765, L27}

\bibitem[\protect\citeauthoryear{{Hayes}, {Runnholm}, {Gronke}  \&
  {Scarlata}}{{Hayes} et~al.}{2021}]{Hayes2021}
{Hayes} M.~J.,  {Runnholm} A.,  {Gronke} M.,   {Scarlata} C.,  2021, \mn@doi
  [\apj] {10.3847/1538-4357/abd246}, \href
  {https://ui.adsabs.harvard.edu/abs/2021ApJ...908...36H} {908, 36}

\bibitem[\protect\citeauthoryear{{Heckman}, {Sembach}, {Meurer}, {Leitherer},
  {Calzetti}  \& {Martin}}{{Heckman} et~al.}{2001}]{Heckman2001}
{Heckman} T.~M.,  {Sembach} K.~R.,  {Meurer} G.~R.,  {Leitherer} C.,
  {Calzetti} D.,   {Martin} C.~L.,  2001, \mn@doi [\apj] {10.1086/322475},
  \href {https://ui.adsabs.harvard.edu/abs/2001ApJ...558...56H} {558, 56}

\bibitem[\protect\citeauthoryear{{Henry}, {Scarlata}, {Martin}  \&
  {Erb}}{{Henry} et~al.}{2015}]{Henry2015}
{Henry} A.,  {Scarlata} C.,  {Martin} C.~L.,   {Erb} D.,  2015, \mn@doi [\apj]
  {10.1088/0004-637X/809/1/19}, \href
  {https://ui.adsabs.harvard.edu/abs/2015ApJ...809...19H} {809, 19}

\bibitem[\protect\citeauthoryear{{Henry}, {Berg}, {Scarlata}, {Verhamme}  \&
  {Erb}}{{Henry} et~al.}{2018}]{Henry2018}
{Henry} A.,  {Berg} D.~A.,  {Scarlata} C.,  {Verhamme} A.,   {Erb} D.,  2018,
  \mn@doi [\apj] {10.3847/1538-4357/aab099}, \href
  {https://ui.adsabs.harvard.edu/abs/2018ApJ...855...96H} {855, 96}

\bibitem[\protect\citeauthoryear{{Herenz} \& {Wisotzki}}{{Herenz} \&
  {Wisotzki}}{2017}]{Herenz2017a}
{Herenz} E.~C.,  {Wisotzki} L.,  2017, \mn@doi [\aap]
  {10.1051/0004-6361/201629507}, \href
  {https://ui.adsabs.harvard.edu/abs/2017A&A...602A.111H} {602, A111}

\bibitem[\protect\citeauthoryear{{Herenz} et~al.,}{{Herenz}
  et~al.}{2017}]{Herenz2017b}
{Herenz} E.~C.,  et~al., 2017, \mn@doi [\aap] {10.1051/0004-6361/201731055},
  \href {https://ui.adsabs.harvard.edu/abs/2017A&A...606A..12H} {606, A12}

\bibitem[\protect\citeauthoryear{{Herenz} et~al.,}{{Herenz}
  et~al.}{2019}]{Herenz2019}
{Herenz} E.~C.,  et~al., 2019, \mn@doi [\aap] {10.1051/0004-6361/201834164},
  \href {https://ui.adsabs.harvard.edu/abs/2019A&A...621A.107H} {621, A107}

\bibitem[\protect\citeauthoryear{{Herrero Alonso}, {Wisotzki}, {Miyaji},
  {Schaye}, {Pharo}  \& {Krumpe}}{{Herrero Alonso}
  et~al.}{2023}]{HerreroAlonso2023}
{Herrero Alonso} Y.,  {Wisotzki} L.,  {Miyaji} T.,  {Schaye} J.,  {Pharo} J.,
  {Krumpe} M.,  2023, \mn@doi [\aap] {10.1051/0004-6361/202347294}, \href
  {https://ui.adsabs.harvard.edu/abs/2023A&A...677A.125H} {677, A125}

\bibitem[\protect\citeauthoryear{{Hoag} et~al.,}{{Hoag}
  et~al.}{2019}]{Hoag2019}
{Hoag} A.,  et~al., 2019, \mn@doi [\apj] {10.3847/1538-4357/ab1de7}, \href
  {https://ui.adsabs.harvard.edu/abs/2019ApJ...878...12H} {878, 12}

\bibitem[\protect\citeauthoryear{{Horne}}{{Horne}}{1986}]{Horne1986}
{Horne} K.,  1986, \mn@doi [\pasp] {10.1086/131801}, \href
  {https://ui.adsabs.harvard.edu/abs/1986PASP...98..609H} {98, 609}

\bibitem[\protect\citeauthoryear{{Hu}, {Cowie}, {Songaila}, {Barger},
  {Rosenwasser}  \& {Wold}}{{Hu} et~al.}{2016}]{Hu2016}
{Hu} E.~M.,  {Cowie} L.~L.,  {Songaila} A.,  {Barger} A.~J.,  {Rosenwasser} B.,
    {Wold} I.~G.~B.,  2016, \mn@doi [\apjl] {10.3847/2041-8205/825/1/L7}, \href
  {https://ui.adsabs.harvard.edu/abs/2016ApJ...825L...7H} {825, L7}

\bibitem[\protect\citeauthoryear{{Hunter}}{{Hunter}}{2007}]{Hunter2007}
{Hunter} J.~D.,  2007, \mn@doi [Computing in Science and Engineering]
  {10.1109/MCSE.2007.55}, \href
  {https://ui.adsabs.harvard.edu/abs/2007CSE.....9...90H} {9, 90}

\bibitem[\protect\citeauthoryear{{Hutchison} et~al.,}{{Hutchison}
  et~al.}{2019}]{Hutchison2019}
{Hutchison} T.~A.,  et~al., 2019, \mn@doi [\apj] {10.3847/1538-4357/ab22a2},
  \href {https://ui.adsabs.harvard.edu/abs/2019ApJ...879...70H} {879, 70}

\bibitem[\protect\citeauthoryear{{Ichikawa} et~al.,}{{Ichikawa}
  et~al.}{2006}]{Ichikawa2006}
{Ichikawa} T.,  et~al., 2006, in {McLean} I.~S.,  {Iye} M.,  eds,  Society of
  Photo-Optical Instrumentation Engineers (SPIE) Conference Series Vol. 6269,
  Ground-based and Airborne Instrumentation for Astronomy. p. 626916,
  \mn@doi{10.1117/12.670078}

\bibitem[\protect\citeauthoryear{{Illingworth} et~al.,}{{Illingworth}
  et~al.}{2013}]{Illingworth2013}
{Illingworth} G.~D.,  et~al., 2013, \mn@doi [\apjs]
  {10.1088/0067-0049/209/1/6}, \href
  {https://ui.adsabs.harvard.edu/abs/2013ApJS..209....6I} {209, 6}

\bibitem[\protect\citeauthoryear{{Inami} et~al.,}{{Inami}
  et~al.}{2017}]{Inami2017}
{Inami} H.,  et~al., 2017, \mn@doi [\aap] {10.1051/0004-6361/201731195}, \href
  {https://ui.adsabs.harvard.edu/abs/2017A&A...608A...2I} {608, A2}

\bibitem[\protect\citeauthoryear{{Inoue}, {Shimizu}, {Iwata}  \&
  {Tanaka}}{{Inoue} et~al.}{2014}]{Inoue2014}
{Inoue} A.~K.,  {Shimizu} I.,  {Iwata} I.,   {Tanaka} M.,  2014, \mn@doi
  [\mnras] {10.1093/mnras/stu936}, \href
  {https://ui.adsabs.harvard.edu/abs/2014MNRAS.442.1805I} {442, 1805}

\bibitem[\protect\citeauthoryear{{Inoue} et~al.,}{{Inoue}
  et~al.}{2016}]{Inoue2016}
{Inoue} A.~K.,  et~al., 2016, \mn@doi [Science] {10.1126/science.aaf0714},
  \href {https://ui.adsabs.harvard.edu/abs/2016Sci...352.1559I} {352, 1559}

\bibitem[\protect\citeauthoryear{{Itoh} et~al.,}{{Itoh}
  et~al.}{2018}]{Itoh2018}
{Itoh} R.,  et~al., 2018, \mn@doi [\apj] {10.3847/1538-4357/aadfe4}, \href
  {https://ui.adsabs.harvard.edu/abs/2018ApJ...867...46I} {867, 46}

\bibitem[\protect\citeauthoryear{{Izotov}, {Stasi{\'n}ska}, {Meynet}, {Guseva}
  \& {Thuan}}{{Izotov} et~al.}{2006}]{Izotov2006}
{Izotov} Y.~I.,  {Stasi{\'n}ska} G.,  {Meynet} G.,  {Guseva} N.~G.,   {Thuan}
  T.~X.,  2006, \mn@doi [\aap] {10.1051/0004-6361:20053763}, \href
  {https://ui.adsabs.harvard.edu/abs/2006A&A...448..955I} {448, 955}

\bibitem[\protect\citeauthoryear{{Izotov}, {Schaerer}, {Worseck}, {Verhamme},
  {Guseva}, {Thuan}, {Orlitov{\'a}}  \& {Fricke}}{{Izotov}
  et~al.}{2020}]{Izotov2020}
{Izotov} Y.~I.,  {Schaerer} D.,  {Worseck} G.,  {Verhamme} A.,  {Guseva} N.~G.,
   {Thuan} T.~X.,  {Orlitov{\'a}} I.,   {Fricke} K.~J.,  2020, \mn@doi [\mnras]
  {10.1093/mnras/stz3041}, \href
  {https://ui.adsabs.harvard.edu/abs/2020MNRAS.491..468I} {491, 468}

\bibitem[\protect\citeauthoryear{{Izotov}, {Worseck}, {Schaerer}, {Guseva},
  {Chisholm}, {Thuan}, {Fricke}  \& {Verhamme}}{{Izotov}
  et~al.}{2021}]{Izotov2021}
{Izotov} Y.~I.,  {Worseck} G.,  {Schaerer} D.,  {Guseva} N.~G.,  {Chisholm} J.,
   {Thuan} T.~X.,  {Fricke} K.~J.,   {Verhamme} A.,  2021, \mn@doi [\mnras]
  {10.1093/mnras/stab612}, \href
  {https://ui.adsabs.harvard.edu/abs/2021MNRAS.503.1734I} {503, 1734}

\bibitem[\protect\citeauthoryear{{Izotov}, {Thuan}, {Guseva}, {Schaerer},
  {Worseck}  \& {Verhamme}}{{Izotov} et~al.}{2024}]{Izotov2024}
{Izotov} Y.~I.,  {Thuan} T.~X.,  {Guseva} N.~G.,  {Schaerer} D.,  {Worseck} G.,
    {Verhamme} A.,  2024, \mn@doi [\mnras] {10.1093/mnras/stad3151}, \href
  {https://ui.adsabs.harvard.edu/abs/2024MNRAS.527..281I} {527, 281}

\bibitem[\protect\citeauthoryear{{Jakobsen} et~al.,}{{Jakobsen}
  et~al.}{2022}]{Jakobsen2022}
{Jakobsen} P.,  et~al., 2022, \mn@doi [\aap] {10.1051/0004-6361/202142663},
  \href {https://ui.adsabs.harvard.edu/abs/2022A&A...661A..80J} {661, A80}

\bibitem[\protect\citeauthoryear{{Jaskot} \& {Oey}}{{Jaskot} \&
  {Oey}}{2014}]{Jaskot2014}
{Jaskot} A.~E.,  {Oey} M.~S.,  2014, \mn@doi [\apjl]
  {10.1088/2041-8205/791/2/L19}, \href
  {https://ui.adsabs.harvard.edu/abs/2014ApJ...791L..19J} {791, L19}

\bibitem[\protect\citeauthoryear{{Jaskot}, {Dowd}, {Oey}, {Scarlata}  \&
  {McKinney}}{{Jaskot} et~al.}{2019}]{Jaskot2019}
{Jaskot} A.~E.,  {Dowd} T.,  {Oey} M.~S.,  {Scarlata} C.,   {McKinney} J.,
  2019, \mn@doi [\apj] {10.3847/1538-4357/ab3d3b}, \href
  {https://ui.adsabs.harvard.edu/abs/2019ApJ...885...96J} {885, 96}

\bibitem[\protect\citeauthoryear{{Jones}, {Stark}  \& {Ellis}}{{Jones}
  et~al.}{2012}]{Jones2012}
{Jones} T.,  {Stark} D.~P.,   {Ellis} R.~S.,  2012, \mn@doi [\apj]
  {10.1088/0004-637X/751/1/51}, \href
  {https://ui.adsabs.harvard.edu/abs/2012ApJ...751...51J} {751, 51}

\bibitem[\protect\citeauthoryear{{Jones} et~al.,}{{Jones}
  et~al.}{2024}]{Jones2024}
{Jones} G.~C.,  et~al., 2024, \mn@doi [\aap] {10.1051/0004-6361/202347099},
  \href {https://ui.adsabs.harvard.edu/abs/2024A&A...683A.238J} {683, A238}

\bibitem[\protect\citeauthoryear{{Jung} et~al.,}{{Jung}
  et~al.}{2018}]{Jung2018}
{Jung} I.,  et~al., 2018, \mn@doi [\apj] {10.3847/1538-4357/aad686}, \href
  {https://ui.adsabs.harvard.edu/abs/2018ApJ...864..103J} {864, 103}

\bibitem[\protect\citeauthoryear{{Jung} et~al.,}{{Jung}
  et~al.}{2022}]{Jung2022}
{Jung} I.,  et~al., 2022, arXiv e-prints, \href
  {https://ui.adsabs.harvard.edu/abs/2022arXiv221209850J} {p. arXiv:2212.09850}

\bibitem[\protect\citeauthoryear{{Jung} et~al.,}{{Jung}
  et~al.}{2023}]{Jung2023}
{Jung} I.,  et~al., 2023, \mn@doi [arXiv e-prints] {10.48550/arXiv.2304.05385},
  \href {https://ui.adsabs.harvard.edu/abs/2023arXiv230405385J} {p.
  arXiv:2304.05385}

\bibitem[\protect\citeauthoryear{{Kakiichi} \& {Gronke}}{{Kakiichi} \&
  {Gronke}}{2021}]{Kakiichi2021}
{Kakiichi} K.,  {Gronke} M.,  2021, \mn@doi [\apj] {10.3847/1538-4357/abc2d9},
  \href {https://ui.adsabs.harvard.edu/abs/2021ApJ...908...30K} {908, 30}

\bibitem[\protect\citeauthoryear{{Kim} et~al.,}{{Kim} et~al.}{2023}]{Kim2023}
{Kim} K.~J.,  et~al., 2023, \mn@doi [\apjl] {10.3847/2041-8213/acf0c5}, \href
  {https://ui.adsabs.harvard.edu/abs/2023ApJ...955L..17K} {955, L17}

\bibitem[\protect\citeauthoryear{{Kimm}, {Blaizot}, {Garel}, {Michel-Dansac},
  {Katz}, {Rosdahl}, {Verhamme}  \& {Haehnelt}}{{Kimm} et~al.}{2019}]{Kimm2019}
{Kimm} T.,  {Blaizot} J.,  {Garel} T.,  {Michel-Dansac} L.,  {Katz} H.,
  {Rosdahl} J.,  {Verhamme} A.,   {Haehnelt} M.,  2019, \mn@doi [\mnras]
  {10.1093/mnras/stz989}, \href
  {https://ui.adsabs.harvard.edu/abs/2019MNRAS.486.2215K} {486, 2215}

\bibitem[\protect\citeauthoryear{{Konno} et~al.,}{{Konno}
  et~al.}{2018}]{Konno2018}
{Konno} A.,  et~al., 2018, \mn@doi [\pasj] {10.1093/pasj/psx131}, \href
  {https://ui.adsabs.harvard.edu/abs/2018PASJ...70S..16K} {70, S16}

\bibitem[\protect\citeauthoryear{{Kornei}, {Shapley}, {Erb}, {Steidel},
  {Reddy}, {Pettini}  \& {Bogosavljevi{\'c}}}{{Kornei}
  et~al.}{2010}]{Kornei2010}
{Kornei} K.~A.,  {Shapley} A.~E.,  {Erb} D.~K.,  {Steidel} C.~C.,  {Reddy}
  N.~A.,  {Pettini} M.,   {Bogosavljevi{\'c}} M.,  2010, \mn@doi [\apj]
  {10.1088/0004-637X/711/2/693}, \href
  {https://ui.adsabs.harvard.edu/abs/2010ApJ...711..693K} {711, 693}

\bibitem[\protect\citeauthoryear{{Kron}}{{Kron}}{1980}]{Kron1980}
{Kron} R.~G.,  1980, \mn@doi [\apjs] {10.1086/190669}, \href
  {https://ui.adsabs.harvard.edu/abs/1980ApJS...43..305K} {43, 305}

\bibitem[\protect\citeauthoryear{{Kusakabe} et~al.,}{{Kusakabe}
  et~al.}{2020}]{Kusakabe2020}
{Kusakabe} H.,  et~al., 2020, \mn@doi [\aap] {10.1051/0004-6361/201937340},
  \href {https://ui.adsabs.harvard.edu/abs/2020A&A...638A..12K} {638, A12}

\bibitem[\protect\citeauthoryear{{Labb{\'e}} et~al.,}{{Labb{\'e}}
  et~al.}{2013}]{Labbe2013}
{Labb{\'e}} I.,  et~al., 2013, \mn@doi [\apjl] {10.1088/2041-8205/777/2/L19},
  \href {https://ui.adsabs.harvard.edu/abs/2013ApJ...777L..19L} {777, L19}

\bibitem[\protect\citeauthoryear{{Laporte}, {Nakajima}, {Ellis}, {Zitrin},
  {Stark}, {Mainali}  \& {Roberts-Borsani}}{{Laporte}
  et~al.}{2017}]{Laporte2017}
{Laporte} N.,  {Nakajima} K.,  {Ellis} R.~S.,  {Zitrin} A.,  {Stark} D.~P.,
  {Mainali} R.,   {Roberts-Borsani} G.~W.,  2017, \mn@doi [\apj]
  {10.3847/1538-4357/aa96a8}, \href
  {https://ui.adsabs.harvard.edu/abs/2017ApJ...851...40L} {851, 40}

\bibitem[\protect\citeauthoryear{{Laporte}, {Meyer}, {Ellis}, {Robertson},
  {Chisholm}  \& {Roberts-Borsani}}{{Laporte} et~al.}{2021}]{Laporte2021}
{Laporte} N.,  {Meyer} R.~A.,  {Ellis} R.~S.,  {Robertson} B.~E.,  {Chisholm}
  J.,   {Roberts-Borsani} G.~W.,  2021, \mn@doi [\mnras]
  {10.1093/mnras/stab1239}, \href
  {https://ui.adsabs.harvard.edu/abs/2021MNRAS.505.3336L} {505, 3336}

\bibitem[\protect\citeauthoryear{{Larson} et~al.,}{{Larson}
  et~al.}{2022}]{Larson2022}
{Larson} R.~L.,  et~al., 2022, \mn@doi [\apj] {10.3847/1538-4357/ac5dbd}, \href
  {https://ui.adsabs.harvard.edu/abs/2022ApJ...930..104L} {930, 104}

\bibitem[\protect\citeauthoryear{{Larson} et~al.,}{{Larson}
  et~al.}{2023}]{Larson2023}
{Larson} R.~L.,  et~al., 2023, \mn@doi [\apjl] {10.3847/2041-8213/ace619},
  \href {https://ui.adsabs.harvard.edu/abs/2023ApJ...953L..29L} {953, L29}

\bibitem[\protect\citeauthoryear{{Laursen}, {Sommer-Larsen}  \&
  {Razoumov}}{{Laursen} et~al.}{2011}]{Laursen2011}
{Laursen} P.,  {Sommer-Larsen} J.,   {Razoumov} A.~O.,  2011, \mn@doi [\apj]
  {10.1088/0004-637X/728/1/52}, \href
  {https://ui.adsabs.harvard.edu/abs/2011ApJ...728...52L} {728, 52}

\bibitem[\protect\citeauthoryear{{Leclercq} et~al.,}{{Leclercq}
  et~al.}{2017}]{Leclercq2017}
{Leclercq} F.,  et~al., 2017, \mn@doi [\aap] {10.1051/0004-6361/201731480},
  \href {https://ui.adsabs.harvard.edu/abs/2017A&A...608A...8L} {608, A8}

\bibitem[\protect\citeauthoryear{{Leclercq} et~al.,}{{Leclercq}
  et~al.}{2020}]{Leclercq2020}
{Leclercq} F.,  et~al., 2020, \mn@doi [\aap] {10.1051/0004-6361/201937339},
  \href {https://ui.adsabs.harvard.edu/abs/2020A&A...635A..82L} {635, A82}

\bibitem[\protect\citeauthoryear{{Lehnert} et~al.,}{{Lehnert}
  et~al.}{2010}]{Lehnert2010}
{Lehnert} M.~D.,  et~al., 2010, \mn@doi [\nat] {10.1038/nature09462}, \href
  {https://ui.adsabs.harvard.edu/abs/2010Natur.467..940L} {467, 940}

\bibitem[\protect\citeauthoryear{{Leonova} et~al.,}{{Leonova}
  et~al.}{2022}]{Leonova2022}
{Leonova} E.,  et~al., 2022, \mn@doi [\mnras] {10.1093/mnras/stac1908}, \href
  {https://ui.adsabs.harvard.edu/abs/2022MNRAS.515.5790L} {515, 5790}

\bibitem[\protect\citeauthoryear{{Li} \& {Gronke}}{{Li} \&
  {Gronke}}{2022}]{Li2022b}
{Li} Z.,  {Gronke} M.,  2022, \mn@doi [\mnras] {10.1093/mnras/stac1207}, \href
  {https://ui.adsabs.harvard.edu/abs/2022MNRAS.513.5034L} {513, 5034}

\bibitem[\protect\citeauthoryear{{Li}, {Steidel}, {Gronke}, {Chen}  \&
  {Matsuda}}{{Li} et~al.}{2022}]{Li2022a}
{Li} Z.,  {Steidel} C.~C.,  {Gronke} M.,  {Chen} Y.,   {Matsuda} Y.,  2022,
  \mn@doi [\mnras] {10.1093/mnras/stac958}, \href
  {https://ui.adsabs.harvard.edu/abs/2022MNRAS.513.3414L} {513, 3414}

\bibitem[\protect\citeauthoryear{{Lin} et~al.,}{{Lin} et~al.}{2024}]{Lin2024}
{Lin} X.,  et~al., 2024, \mn@doi [arXiv e-prints] {10.48550/arXiv.2401.09532},
  \href {https://ui.adsabs.harvard.edu/abs/2024arXiv240109532L} {p.
  arXiv:2401.09532}

\bibitem[\protect\citeauthoryear{{Lu}, {Mason}, {Hutter}, {Mesinger}, {Qin},
  {Stark}  \& {Endsley}}{{Lu} et~al.}{2024}]{Lu2024}
{Lu} T.-Y.,  {Mason} C.~A.,  {Hutter} A.,  {Mesinger} A.,  {Qin} Y.,  {Stark}
  D.~P.,   {Endsley} R.,  2024, \mn@doi [\mnras] {10.1093/mnras/stae266}, \href
  {https://ui.adsabs.harvard.edu/abs/2024MNRAS.528.4872L} {528, 4872}

\bibitem[\protect\citeauthoryear{{Lyu} et~al.,}{{Lyu} et~al.}{2024}]{Lyu2024}
{Lyu} J.,  et~al., 2024, \mn@doi [\apj] {10.3847/1538-4357/ad3643}, \href
  {https://ui.adsabs.harvard.edu/abs/2024ApJ...966..229L} {966, 229}

\bibitem[\protect\citeauthoryear{{Ma}, {Quataert}, {Wetzel}, {Hopkins},
  {Faucher-Gigu{\`e}re}  \& {Kere{\v{s}}}}{{Ma} et~al.}{2020}]{Ma2020}
{Ma} X.,  {Quataert} E.,  {Wetzel} A.,  {Hopkins} P.~F.,  {Faucher-Gigu{\`e}re}
  C.-A.,   {Kere{\v{s}}} D.,  2020, \mn@doi [\mnras] {10.1093/mnras/staa2404},
  \href {https://ui.adsabs.harvard.edu/abs/2020MNRAS.498.2001M} {498, 2001}

\bibitem[\protect\citeauthoryear{{Madau}}{{Madau}}{1995}]{Madau1995}
{Madau} P.,  1995, \mn@doi [\apj] {10.1086/175332}, \href
  {https://ui.adsabs.harvard.edu/abs/1995ApJ...441...18M} {441, 18}

\bibitem[\protect\citeauthoryear{{Mainali}, {Kollmeier}, {Stark}, {Simcoe},
  {Walth}, {Newman}  \& {Miller}}{{Mainali} et~al.}{2017}]{Mainali2017}
{Mainali} R.,  {Kollmeier} J.~A.,  {Stark} D.~P.,  {Simcoe} R.~A.,  {Walth} G.,
   {Newman} A.~B.,   {Miller} D.~R.,  2017, \mn@doi [\apjl]
  {10.3847/2041-8213/836/1/L14}, \href
  {https://ui.adsabs.harvard.edu/abs/2017ApJ...836L..14M} {836, L14}

\bibitem[\protect\citeauthoryear{{Maiolino} et~al.,}{{Maiolino}
  et~al.}{2015}]{Maiolino2015}
{Maiolino} R.,  et~al., 2015, \mn@doi [\mnras] {10.1093/mnras/stv1194}, \href
  {https://ui.adsabs.harvard.edu/abs/2015MNRAS.452...54M} {452, 54}

\bibitem[\protect\citeauthoryear{{Maiolino} et~al.,}{{Maiolino}
  et~al.}{2023}]{Maiolino2023}
{Maiolino} R.,  et~al., 2023, \mn@doi [arXiv e-prints]
  {10.48550/arXiv.2308.01230}, \href
  {https://ui.adsabs.harvard.edu/abs/2023arXiv230801230M} {p. arXiv:2308.01230}

\bibitem[\protect\citeauthoryear{{Mary}, {Bacon}, {Conseil}, {Piqueras}  \&
  {Schutz}}{{Mary} et~al.}{2020}]{Mary2020}
{Mary} D.,  {Bacon} R.,  {Conseil} S.,  {Piqueras} L.,   {Schutz} A.,  2020,
  \mn@doi [\aap] {10.1051/0004-6361/201937001}, \href
  {https://ui.adsabs.harvard.edu/abs/2020A&A...635A.194M} {635, A194}

\bibitem[\protect\citeauthoryear{{Maseda} et~al.,}{{Maseda}
  et~al.}{2020}]{Maseda2020}
{Maseda} M.~V.,  et~al., 2020, \mn@doi [\mnras] {10.1093/mnras/staa622}, \href
  {https://ui.adsabs.harvard.edu/abs/2020MNRAS.493.5120M} {493, 5120}

\bibitem[\protect\citeauthoryear{{Maseda} et~al.,}{{Maseda}
  et~al.}{2023}]{Maseda2023}
{Maseda} M.~V.,  et~al., 2023, \mn@doi [\apj] {10.3847/1538-4357/acf12b}, \href
  {https://ui.adsabs.harvard.edu/abs/2023ApJ...956...11M} {956, 11}

\bibitem[\protect\citeauthoryear{{Mason} \& {Gronke}}{{Mason} \&
  {Gronke}}{2020}]{Mason2020}
{Mason} C.~A.,  {Gronke} M.,  2020, \mn@doi [\mnras] {10.1093/mnras/staa2910},
  \href {https://ui.adsabs.harvard.edu/abs/2020MNRAS.499.1395M} {499, 1395}

\bibitem[\protect\citeauthoryear{{Mason}, {Treu}, {Dijkstra}, {Mesinger},
  {Trenti}, {Pentericci}, {de Barros}  \& {Vanzella}}{{Mason}
  et~al.}{2018a}]{Mason2018a}
{Mason} C.~A.,  {Treu} T.,  {Dijkstra} M.,  {Mesinger} A.,  {Trenti} M.,
  {Pentericci} L.,  {de Barros} S.,   {Vanzella} E.,  2018a, \mn@doi [\apj]
  {10.3847/1538-4357/aab0a7}, \href
  {https://ui.adsabs.harvard.edu/abs/2018ApJ...856....2M} {856, 2}

\bibitem[\protect\citeauthoryear{{Mason} et~al.,}{{Mason}
  et~al.}{2018b}]{Mason2018b}
{Mason} C.~A.,  et~al., 2018b, \mn@doi [\apjl] {10.3847/2041-8213/aabbab},
  \href {https://ui.adsabs.harvard.edu/abs/2018ApJ...857L..11M} {857, L11}

\bibitem[\protect\citeauthoryear{{Mason} et~al.,}{{Mason}
  et~al.}{2019}]{Mason2019}
{Mason} C.~A.,  et~al., 2019, \mn@doi [\mnras] {10.1093/mnras/stz632}, \href
  {https://ui.adsabs.harvard.edu/abs/2019MNRAS.485.3947M} {485, 3947}

\bibitem[\protect\citeauthoryear{{Matsuda} et~al.,}{{Matsuda}
  et~al.}{2012}]{Matsuda2012}
{Matsuda} Y.,  et~al., 2012, \mn@doi [\mnras]
  {10.1111/j.1365-2966.2012.21143.x}, \href
  {https://ui.adsabs.harvard.edu/abs/2012MNRAS.425..878M} {425, 878}

\bibitem[\protect\citeauthoryear{{Matthee}, {Sobral}, {Oteo}, {Best}, {Smail},
  {R{\"o}ttgering}  \& {Paulino-Afonso}}{{Matthee} et~al.}{2016}]{Matthee2016}
{Matthee} J.,  {Sobral} D.,  {Oteo} I.,  {Best} P.,  {Smail} I.,
  {R{\"o}ttgering} H.,   {Paulino-Afonso} A.,  2016, \mn@doi [\mnras]
  {10.1093/mnras/stw322}, \href
  {https://ui.adsabs.harvard.edu/abs/2016MNRAS.458..449M} {458, 449}

\bibitem[\protect\citeauthoryear{{Matthee}, {Sobral}, {Gronke},
  {Paulino-Afonso}, {Stefanon}  \& {R{\"o}ttgering}}{{Matthee}
  et~al.}{2018}]{Matthee2018}
{Matthee} J.,  {Sobral} D.,  {Gronke} M.,  {Paulino-Afonso} A.,  {Stefanon} M.,
    {R{\"o}ttgering} H.,  2018, \mn@doi [\aap] {10.1051/0004-6361/201833528},
  \href {https://ui.adsabs.harvard.edu/abs/2018A&A...619A.136M} {619, A136}

\bibitem[\protect\citeauthoryear{{Matthee}, {Sobral}, {Gronke}, {Pezzulli},
  {Cantalupo}, {R{\"o}ttgering}, {Darvish}  \& {Santos}}{{Matthee}
  et~al.}{2020}]{Matthee2020}
{Matthee} J.,  {Sobral} D.,  {Gronke} M.,  {Pezzulli} G.,  {Cantalupo} S.,
  {R{\"o}ttgering} H.,  {Darvish} B.,   {Santos} S.,  2020, \mn@doi [\mnras]
  {10.1093/mnras/stz3554}, \href
  {https://ui.adsabs.harvard.edu/abs/2020MNRAS.492.1778M} {492, 1778}

\bibitem[\protect\citeauthoryear{{Matthee} et~al.,}{{Matthee}
  et~al.}{2021}]{Matthee2021}
{Matthee} J.,  et~al., 2021, \mn@doi [\mnras] {10.1093/mnras/stab1304}, \href
  {https://ui.adsabs.harvard.edu/abs/2021MNRAS.505.1382M} {505, 1382}

\bibitem[\protect\citeauthoryear{{Matthee}, {Mackenzie}, {Simcoe}, {Kashino},
  {Lilly}, {Bordoloi}  \& {Eilers}}{{Matthee} et~al.}{2023}]{Matthee2023}
{Matthee} J.,  {Mackenzie} R.,  {Simcoe} R.~A.,  {Kashino} D.,  {Lilly} S.~J.,
  {Bordoloi} R.,   {Eilers} A.-C.,  2023, \mn@doi [\apj]
  {10.3847/1538-4357/acc846}, \href
  {https://ui.adsabs.harvard.edu/abs/2023ApJ...950...67M} {950, 67}

\bibitem[\protect\citeauthoryear{{Matthee} et~al.,}{{Matthee}
  et~al.}{2024}]{Matthee2024}
{Matthee} J.,  et~al., 2024, \mn@doi [\apj] {10.3847/1538-4357/ad2345}, \href
  {https://ui.adsabs.harvard.edu/abs/2024ApJ...963..129M} {963, 129}

\bibitem[\protect\citeauthoryear{{McLinden} et~al.,}{{McLinden}
  et~al.}{2011}]{McLinden2011}
{McLinden} E.~M.,  et~al., 2011, \mn@doi [\apj] {10.1088/0004-637X/730/2/136},
  \href {https://ui.adsabs.harvard.edu/abs/2011ApJ...730..136M} {730, 136}

\bibitem[\protect\citeauthoryear{{McLinden}, {Rhoads}, {Malhotra},
  {Finkelstein}, {Richardson}, {Smith}  \& {Tilvi}}{{McLinden}
  et~al.}{2014}]{McLinden2014}
{McLinden} E.~M.,  {Rhoads} J.~E.,  {Malhotra} S.,  {Finkelstein} S.~L.,
  {Richardson} M.~L.~A.,  {Smith} B.,   {Tilvi} V.~S.,  2014, \mn@doi [\mnras]
  {10.1093/mnras/stu023}, \href
  {https://ui.adsabs.harvard.edu/abs/2014MNRAS.439..446M} {439, 446}

\bibitem[\protect\citeauthoryear{{McLure}, {Dunlop}, {Cirasuolo}, {Koekemoer},
  {Sabbi}, {Stark}, {Targett}  \& {Ellis}}{{McLure} et~al.}{2010}]{McLure2010}
{McLure} R.~J.,  {Dunlop} J.~S.,  {Cirasuolo} M.,  {Koekemoer} A.~M.,  {Sabbi}
  E.,  {Stark} D.~P.,  {Targett} T.~A.,   {Ellis} R.~S.,  2010, \mn@doi
  [\mnras] {10.1111/j.1365-2966.2009.16176.x}, \href
  {https://ui.adsabs.harvard.edu/abs/2010MNRAS.403..960M} {403, 960}

\bibitem[\protect\citeauthoryear{{Mesinger}, {Aykutalp}, {Vanzella},
  {Pentericci}, {Ferrara}  \& {Dijkstra}}{{Mesinger}
  et~al.}{2015}]{Mesinger2015}
{Mesinger} A.,  {Aykutalp} A.,  {Vanzella} E.,  {Pentericci} L.,  {Ferrara} A.,
    {Dijkstra} M.,  2015, \mn@doi [\mnras] {10.1093/mnras/stu2089}, \href
  {https://ui.adsabs.harvard.edu/abs/2015MNRAS.446..566M} {446, 566}

\bibitem[\protect\citeauthoryear{{Meurer}, {Heckman}  \& {Calzetti}}{{Meurer}
  et~al.}{1999}]{Meurer1999}
{Meurer} G.~R.,  {Heckman} T.~M.,   {Calzetti} D.,  1999, \mn@doi [\apj]
  {10.1086/307523}, \href
  {https://ui.adsabs.harvard.edu/abs/1999ApJ...521...64M} {521, 64}

\bibitem[\protect\citeauthoryear{{Meyer}, {Laporte}, {Ellis}, {Verhamme}  \&
  {Garel}}{{Meyer} et~al.}{2021}]{Meyer2021}
{Meyer} R.~A.,  {Laporte} N.,  {Ellis} R.~S.,  {Verhamme} A.,   {Garel} T.,
  2021, \mn@doi [\mnras] {10.1093/mnras/staa3216}, \href
  {https://ui.adsabs.harvard.edu/abs/2021MNRAS.500..558M} {500, 558}

\bibitem[\protect\citeauthoryear{{Miralda-Escud{\'e}}}{{Miralda-Escud{\'e}}}{1998}]{Miralda-Escude1998}
{Miralda-Escud{\'e}} J.,  1998, \mn@doi [\apj] {10.1086/305799}, \href
  {https://ui.adsabs.harvard.edu/abs/1998ApJ...501...15M} {501, 15}

\bibitem[\protect\citeauthoryear{{Mitchell}, {Blaizot}, {Cadiou}, {Dubois},
  {Garel}  \& {Rosdahl}}{{Mitchell} et~al.}{2021}]{Mitchell2021}
{Mitchell} P.~D.,  {Blaizot} J.,  {Cadiou} C.,  {Dubois} Y.,  {Garel} T.,
  {Rosdahl} J.,  2021, \mn@doi [\mnras] {10.1093/mnras/stab035}, \href
  {https://ui.adsabs.harvard.edu/abs/2021MNRAS.501.5757M} {501, 5757}

\bibitem[\protect\citeauthoryear{{Momose} et~al.,}{{Momose}
  et~al.}{2014}]{Momose2014}
{Momose} R.,  et~al., 2014, \mn@doi [\mnras] {10.1093/mnras/stu825}, \href
  {https://ui.adsabs.harvard.edu/abs/2014MNRAS.442..110M} {442, 110}

\bibitem[\protect\citeauthoryear{{Morales}, {Mason}, {Bruton}, {Gronke},
  {Haardt}  \& {Scarlata}}{{Morales} et~al.}{2021}]{Morales2021}
{Morales} A.~M.,  {Mason} C.~A.,  {Bruton} S.,  {Gronke} M.,  {Haardt} F.,
  {Scarlata} C.,  2021, \mn@doi [\apj] {10.3847/1538-4357/ac1104}, \href
  {https://ui.adsabs.harvard.edu/abs/2021ApJ...919..120M} {919, 120}

\bibitem[\protect\citeauthoryear{{Naidu} et~al.,}{{Naidu}
  et~al.}{2017}]{Naidu2017}
{Naidu} R.~P.,  et~al., 2017, \mn@doi [\apj] {10.3847/1538-4357/aa8863}, \href
  {https://ui.adsabs.harvard.edu/abs/2017ApJ...847...12N} {847, 12}

\bibitem[\protect\citeauthoryear{{Naidu}, {Tacchella}, {Mason}, {Bose}, {Oesch}
   \& {Conroy}}{{Naidu} et~al.}{2020}]{Naidu2020}
{Naidu} R.~P.,  {Tacchella} S.,  {Mason} C.~A.,  {Bose} S.,  {Oesch} P.~A.,
  {Conroy} C.,  2020, \mn@doi [\apj] {10.3847/1538-4357/ab7cc9}, \href
  {https://ui.adsabs.harvard.edu/abs/2020ApJ...892..109N} {892, 109}

\bibitem[\protect\citeauthoryear{{Naidu} et~al.,}{{Naidu}
  et~al.}{2022}]{Naidu2022}
{Naidu} R.~P.,  et~al., 2022, \mn@doi [\mnras] {10.1093/mnras/stab3601}, \href
  {https://ui.adsabs.harvard.edu/abs/2022MNRAS.510.4582N} {510, 4582}

\bibitem[\protect\citeauthoryear{{Nakajima}, {Ouchi}, {Isobe}, {Harikane},
  {Zhang}, {Ono}, {Umeda}  \& {Oguri}}{{Nakajima} et~al.}{2023}]{Nakajima2023}
{Nakajima} K.,  {Ouchi} M.,  {Isobe} Y.,  {Harikane} Y.,  {Zhang} Y.,  {Ono}
  Y.,  {Umeda} H.,   {Oguri} M.,  2023, \mn@doi [\apjs]
  {10.3847/1538-4365/acd556}, \href
  {https://ui.adsabs.harvard.edu/abs/2023ApJS..269...33N} {269, 33}

\bibitem[\protect\citeauthoryear{{Nakane} et~al.,}{{Nakane}
  et~al.}{2024}]{Nakane2024}
{Nakane} M.,  et~al., 2024, \mn@doi [\apj] {10.3847/1538-4357/ad38c2}, \href
  {https://ui.adsabs.harvard.edu/abs/2024ApJ...967...28N} {967, 28}

\bibitem[\protect\citeauthoryear{{Oesch} et~al.,}{{Oesch}
  et~al.}{2015}]{Oesch2015}
{Oesch} P.~A.,  et~al., 2015, \mn@doi [\apjl] {10.1088/2041-8205/804/2/L30},
  \href {https://ui.adsabs.harvard.edu/abs/2015ApJ...804L..30O} {804, L30}

\bibitem[\protect\citeauthoryear{{Oesch} et~al.,}{{Oesch}
  et~al.}{2023}]{Oesch2023}
{Oesch} P.~A.,  et~al., 2023, \mn@doi [\mnras] {10.1093/mnras/stad2411}, \href
  {https://ui.adsabs.harvard.edu/abs/2023MNRAS.525.2864O} {525, 2864}

\bibitem[\protect\citeauthoryear{{Oke} \& {Gunn}}{{Oke} \&
  {Gunn}}{1983}]{Oke1983}
{Oke} J.~B.,  {Gunn} J.~E.,  1983, \mn@doi [\apj] {10.1086/160817}, \href
  {https://ui.adsabs.harvard.edu/abs/1983ApJ...266..713O} {266, 713}

\bibitem[\protect\citeauthoryear{{Ono}, {Ouchi}, {Shimasaku}, {Dunlop},
  {Farrah}, {McLure}  \& {Okamura}}{{Ono} et~al.}{2010}]{Ono2010}
{Ono} Y.,  {Ouchi} M.,  {Shimasaku} K.,  {Dunlop} J.,  {Farrah} D.,  {McLure}
  R.,   {Okamura} S.,  2010, \mn@doi [\apj] {10.1088/0004-637X/724/2/1524},
  \href {https://ui.adsabs.harvard.edu/abs/2010ApJ...724.1524O} {724, 1524}

\bibitem[\protect\citeauthoryear{{Ono} et~al.,}{{Ono} et~al.}{2012}]{Ono2012}
{Ono} Y.,  et~al., 2012, \mn@doi [\apj] {10.1088/0004-637X/744/2/83}, \href
  {https://ui.adsabs.harvard.edu/abs/2012ApJ...744...83O} {744, 83}

\bibitem[\protect\citeauthoryear{{Osterbrock} \& {Ferland}}{{Osterbrock} \&
  {Ferland}}{2006}]{Osterbrock2006}
{Osterbrock} D.~E.,  {Ferland} G.~J.,  2006, {Astrophysics of gaseous nebulae
  and active galactic nuclei}

\bibitem[\protect\citeauthoryear{{Ota} et~al.,}{{Ota} et~al.}{2017}]{Ota2017}
{Ota} K.,  et~al., 2017, \mn@doi [\apj] {10.3847/1538-4357/aa7a0a}, \href
  {https://ui.adsabs.harvard.edu/abs/2017ApJ...844...85O} {844, 85}

\bibitem[\protect\citeauthoryear{{Ouchi} et~al.,}{{Ouchi}
  et~al.}{2010}]{Ouchi2010}
{Ouchi} M.,  et~al., 2010, \mn@doi [\apj] {10.1088/0004-637X/723/1/869}, \href
  {https://ui.adsabs.harvard.edu/abs/2010ApJ...723..869O} {723, 869}

\bibitem[\protect\citeauthoryear{{Ouchi}, {Ono}  \& {Shibuya}}{{Ouchi}
  et~al.}{2020}]{Ouchi2020}
{Ouchi} M.,  {Ono} Y.,   {Shibuya} T.,  2020, \mn@doi [\araa]
  {10.1146/annurev-astro-032620-021859}, \href
  {https://ui.adsabs.harvard.edu/abs/2020ARA&A..58..617O} {58, 617}

\bibitem[\protect\citeauthoryear{{Pahl}, {Shapley}, {Steidel}, {Reddy}, {Chen},
  {Rudie}  \& {Strom}}{{Pahl} et~al.}{2023}]{Pahl2023}
{Pahl} A.~J.,  {Shapley} A.,  {Steidel} C.~C.,  {Reddy} N.~A.,  {Chen} Y.,
  {Rudie} G.~C.,   {Strom} A.~L.,  2023, \mn@doi [\mnras]
  {10.1093/mnras/stad774}, \href
  {https://ui.adsabs.harvard.edu/abs/2023MNRAS.521.3247P} {521, 3247}

\bibitem[\protect\citeauthoryear{{Pahl}, {Shapley}, {Steidel}, {Reddy}, {Chen}
  \& {Rudie}}{{Pahl} et~al.}{2024}]{Pahl2024}
{Pahl} A.~J.,  {Shapley} A.~E.,  {Steidel} C.~C.,  {Reddy} N.~A.,  {Chen} Y.,
  {Rudie} G.~C.,  2024, \mn@doi [arXiv e-prints] {10.48550/arXiv.2401.09526},
  \href {https://ui.adsabs.harvard.edu/abs/2024arXiv240109526P} {p.
  arXiv:2401.09526}

\bibitem[\protect\citeauthoryear{{Pei}}{{Pei}}{1992}]{Pei1992}
{Pei} Y.~C.,  1992, \mn@doi [\apj] {10.1086/171637}, \href
  {https://ui.adsabs.harvard.edu/abs/1992ApJ...395..130P} {395, 130}

\bibitem[\protect\citeauthoryear{{Pentericci} et~al.,}{{Pentericci}
  et~al.}{2016}]{Pentericci2016}
{Pentericci} L.,  et~al., 2016, \mn@doi [\apjl] {10.3847/2041-8205/829/1/L11},
  \href {https://ui.adsabs.harvard.edu/abs/2016ApJ...829L..11P} {829, L11}

\bibitem[\protect\citeauthoryear{{Pentericci} et~al.,}{{Pentericci}
  et~al.}{2018}]{Pentericci2018}
{Pentericci} L.,  et~al., 2018, \mn@doi [\aap] {10.1051/0004-6361/201732465},
  \href {https://ui.adsabs.harvard.edu/abs/2018A&A...619A.147P} {619, A147}

\bibitem[\protect\citeauthoryear{{Perrin}, {Sivaramakrishnan}, {Lajoie},
  {Elliott}, {Pueyo}, {Ravindranath}  \& {Albert}}{{Perrin}
  et~al.}{2014}]{Perrin2014}
{Perrin} M.~D.,  {Sivaramakrishnan} A.,  {Lajoie} C.-P.,  {Elliott} E.,
  {Pueyo} L.,  {Ravindranath} S.,   {Albert} L.,  2014, in {Oschmann}
  Jacobus~M. J.,  {Clampin} M.,  {Fazio} G.~G.,   {MacEwen} H.~A.,  eds,
  Society of Photo-Optical Instrumentation Engineers (SPIE) Conference Series
  Vol. 9143, Space Telescopes and Instrumentation 2014: Optical, Infrared, and
  Millimeter Wave. p. 91433X, \mn@doi{10.1117/12.2056689}

\bibitem[\protect\citeauthoryear{{Planck Collaboration} et~al.,}{{Planck
  Collaboration} et~al.}{2020}]{Planck2020}
{Planck Collaboration} et~al., 2020, \mn@doi [\aap]
  {10.1051/0004-6361/201833910}, \href
  {https://ui.adsabs.harvard.edu/abs/2020A&A...641A...6P} {641, A6}

\bibitem[\protect\citeauthoryear{{Prieto-Lyon} et~al.,}{{Prieto-Lyon}
  et~al.}{2023}]{Prieto-Lyon2023}
{Prieto-Lyon} G.,  et~al., 2023, \mn@doi [\apj] {10.3847/1538-4357/acf715},
  \href {https://ui.adsabs.harvard.edu/abs/2023ApJ...956..136P} {956, 136}

\bibitem[\protect\citeauthoryear{{Raiter}, {Schaerer}  \& {Fosbury}}{{Raiter}
  et~al.}{2010}]{Raiter2010}
{Raiter} A.,  {Schaerer} D.,   {Fosbury} R.~A.~E.,  2010, \mn@doi [\aap]
  {10.1051/0004-6361/201015236}, \href
  {https://ui.adsabs.harvard.edu/abs/2010A&A...523A..64R} {523, A64}

\bibitem[\protect\citeauthoryear{{Reddy} \& {Steidel}}{{Reddy} \&
  {Steidel}}{2009}]{Reddy2009}
{Reddy} N.~A.,  {Steidel} C.~C.,  2009, \mn@doi [\apj]
  {10.1088/0004-637X/692/1/778}, \href
  {https://ui.adsabs.harvard.edu/abs/2009ApJ...692..778R} {692, 778}

\bibitem[\protect\citeauthoryear{{Reddy}, {Steidel}, {Pettini},
  {Bogosavljevi{\'c}}  \& {Shapley}}{{Reddy} et~al.}{2016}]{Reddy2016}
{Reddy} N.~A.,  {Steidel} C.~C.,  {Pettini} M.,  {Bogosavljevi{\'c}} M.,
  {Shapley} A.~E.,  2016, \mn@doi [\apj] {10.3847/0004-637X/828/2/108}, \href
  {https://ui.adsabs.harvard.edu/abs/2016ApJ...828..108R} {828, 108}

\bibitem[\protect\citeauthoryear{{Rieke} et~al.,}{{Rieke}
  et~al.}{2023a}]{Rieke2023a}
{Rieke} M.~J.,  et~al., 2023a, \mn@doi [\pasp] {10.1088/1538-3873/acac53},
  \href {https://ui.adsabs.harvard.edu/abs/2023PASP..135b8001R} {135, 028001}

\bibitem[\protect\citeauthoryear{{Rieke} et~al.,}{{Rieke}
  et~al.}{2023b}]{Rieke2023b}
{Rieke} M.~J.,  et~al., 2023b, \mn@doi [\apjs] {10.3847/1538-4365/acf44d},
  \href {https://ui.adsabs.harvard.edu/abs/2023ApJS..269...16R} {269, 16}

\bibitem[\protect\citeauthoryear{{Rinaldi} et~al.,}{{Rinaldi}
  et~al.}{2023}]{Rinaldi2023}
{Rinaldi} P.,  et~al., 2023, \mn@doi [\apj] {10.3847/1538-4357/acdc27}, \href
  {https://ui.adsabs.harvard.edu/abs/2023ApJ...952..143R} {952, 143}

\bibitem[\protect\citeauthoryear{{Rivera-Thorsen} et~al.,}{{Rivera-Thorsen}
  et~al.}{2017}]{Rivera-Thorsen2017}
{Rivera-Thorsen} T.~E.,  et~al., 2017, \mn@doi [\aap]
  {10.1051/0004-6361/201732173}, \href
  {https://ui.adsabs.harvard.edu/abs/2017A&A...608L...4R} {608, L4}

\bibitem[\protect\citeauthoryear{{Roberts-Borsani} et~al.,}{{Roberts-Borsani}
  et~al.}{2016}]{Roberts-Borsani2016}
{Roberts-Borsani} G.~W.,  et~al., 2016, \mn@doi [\apj]
  {10.3847/0004-637X/823/2/143}, \href
  {https://ui.adsabs.harvard.edu/abs/2016ApJ...823..143R} {823, 143}

\bibitem[\protect\citeauthoryear{{Roberts-Borsani}, {Ellis}  \&
  {Laporte}}{{Roberts-Borsani} et~al.}{2020}]{Roberts-Borsani2020}
{Roberts-Borsani} G.~W.,  {Ellis} R.~S.,   {Laporte} N.,  2020, \mn@doi
  [\mnras] {10.1093/mnras/staa2085}, \href
  {https://ui.adsabs.harvard.edu/abs/2020MNRAS.497.3440R} {497, 3440}

\bibitem[\protect\citeauthoryear{{Robertson}}{{Robertson}}{2022}]{Robertson2022}
{Robertson} B.~E.,  2022, \mn@doi [\araa]
  {10.1146/annurev-astro-120221-044656}, \href
  {https://ui.adsabs.harvard.edu/abs/2022ARA&A..60..121R} {60, 121}

\bibitem[\protect\citeauthoryear{{Robertson}, {Ellis}, {Dunlop}, {McLure}  \&
  {Stark}}{{Robertson} et~al.}{2010}]{Robertson2010}
{Robertson} B.~E.,  {Ellis} R.~S.,  {Dunlop} J.~S.,  {McLure} R.~J.,   {Stark}
  D.~P.,  2010, \mn@doi [\nat] {10.1038/nature09527}, \href
  {https://ui.adsabs.harvard.edu/abs/2010Natur.468...49R} {468, 49}

\bibitem[\protect\citeauthoryear{{Robertson} et~al.,}{{Robertson}
  et~al.}{2023}]{Robertson2023}
{Robertson} B.~E.,  et~al., 2023, \mn@doi [Nature Astronomy]
  {10.1038/s41550-023-01921-1}, \href
  {https://ui.adsabs.harvard.edu/abs/2023NatAs.tmp...67R} {}

\bibitem[\protect\citeauthoryear{{Roy} et~al.,}{{Roy} et~al.}{2023}]{Roy2023}
{Roy} N.,  et~al., 2023, \mn@doi [\apjl] {10.3847/2041-8213/acdbce}, \href
  {https://ui.adsabs.harvard.edu/abs/2023ApJ...952L..14R} {952, L14}

\bibitem[\protect\citeauthoryear{{Ryabchikova}, {Piskunov}, {Kurucz},
  {Stempels}, {Heiter}, {Pakhomov}  \& {Barklem}}{{Ryabchikova}
  et~al.}{2015}]{Ryabchikova2015}
{Ryabchikova} T.,  {Piskunov} N.,  {Kurucz} R.~L.,  {Stempels} H.~C.,  {Heiter}
  U.,  {Pakhomov} Y.,   {Barklem} P.~S.,  2015, \mn@doi [\physscr]
  {10.1088/0031-8949/90/5/054005}, \href
  {https://ui.adsabs.harvard.edu/abs/2015PhyS...90e4005R} {90, 054005}

\bibitem[\protect\citeauthoryear{{Saldana-Lopez} et~al.,}{{Saldana-Lopez}
  et~al.}{2022}]{Saldana-Lopez2022}
{Saldana-Lopez} A.,  et~al., 2022, \mn@doi [\aap]
  {10.1051/0004-6361/202141864}, \href
  {https://ui.adsabs.harvard.edu/abs/2022A&A...663A..59S} {663, A59}

\bibitem[\protect\citeauthoryear{{Salmon} et~al.,}{{Salmon}
  et~al.}{2015}]{Salmon2015}
{Salmon} B.,  et~al., 2015, \mn@doi [\apj] {10.1088/0004-637X/799/2/183}, \href
  {https://ui.adsabs.harvard.edu/abs/2015ApJ...799..183S} {799, 183}

\bibitem[\protect\citeauthoryear{{Sanders} et~al.,}{{Sanders}
  et~al.}{2016}]{Sanders2016}
{Sanders} R.~L.,  et~al., 2016, \mn@doi [\apj] {10.3847/0004-637X/816/1/23},
  \href {https://ui.adsabs.harvard.edu/abs/2016ApJ...816...23S} {816, 23}

\bibitem[\protect\citeauthoryear{{Sanders}, {Shapley}, {Topping}, {Reddy}  \&
  {Brammer}}{{Sanders} et~al.}{2023}]{Sanders2023}
{Sanders} R.~L.,  {Shapley} A.~E.,  {Topping} M.~W.,  {Reddy} N.~A.,
  {Brammer} G.~B.,  2023, \mn@doi [\apj] {10.3847/1538-4357/acedad}, \href
  {https://ui.adsabs.harvard.edu/abs/2023ApJ...955...54S} {955, 54}

\bibitem[\protect\citeauthoryear{{Sanders}, {Shapley}, {Topping}, {Reddy}  \&
  {Brammer}}{{Sanders} et~al.}{2024}]{Sanders2024}
{Sanders} R.~L.,  {Shapley} A.~E.,  {Topping} M.~W.,  {Reddy} N.~A.,
  {Brammer} G.~B.,  2024, \mn@doi [\apj] {10.3847/1538-4357/ad15fc}, \href
  {https://ui.adsabs.harvard.edu/abs/2024ApJ...962...24S} {962, 24}

\bibitem[\protect\citeauthoryear{{Santos}}{{Santos}}{2004}]{Santos2004}
{Santos} M.~R.,  2004, \mn@doi [\mnras] {10.1111/j.1365-2966.2004.07594.x},
  \href {https://ui.adsabs.harvard.edu/abs/2004MNRAS.349.1137S} {349, 1137}

\bibitem[\protect\citeauthoryear{{Santos}, {Sobral}  \& {Matthee}}{{Santos}
  et~al.}{2016}]{Santos2016}
{Santos} S.,  {Sobral} D.,   {Matthee} J.,  2016, \mn@doi [\mnras]
  {10.1093/mnras/stw2076}, \href
  {https://ui.adsabs.harvard.edu/abs/2016MNRAS.463.1678S} {463, 1678}

\bibitem[\protect\citeauthoryear{{Saxena} et~al.,}{{Saxena}
  et~al.}{2023}]{Saxena2023}
{Saxena} A.,  et~al., 2023, \mn@doi [\aap] {10.1051/0004-6361/202346245}, \href
  {https://ui.adsabs.harvard.edu/abs/2023A&A...678A..68S} {678, A68}

\bibitem[\protect\citeauthoryear{{Saxena} et~al.,}{{Saxena}
  et~al.}{2024}]{Saxena2024}
{Saxena} A.,  et~al., 2024, \mn@doi [\aap] {10.1051/0004-6361/202347132}, \href
  {https://ui.adsabs.harvard.edu/abs/2024A&A...684A..84S} {684, A84}

\bibitem[\protect\citeauthoryear{{Schenker}, {Ellis}, {Konidaris}  \&
  {Stark}}{{Schenker} et~al.}{2014}]{Schenker2014}
{Schenker} M.~A.,  {Ellis} R.~S.,  {Konidaris} N.~P.,   {Stark} D.~P.,  2014,
  \mn@doi [\apj] {10.1088/0004-637X/795/1/20}, \href
  {https://ui.adsabs.harvard.edu/abs/2014ApJ...795...20S} {795, 20}

\bibitem[\protect\citeauthoryear{{Shapley}, {Steidel}, {Pettini}  \&
  {Adelberger}}{{Shapley} et~al.}{2003}]{Shapley2003}
{Shapley} A.~E.,  {Steidel} C.~C.,  {Pettini} M.,   {Adelberger} K.~L.,  2003,
  \mn@doi [\apj] {10.1086/373922}, \href
  {https://ui.adsabs.harvard.edu/abs/2003ApJ...588...65S} {588, 65}

\bibitem[\protect\citeauthoryear{{Shapley}, {Reddy}, {Sanders}, {Topping}  \&
  {Brammer}}{{Shapley} et~al.}{2023}]{Shapley2023}
{Shapley} A.~E.,  {Reddy} N.~A.,  {Sanders} R.~L.,  {Topping} M.~W.,
  {Brammer} G.~B.,  2023, \mn@doi [\apjl] {10.3847/2041-8213/acd939}, \href
  {https://ui.adsabs.harvard.edu/abs/2023ApJ...950L...1S} {950, L1}

\bibitem[\protect\citeauthoryear{{Shibuya} et~al.,}{{Shibuya}
  et~al.}{2014}]{Shibuya2014}
{Shibuya} T.,  et~al., 2014, \mn@doi [\apj] {10.1088/0004-637X/788/1/74}, \href
  {https://ui.adsabs.harvard.edu/abs/2014ApJ...788...74S} {788, 74}

\bibitem[\protect\citeauthoryear{{Simmonds} et~al.,}{{Simmonds}
  et~al.}{2023}]{Simmonds2023}
{Simmonds} C.,  et~al., 2023, \mn@doi [\mnras] {10.1093/mnras/stad1749}, \href
  {https://ui.adsabs.harvard.edu/abs/2023MNRAS.523.5468S} {523, 5468}

\bibitem[\protect\citeauthoryear{{Simmonds} et~al.,}{{Simmonds}
  et~al.}{2024}]{Simmonds2024}
{Simmonds} C.,  et~al., 2024, \mn@doi [\mnras] {10.1093/mnras/stad3605}, \href
  {https://ui.adsabs.harvard.edu/abs/2024MNRAS.527.6139S} {527, 6139}

\bibitem[\protect\citeauthoryear{{Songaila}, {Hu}, {Barger}, {Cowie},
  {Hasinger}, {Rosenwasser}  \& {Waters}}{{Songaila}
  et~al.}{2018}]{Songaila2018}
{Songaila} A.,  {Hu} E.~M.,  {Barger} A.~J.,  {Cowie} L.~L.,  {Hasinger} G.,
  {Rosenwasser} B.,   {Waters} C.,  2018, \mn@doi [\apj]
  {10.3847/1538-4357/aac021}, \href
  {https://ui.adsabs.harvard.edu/abs/2018ApJ...859...91S} {859, 91}

\bibitem[\protect\citeauthoryear{{Stanway}, {Bremer}  \& {Lehnert}}{{Stanway}
  et~al.}{2008}]{Stanway2008}
{Stanway} E.~R.,  {Bremer} M.~N.,   {Lehnert} M.~D.,  2008, \mn@doi [\mnras]
  {10.1111/j.1365-2966.2008.12853.x}, \href
  {https://ui.adsabs.harvard.edu/abs/2008MNRAS.385..493S} {385, 493}

\bibitem[\protect\citeauthoryear{{Stark}}{{Stark}}{2016}]{Stark2016}
{Stark} D.~P.,  2016, \mn@doi [\araa] {10.1146/annurev-astro-081915-023417},
  \href {https://ui.adsabs.harvard.edu/abs/2016ARA&A..54..761S} {54, 761}

\bibitem[\protect\citeauthoryear{{Stark}, {Ellis}, {Bunker}, {Bundy},
  {Targett}, {Benson}  \& {Lacy}}{{Stark} et~al.}{2009}]{Stark2009}
{Stark} D.~P.,  {Ellis} R.~S.,  {Bunker} A.,  {Bundy} K.,  {Targett} T.,
  {Benson} A.,   {Lacy} M.,  2009, \mn@doi [\apj]
  {10.1088/0004-637X/697/2/1493}, \href
  {https://ui.adsabs.harvard.edu/abs/2009ApJ...697.1493S} {697, 1493}

\bibitem[\protect\citeauthoryear{{Stark}, {Ellis}, {Chiu}, {Ouchi}  \&
  {Bunker}}{{Stark} et~al.}{2010}]{Stark2010}
{Stark} D.~P.,  {Ellis} R.~S.,  {Chiu} K.,  {Ouchi} M.,   {Bunker} A.,  2010,
  \mn@doi [\mnras] {10.1111/j.1365-2966.2010.17227.x}, \href
  {https://ui.adsabs.harvard.edu/abs/2010MNRAS.408.1628S} {408, 1628}

\bibitem[\protect\citeauthoryear{{Stark}, {Ellis}  \& {Ouchi}}{{Stark}
  et~al.}{2011}]{Stark2011}
{Stark} D.~P.,  {Ellis} R.~S.,   {Ouchi} M.,  2011, \mn@doi [\apjl]
  {10.1088/2041-8205/728/1/L2}, \href
  {https://ui.adsabs.harvard.edu/abs/2011ApJ...728L...2S} {728, L2}

\bibitem[\protect\citeauthoryear{{Stark} et~al.,}{{Stark}
  et~al.}{2015}]{Stark2015}
{Stark} D.~P.,  et~al., 2015, \mn@doi [\mnras] {10.1093/mnras/stv688}, \href
  {https://ui.adsabs.harvard.edu/abs/2015MNRAS.450.1846S} {450, 1846}

\bibitem[\protect\citeauthoryear{{Stark} et~al.,}{{Stark}
  et~al.}{2017}]{Stark2017}
{Stark} D.~P.,  et~al., 2017, \mn@doi [\mnras] {10.1093/mnras/stw2233}, \href
  {https://ui.adsabs.harvard.edu/abs/2017MNRAS.464..469S} {464, 469}

\bibitem[\protect\citeauthoryear{{Steidel}, {Strom}, {Pettini}, {Rudie},
  {Reddy}  \& {Trainor}}{{Steidel} et~al.}{2016}]{Steidel2016}
{Steidel} C.~C.,  {Strom} A.~L.,  {Pettini} M.,  {Rudie} G.~C.,  {Reddy} N.~A.,
    {Trainor} R.~F.,  2016, \mn@doi [\apj] {10.3847/0004-637X/826/2/159}, \href
  {https://ui.adsabs.harvard.edu/abs/2016ApJ...826..159S} {826, 159}

\bibitem[\protect\citeauthoryear{{Steidel}, {Bogosavljevi{\'c}}, {Shapley},
  {Reddy}, {Rudie}, {Pettini}, {Trainor}  \& {Strom}}{{Steidel}
  et~al.}{2018}]{Steidel2018}
{Steidel} C.~C.,  {Bogosavljevi{\'c}} M.,  {Shapley} A.~E.,  {Reddy} N.~A.,
  {Rudie} G.~C.,  {Pettini} M.,  {Trainor} R.~F.,   {Strom} A.~L.,  2018,
  \mn@doi [\apj] {10.3847/1538-4357/aaed28}, \href
  {https://ui.adsabs.harvard.edu/abs/2018ApJ...869..123S} {869, 123}

\bibitem[\protect\citeauthoryear{{Tacchella} et~al.,}{{Tacchella}
  et~al.}{2022}]{Tacchella2022}
{Tacchella} S.,  et~al., 2022, \mn@doi [\apj] {10.3847/1538-4357/ac4cad}, \href
  {https://ui.adsabs.harvard.edu/abs/2022ApJ...927..170T} {927, 170}

\bibitem[\protect\citeauthoryear{{Tacchella} et~al.,}{{Tacchella}
  et~al.}{2023a}]{Tacchella2023b}
{Tacchella} S.,  et~al., 2023a, \mn@doi [\mnras] {10.1093/mnras/stad1408},
  \href {https://ui.adsabs.harvard.edu/abs/2023MNRAS.522.6236T} {522, 6236}

\bibitem[\protect\citeauthoryear{{Tacchella} et~al.,}{{Tacchella}
  et~al.}{2023b}]{Tacchella2023a}
{Tacchella} S.,  et~al., 2023b, \mn@doi [\apj] {10.3847/1538-4357/acdbc6},
  \href {https://ui.adsabs.harvard.edu/abs/2023ApJ...952...74T} {952, 74}

\bibitem[\protect\citeauthoryear{{Tang}, {Stark}, {Chevallard}  \&
  {Charlot}}{{Tang} et~al.}{2019}]{Tang2019}
{Tang} M.,  {Stark} D.~P.,  {Chevallard} J.,   {Charlot} S.,  2019, \mn@doi
  [\mnras] {10.1093/mnras/stz2236}, \href
  {https://ui.adsabs.harvard.edu/abs/2019MNRAS.489.2572T} {489, 2572}

\bibitem[\protect\citeauthoryear{{Tang}, {Stark}, {Chevallard}, {Charlot},
  {Endsley}  \& {Congiu}}{{Tang} et~al.}{2021}]{Tang2021}
{Tang} M.,  {Stark} D.~P.,  {Chevallard} J.,  {Charlot} S.,  {Endsley} R.,
  {Congiu} E.,  2021, \mn@doi [\mnras] {10.1093/mnras/stab705}, \href
  {https://ui.adsabs.harvard.edu/abs/2021MNRAS.503.4105T} {503, 4105}

\bibitem[\protect\citeauthoryear{{Tang}, {Stark}  \& {Ellis}}{{Tang}
  et~al.}{2022}]{Tang2022}
{Tang} M.,  {Stark} D.~P.,   {Ellis} R.~S.,  2022, \mn@doi [\mnras]
  {10.1093/mnras/stac1280}, \href
  {https://ui.adsabs.harvard.edu/abs/2022MNRAS.513.5211T} {513, 5211}

\bibitem[\protect\citeauthoryear{{Tang} et~al.,}{{Tang}
  et~al.}{2023}]{Tang2023}
{Tang} M.,  et~al., 2023, \mn@doi [\mnras] {10.1093/mnras/stad2763}, \href
  {https://ui.adsabs.harvard.edu/abs/2023MNRAS.526.1657T} {526, 1657}

\bibitem[\protect\citeauthoryear{{Tang}, {Stark}, {Ellis}, {Topping}, {Mason},
  {Li}  \& {Plat}}{{Tang} et~al.}{2024}]{Tang2024}
{Tang} M.,  {Stark} D.~P.,  {Ellis} R.~S.,  {Topping} M.~W.,  {Mason} C.,  {Li}
  Z.,   {Plat} A.,  2024, \mn@doi [arXiv e-prints] {10.48550/arXiv.2404.06569},
  \href {https://ui.adsabs.harvard.edu/abs/2024arXiv240406569T} {p.
  arXiv:2404.06569}

\bibitem[\protect\citeauthoryear{{Tilvi} et~al.,}{{Tilvi}
  et~al.}{2020}]{Tilvi2020}
{Tilvi} V.,  et~al., 2020, \mn@doi [\apjl] {10.3847/2041-8213/ab75ec}, \href
  {https://ui.adsabs.harvard.edu/abs/2020ApJ...891L..10T} {891, L10}

\bibitem[\protect\citeauthoryear{{Topping}, {Stark}, {Endsley}, {Plat},
  {Whitler}, {Chen}  \& {Charlot}}{{Topping} et~al.}{2022}]{Topping2022}
{Topping} M.~W.,  {Stark} D.~P.,  {Endsley} R.,  {Plat} A.,  {Whitler} L.,
  {Chen} Z.,   {Charlot} S.,  2022, \mn@doi [\apj] {10.3847/1538-4357/aca522},
  \href {https://ui.adsabs.harvard.edu/abs/2022ApJ...941..153T} {941, 153}

\bibitem[\protect\citeauthoryear{{Topping} et~al.,}{{Topping}
  et~al.}{2024}]{Topping2024}
{Topping} M.~W.,  et~al., 2024, \mn@doi [\mnras] {10.1093/mnras/stae800}, \href
  {https://ui.adsabs.harvard.edu/abs/2024MNRAS.529.4087T} {529, 4087}

\bibitem[\protect\citeauthoryear{{Torralba-Torregrosa}
  et~al.,}{{Torralba-Torregrosa} et~al.}{2024}]{Torralba-Torregrosa2024}
{Torralba-Torregrosa} A.,  et~al., 2024, \mn@doi [arXiv e-prints]
  {10.48550/arXiv.2404.10040}, \href
  {https://ui.adsabs.harvard.edu/abs/2024arXiv240410040T} {p. arXiv:2404.10040}

\bibitem[\protect\citeauthoryear{{Trebitsch}, {Verhamme}, {Blaizot}  \&
  {Rosdahl}}{{Trebitsch} et~al.}{2016}]{Trebitsch2016}
{Trebitsch} M.,  {Verhamme} A.,  {Blaizot} J.,   {Rosdahl} J.,  2016, \mn@doi
  [\aap] {10.1051/0004-6361/201527024}, \href
  {https://ui.adsabs.harvard.edu/abs/2016A&A...593A.122T} {593, A122}

\bibitem[\protect\citeauthoryear{{Treu}, {Schmidt}, {Trenti}, {Bradley}  \&
  {Stiavelli}}{{Treu} et~al.}{2013}]{Treu2013}
{Treu} T.,  {Schmidt} K.~B.,  {Trenti} M.,  {Bradley} L.~D.,   {Stiavelli} M.,
  2013, \mn@doi [\apjl] {10.1088/2041-8205/775/1/L29}, \href
  {https://ui.adsabs.harvard.edu/abs/2013ApJ...775L..29T} {775, L29}

\bibitem[\protect\citeauthoryear{{Umeda}, {Ouchi}, {Nakajima}, {Harikane},
  {Ono}, {Xu}, {Isobe}  \& {Zhang}}{{Umeda} et~al.}{2023}]{Umeda2023}
{Umeda} H.,  {Ouchi} M.,  {Nakajima} K.,  {Harikane} Y.,  {Ono} Y.,  {Xu} Y.,
  {Isobe} Y.,   {Zhang} Y.,  2023, \mn@doi [arXiv e-prints]
  {10.48550/arXiv.2306.00487}, \href
  {https://ui.adsabs.harvard.edu/abs/2023arXiv230600487U} {p. arXiv:2306.00487}

\bibitem[\protect\citeauthoryear{{Urrutia} et~al.,}{{Urrutia}
  et~al.}{2019}]{Urrutia2019}
{Urrutia} T.,  et~al., 2019, \mn@doi [\aap] {10.1051/0004-6361/201834656},
  \href {https://ui.adsabs.harvard.edu/abs/2019A&A...624A.141U} {624, A141}

\bibitem[\protect\citeauthoryear{{Vanzella} et~al.,}{{Vanzella}
  et~al.}{2009}]{Vanzella2009}
{Vanzella} E.,  et~al., 2009, \mn@doi [\apj] {10.1088/0004-637X/695/2/1163},
  \href {https://ui.adsabs.harvard.edu/abs/2009ApJ...695.1163V} {695, 1163}

\bibitem[\protect\citeauthoryear{{Vanzella} et~al.,}{{Vanzella}
  et~al.}{2018}]{Vanzella2018}
{Vanzella} E.,  et~al., 2018, \mn@doi [\mnras] {10.1093/mnrasl/sly023}, \href
  {https://ui.adsabs.harvard.edu/abs/2018MNRAS.476L..15V} {476, L15}

\bibitem[\protect\citeauthoryear{{Verhamme}, {Schaerer}  \&
  {Maselli}}{{Verhamme} et~al.}{2006}]{Verhamme2006}
{Verhamme} A.,  {Schaerer} D.,   {Maselli} A.,  2006, \mn@doi [\aap]
  {10.1051/0004-6361:20065554}, \href
  {https://ui.adsabs.harvard.edu/abs/2006A&A...460..397V} {460, 397}

\bibitem[\protect\citeauthoryear{{Verhamme}, {Schaerer}, {Atek}  \&
  {Tapken}}{{Verhamme} et~al.}{2008}]{Verhamme2008}
{Verhamme} A.,  {Schaerer} D.,  {Atek} H.,   {Tapken} C.,  2008, \mn@doi [\aap]
  {10.1051/0004-6361:200809648}, \href
  {https://ui.adsabs.harvard.edu/abs/2008A&A...491...89V} {491, 89}

\bibitem[\protect\citeauthoryear{{Verhamme}, {Orlitov{\'a}}, {Schaerer}  \&
  {Hayes}}{{Verhamme} et~al.}{2015}]{Verhamme2015}
{Verhamme} A.,  {Orlitov{\'a}} I.,  {Schaerer} D.,   {Hayes} M.,  2015, \mn@doi
  [\aap] {10.1051/0004-6361/201423978}, \href
  {https://ui.adsabs.harvard.edu/abs/2015A&A...578A...7V} {578, A7}

\bibitem[\protect\citeauthoryear{{Virtanen} et~al.,}{{Virtanen}
  et~al.}{2020}]{Virtanen2020}
{Virtanen} P.,  et~al., 2020, \mn@doi [Nature Methods]
  {10.1038/s41592-019-0686-2}, \href
  {https://ui.adsabs.harvard.edu/abs/2020NatMe..17..261V} {17, 261}

\bibitem[\protect\citeauthoryear{{Wang} et~al.,}{{Wang}
  et~al.}{2020}]{Wang2020}
{Wang} F.,  et~al., 2020, \mn@doi [\apj] {10.3847/1538-4357/ab8c45}, \href
  {https://ui.adsabs.harvard.edu/abs/2020ApJ...896...23W} {896, 23}

\bibitem[\protect\citeauthoryear{{Whitaker} et~al.,}{{Whitaker}
  et~al.}{2019}]{Whitaker2019}
{Whitaker} K.~E.,  et~al., 2019, \mn@doi [\apjs] {10.3847/1538-4365/ab3853},
  \href {https://ui.adsabs.harvard.edu/abs/2019ApJS..244...16W} {244, 16}

\bibitem[\protect\citeauthoryear{{Whitler}, {Mason}, {Ren}, {Dijkstra},
  {Mesinger}, {Pentericci}, {Trenti}  \& {Treu}}{{Whitler}
  et~al.}{2020}]{Whitler2020}
{Whitler} L.~R.,  {Mason} C.~A.,  {Ren} K.,  {Dijkstra} M.,  {Mesinger} A.,
  {Pentericci} L.,  {Trenti} M.,   {Treu} T.,  2020, \mn@doi [\mnras]
  {10.1093/mnras/staa1178}, \href
  {https://ui.adsabs.harvard.edu/abs/2020MNRAS.495.3602W} {495, 3602}

\bibitem[\protect\citeauthoryear{{Whitler}, {Stark}, {Endsley}, {Leja},
  {Charlot}  \& {Chevallard}}{{Whitler} et~al.}{2023}]{Whitler2023}
{Whitler} L.,  {Stark} D.~P.,  {Endsley} R.,  {Leja} J.,  {Charlot} S.,
  {Chevallard} J.,  2023, \mn@doi [\mnras] {10.1093/mnras/stad004}, \href
  {https://ui.adsabs.harvard.edu/abs/2023MNRAS.tmp...35W} {}

\bibitem[\protect\citeauthoryear{{Whitler}, {Stark}, {Endsley}, {Chen},
  {Mason}, {Topping}  \& {Charlot}}{{Whitler} et~al.}{2024}]{Whitler2024}
{Whitler} L.,  {Stark} D.~P.,  {Endsley} R.,  {Chen} Z.,  {Mason} C.,
  {Topping} M.~W.,   {Charlot} S.,  2024, \mn@doi [\mnras]
  {10.1093/mnras/stae516}, \href
  {https://ui.adsabs.harvard.edu/abs/2024MNRAS.529..855W} {529, 855}

\bibitem[\protect\citeauthoryear{{Willott}, {Carilli}, {Wagg}  \&
  {Wang}}{{Willott} et~al.}{2015}]{Willott2015}
{Willott} C.~J.,  {Carilli} C.~L.,  {Wagg} J.,   {Wang} R.,  2015, \mn@doi
  [\apj] {10.1088/0004-637X/807/2/180}, \href
  {https://ui.adsabs.harvard.edu/abs/2015ApJ...807..180W} {807, 180}

\bibitem[\protect\citeauthoryear{{Wisotzki} et~al.,}{{Wisotzki}
  et~al.}{2016}]{Wisotzki2016}
{Wisotzki} L.,  et~al., 2016, \mn@doi [\aap] {10.1051/0004-6361/201527384},
  \href {https://ui.adsabs.harvard.edu/abs/2016A&A...587A..98W} {587, A98}

\bibitem[\protect\citeauthoryear{{Witstok} et~al.,}{{Witstok}
  et~al.}{2024a}]{Witstok2024b}
{Witstok} J.,  et~al., 2024a, \mn@doi [arXiv e-prints]
  {10.48550/arXiv.2404.05724}, \href
  {https://ui.adsabs.harvard.edu/abs/2024arXiv240405724W} {p. arXiv:2404.05724}

\bibitem[\protect\citeauthoryear{{Witstok} et~al.,}{{Witstok}
  et~al.}{2024b}]{Witstok2024a}
{Witstok} J.,  et~al., 2024b, \mn@doi [\aap] {10.1051/0004-6361/202347176},
  \href {https://ui.adsabs.harvard.edu/abs/2024A&A...682A..40W} {682, A40}

\bibitem[\protect\citeauthoryear{{Worseck} et~al.,}{{Worseck}
  et~al.}{2014}]{Worseck2014}
{Worseck} G.,  et~al., 2014, \mn@doi [\mnras] {10.1093/mnras/stu1827}, \href
  {https://ui.adsabs.harvard.edu/abs/2014MNRAS.445.1745W} {445, 1745}

\bibitem[\protect\citeauthoryear{{Wu}, {Jiang}  \& {Ning}}{{Wu}
  et~al.}{2020}]{Wu2020}
{Wu} J.,  {Jiang} L.,   {Ning} Y.,  2020, \mn@doi [\apj]
  {10.3847/1538-4357/ab7333}, \href
  {https://ui.adsabs.harvard.edu/abs/2020ApJ...891..105W} {891, 105}

\bibitem[\protect\citeauthoryear{{Wyithe} \& {Loeb}}{{Wyithe} \&
  {Loeb}}{2005}]{Wyithe2005}
{Wyithe} J. S.~B.,  {Loeb} A.,  2005, \mn@doi [\apj] {10.1086/429529}, \href
  {https://ui.adsabs.harvard.edu/abs/2005ApJ...625....1W} {625, 1}

\bibitem[\protect\citeauthoryear{{Xu} et~al.,}{{Xu} et~al.}{2022}]{Xu2022}
{Xu} X.,  et~al., 2022, \mn@doi [\apj] {10.3847/1538-4357/ac7225}, \href
  {https://ui.adsabs.harvard.edu/abs/2022ApJ...933..202X} {933, 202}

\bibitem[\protect\citeauthoryear{{Xu} et~al.,}{{Xu} et~al.}{2023}]{Xu2023}
{Xu} X.,  et~al., 2023, \mn@doi [\apj] {10.3847/1538-4357/aca89a}, \href
  {https://ui.adsabs.harvard.edu/abs/2023ApJ...943...94X} {943, 94}

\bibitem[\protect\citeauthoryear{{Yang} et~al.,}{{Yang}
  et~al.}{2017}]{Yang2017}
{Yang} H.,  et~al., 2017, \mn@doi [\apj] {10.3847/1538-4357/aa7d4d}, \href
  {https://ui.adsabs.harvard.edu/abs/2017ApJ...844..171Y} {844, 171}

\bibitem[\protect\citeauthoryear{{Yang} et~al.,}{{Yang}
  et~al.}{2020a}]{Yang2020a}
{Yang} J.,  et~al., 2020a, \mn@doi [\apjl] {10.3847/2041-8213/ab9c26}, \href
  {https://ui.adsabs.harvard.edu/abs/2020ApJ...897L..14Y} {897, L14}

\bibitem[\protect\citeauthoryear{{Yang} et~al.,}{{Yang}
  et~al.}{2020b}]{Yang2020b}
{Yang} J.,  et~al., 2020b, \mn@doi [\apj] {10.3847/1538-4357/abbc1b}, \href
  {https://ui.adsabs.harvard.edu/abs/2020ApJ...904...26Y} {904, 26}

\bibitem[\protect\citeauthoryear{{Zhang} et~al.,}{{Zhang}
  et~al.}{2024}]{Zhang2024}
{Zhang} H.,  et~al., 2024, \mn@doi [\apj] {10.3847/1538-4357/ad07d3}, \href
  {https://ui.adsabs.harvard.edu/abs/2024ApJ...961...63Z} {961, 63}

\bibitem[\protect\citeauthoryear{{Zheng} et~al.,}{{Zheng}
  et~al.}{2017}]{Zheng2017}
{Zheng} Z.-Y.,  et~al., 2017, \mn@doi [\apjl] {10.3847/2041-8213/aa794f}, \href
  {https://ui.adsabs.harvard.edu/abs/2017ApJ...842L..22Z} {842, L22}

\bibitem[\protect\citeauthoryear{{Zhu} et~al.,}{{Zhu} et~al.}{2023}]{Zhu2023}
{Zhu} Y.,  et~al., 2023, \mn@doi [\apj] {10.3847/1538-4357/aceef4}, \href
  {https://ui.adsabs.harvard.edu/abs/2023ApJ...955..115Z} {955, 115}

\bibitem[\protect\citeauthoryear{{Zitrin} et~al.,}{{Zitrin}
  et~al.}{2015}]{Zitrin2015}
{Zitrin} A.,  et~al., 2015, \mn@doi [\apjl] {10.1088/2041-8205/810/1/L12},
  \href {https://ui.adsabs.harvard.edu/abs/2015ApJ...810L..12Z} {810, L12}

\bibitem[\protect\citeauthoryear{{de Graaff} et~al.,}{{de Graaff}
  et~al.}{2024}]{deGraaff2024}
{de Graaff} A.,  et~al., 2024, \mn@doi [\aap] {10.1051/0004-6361/202347755},
  \href {https://ui.adsabs.harvard.edu/abs/2024A&A...684A..87D} {684, A87}

\bibitem[\protect\citeauthoryear{{de La Vieuville} et~al.,}{{de La Vieuville}
  et~al.}{2020}]{deLaVieuville2020}
{de La Vieuville} G.,  et~al., 2020, \mn@doi [\aap]
  {10.1051/0004-6361/202037651}, \href
  {https://ui.adsabs.harvard.edu/abs/2020A&A...644A..39D} {644, A39}

\makeatother
\end{thebibliography}




\appendix

\section{Tables of Galaxies with L\lowercase{y}$\alpha$ and H$\alpha$ Detections at $\lowercase{z}\simeq5-6$}


\begin{table*}
\hspace*{-0.9cm}
\begin{tabular}{cccccccccc}
\hline
ID & R.A. & Decl. & M$_{\rm UV}$ & $z_{{\rm Ly}\alpha}$ & $F_{{\rm Ly}\alpha}$ & EW$_{{\rm Ly}\alpha}$ & $\Delta v_{{\rm Ly}\alpha}$ & $f^{\rm case\ B}_{{\rm esc,Ly}\alpha}$ & $f^{\rm case\ A}_{{\rm esc,Ly}\alpha}$ \\
 & (deg) & (deg) & (mag) & & ($10^{-20}$~erg~s$^{-1}$~cm$^{-2}$) & (\AA) & (km~s$^{-1}$) & & \\
\hline
MUSE-102049176 & $53.074469$ & $-27.820237$ & $-18.05\pm0.10$ & $5.1024$ & $2126\pm262$ & $395\pm60$ & $179\pm110$ & $0.760\pm0.393$ & $0.580\pm0.300$ \\
MUSE-107041159 & $53.087996$ & $-27.813147$ & $-19.25\pm0.04$ & $5.3161$ & $4896\pm269$ & $333\pm22$ & $73\pm19$ & $0.847\pm0.172$ & $0.647\pm0.131$ \\
MUSE-116039142 & $53.131702$ & $-27.847032$ & $-19.52\pm0.06$ & $5.3253$ & $2849\pm206$ & $151\pm13$ & $85\pm30$ & $0.829\pm0.211$ & $0.633\pm0.161$ \\
MUSE-117039091 & $53.156207$ & $-27.836079$ & $-18.68\pm0.09$ & $5.7675$ & $1790\pm178$ & $248\pm32$ & $115\pm35$ & $0.678\pm0.168$ & $0.518\pm0.128$ \\
MUSE-118034094 & $53.175266$ & $-27.841127$ & $-19.63\pm0.08$ & $5.2964$ & $4217\pm308$ & $199\pm20$ & $86\pm17$ & $0.783\pm0.172$ & $0.598\pm0.131$ \\
MUSE-119036075 & $53.188854$ & $-27.833476$ & $-18.39\pm0.27$ & $5.3345$ & $1026\pm139$ & $154\pm43$ & $251\pm17$ & $0.413\pm0.120$ & $0.315\pm0.092$ \\
MUSE-119039078 & $53.177428$ & $-27.830966$ & $-18.21\pm0.07$ & $5.6132$ & $1176\pm123$ & $237\pm29$ & $77\pm58$ & $0.494\pm0.185$ & $0.377\pm0.141$ \\
MUSE-119040079 & $53.191824$ & $-27.824896$ & $-19.88\pm0.04$ & $5.7880$ & $2161\pm244$ & $100\pm11$ & $261\pm31$ & $0.383\pm0.082$ & $0.292\pm0.063$ \\
MUSE-122002035 & $53.119796$ & $-27.831269$ & $-17.96\pm0.32$ & $5.3180$ & $1354\pm141$ & $302\pm94$ & $163\pm53$ & $1.470\pm0.997$ & $1.122\pm0.761$ \\
MUSE-123027133 & $53.144620$ & $-27.831002$ & $-18.34\pm0.08$ & $5.8024$ & $1548\pm141$ & $298\pm35$ & $79\pm37$ & $0.766\pm0.345$ & $0.584\pm0.263$ \\
MUSE-124033068 & $53.165978$ & $-27.823859$ & $-18.73\pm0.06$ & $5.0475$ & $2480\pm224$ & $240\pm25$ & $119\pm28$ & $0.778\pm0.234$ & $0.594\pm0.178$ \\
MUSE-125051124 & $53.188380$ & $-27.819461$ & $-20.11\pm0.02$ & $5.5815$ & $2712\pm628$ & $93\pm21$ & $310\pm52$ & $0.309\pm0.092$ & $0.236\pm0.070$ \\
MUSE-125052125 & $53.173472$ & $-27.825016$ & $-17.94\pm0.06$ & $5.6134$ & $900\pm240$ & $232\pm63$ & $140\pm49$ & $0.598\pm0.353$ & $0.457\pm0.270$ \\
MUSE-128044246 & $53.121016$ & $-27.815916$ & $-18.33\pm0.09$ & $4.9347$ & $2444\pm428$ & $326\pm63$ & $60\pm70$ & $1.086\pm0.664$ & $0.828\pm0.507$ \\
MUSE-128045247 & $53.119796$ & $-27.821988$ & $-19.32\pm0.06$ & $4.9666$ & $3064\pm500$ & $166\pm28$ & $193\pm51$ & $1.387\pm0.448$ & $1.059\pm0.342$ \\
MUSE-130033059 & $53.102725$ & $-27.792042$ & $-19.02\pm0.15$ & $5.1138$ & $1494\pm348$ & $113\pm30$ & $326\pm17$ & $0.253\pm0.072$ & $0.193\pm0.055$ \\
MUSE-131016106 & $53.104145$ & $-27.783718$ & $-19.28\pm0.09$ & $6.0641$ & $2196\pm428$ & $197\pm41$ & $273\pm41$ & $0.998\pm0.489$ & $0.762\pm0.373$ \\
MUSE-134036056 & $53.131273$ & $-27.806806$ & $-19.72\pm0.03$ & $5.1251$ & $4024\pm489$ & $162\pm20$ & $113\pm26$ & $0.935\pm0.219$ & $0.713\pm0.167$ \\
MUSE-134037057 & $53.132663$ & $-27.798961$ & $-18.40\pm0.11$ & $5.1302$ & $3906\pm407$ & $533\pm77$ & $201\pm26$ & $0.599\pm0.123$ & $0.457\pm0.094$ \\
MUSE-135049239 & $53.124788$ & $-27.784118$ & $-19.05\pm0.08$ & $5.7854$ & $2252\pm261$ & $223\pm30$ & $147\pm25$ & $0.511\pm0.149$ & $0.390\pm0.114$ \\
MUSE-136041192 & $53.115275$ & $-27.772786$ & $-18.49\pm0.16$ & $5.0806$ & $2014\pm291$ & $246\pm50$ & $234\pm24$ & $0.293\pm0.073$ & $0.224\pm0.056$ \\
MUSE-136044198 & $53.123258$ & $-27.771197$ & $-19.17\pm0.11$ & $5.2338$ & $3847\pm429$ & $270\pm40$ & $111\pm30$ & $0.518\pm0.093$ & $0.395\pm0.071$ \\
MUSE-139047301 & $53.142900$ & $-27.758885$ & $-18.14\pm0.06$ & $4.9374$ & $963\pm196$ & $153\pm32$ & $177\pm46$ & $0.486\pm0.298$ & $0.371\pm0.227$ \\
MUSE-140047114 & $53.134580$ & $-27.756489$ & $-20.16\pm0.06$ & $5.4481$ & $3478\pm447$ & $108\pm15$ & $135\pm16$ & $0.539\pm0.099$ & $0.411\pm0.076$ \\
MUSE-68 & $53.171196$ & $-27.778448$ & $-19.86\pm0.04$ & $4.9397$ & $4855\pm26$ & $158\pm5$ & $258\pm96$ & $0.692\pm0.332$ & $0.528\pm0.254$ \\
MUSE-313 & $53.170944$ & $-27.782429$ & $-17.68\pm0.17$ & $5.1382$ & $530\pm6$ & $140\pm22$ & $186\pm34$ & $0.323\pm0.199$ & $0.247\pm0.152$ \\
MUSE-417 & $53.157867$ & $-27.779982$ & $-18.01\pm0.07$ & $5.1321$ & $224\pm10$ & $43\pm3$ & $156\pm63$ & $0.118\pm0.056$ & $0.090\pm0.042$ \\
MUSE-547 & $53.160609$ & $-27.771537$ & $-18.42\pm0.09$ & $5.9775$ & $581\pm30$ & $111\pm10$ & $236\pm77$ & $0.289\pm0.085$ & $0.221\pm0.065$ \\
MUSE-1478 & $53.153168$ & $-27.766165$ & $-20.14\pm0.04$ & $4.9304$ & $932\pm48$ & $23\pm1$ & $263\pm66$ & $0.096\pm0.035$ & $0.073\pm0.027$ \\
MUSE-1670 & $53.166702$ & $-27.804155$ & $-21.30\pm0.01$ & $5.8325$ & $2348\pm112$ & $29\pm1$ & $470\pm15$ & $0.079\pm0.007$ & $0.061\pm0.005$ \\
MUSE-2069 & $53.151096$ & $-27.782923$ & $-19.39\pm0.03$ & $5.2678$ & $536\pm56$ & $31\pm3$ & $249\pm21$ & $0.165\pm0.047$ & $0.126\pm0.036$ \\
MUSE-2071 & $53.146133$ & $-27.777782$ & $-18.84\pm0.05$ & $4.9304$ & $368\pm32$ & $30\pm3$ & $263\pm203$ & $0.102\pm0.209$ & $0.078\pm0.159$ \\
MUSE-2168 & $53.135960$ & $-27.798378$ & $-18.95\pm0.04$ & $5.7811$ & $1473\pm49$ & $159\pm7$ & $186\pm26$ & $0.263\pm0.066$ & $0.201\pm0.050$ \\
MUSE-2296 & $53.168419$ & $-27.804079$ & $-18.64\pm0.08$ & $4.9500$ & $488\pm55$ & $49\pm6$ & $227\pm27$ & $0.233\pm0.095$ & $0.178\pm0.072$ \\
MUSE-2302 & $53.180447$ & $-27.770603$ & $-17.73\pm0.08$ & $5.0333$ & $370\pm38$ & $89\pm11$ & $318\pm90$ & $0.152\pm0.082$ & $0.116\pm0.063$ \\
MUSE-2307 & $53.183330$ & $-27.795965$ & $-17.86\pm0.06$ & $4.9489$ & $1276\pm50$ & $263\pm17$ & $161\pm30$ & $0.373\pm0.136$ & $0.285\pm0.104$ \\
MUSE-2350 & $53.156380$ & $-27.809574$ & $-18.16\pm0.05$ & $5.0488$ & $1392\pm49$ & $227\pm13$ & $119\pm20$ & $0.595\pm0.169$ & $0.454\pm0.129$ \\
MUSE-2449 & $53.164741$ & $-27.769613$ & $-18.71\pm0.13$ & $5.2709$ & $363\pm61$ & $39\pm8$ & $62\pm39$ & $0.231\pm0.117$ & $0.176\pm0.089$ \\
MUSE-2481 & $53.149803$ & $-27.810652$ & $-18.76\pm0.05$ & $5.0323$ & $254\pm47$ & $23\pm4$ & $219\pm108$ & $0.111\pm0.067$ & $0.085\pm0.051$ \\
MUSE-2502 & $53.158028$ & $-27.817951$ & $-19.24\pm0.04$ & $5.6411$ & $655\pm39$ & $51\pm3$ & $330\pm20$ & $0.178\pm0.039$ & $0.136\pm0.029$ \\
MUSE-2873 & $53.172489$ & $-27.764322$ & $-17.93\pm0.09$ & $5.0508$ & $487\pm32$ & $98\pm10$ & $188\pm40$ & $0.258\pm0.114$ & $0.197\pm0.087$ \\
MUSE-2964 & $53.149456$ & $-27.809727$ & $-17.85\pm0.06$ & $5.0591$ & $313\pm44$ & $68\pm10$ & $253\pm45$ & $0.416\pm0.306$ & $0.318\pm0.233$ \\
MUSE-3090 & $53.155037$ & $-27.762585$ & $-18.16\pm0.10$ & $5.1362$ & $165\pm27$ & $28\pm5$ & $235\pm46$ & $0.155\pm0.085$ & $0.118\pm0.065$ \\
MUSE-3093 & $53.139675$ & $-27.796425$ & $-17.69\pm0.08$ & $5.0518$ & $339\pm24$ & $85\pm8$ & $154\pm35$ & $0.161\pm0.081$ & $0.123\pm0.062$ \\
MUSE-3203 & $53.176548$ & $-27.771025$ & $-18.79\pm0.05$ & $5.8921$ & $1405\pm60$ & $184\pm11$ & $91\pm43$ & $0.610\pm0.274$ & $0.465\pm0.209$ \\
MUSE-3238 & $53.175571$ & $-27.795835$ & $-18.12\pm0.16$ & $5.6185$ & $721\pm43$ & $158\pm25$ & $136\pm41$ & $0.234\pm0.091$ & $0.179\pm0.069$ \\
MUSE-4405 & $53.170544$ & $-27.812431$ & $-18.77\pm0.08$ & $5.8376$ & $419\pm49$ & $55\pm7$ & $285\pm49$ & $0.133\pm0.060$ & $0.102\pm0.046$ \\
MUSE-6231 & $53.178307$ & $-27.800921$ & $-18.96\pm0.04$ & $6.3313$ & $495\pm94$ & $66\pm12$ & $360\pm49$ & $0.147\pm0.081$ & $0.112\pm0.062$ \\
MUSE-6294 & $53.166091$ & $-27.785667$ & $-19.56\pm0.11$ & $5.4715$ & $697\pm23$ & $38\pm4$ & $130\pm10$ & $0.088\pm0.013$ & $0.067\pm0.010$ \\
MUSE-6462 & $53.164040$ & $-27.799646$ & $-20.23\pm0.04$ & $5.4530$ & $819\pm55$ & $23\pm1$ & $335\pm16$ & $0.061\pm0.009$ & $0.046\pm0.007$ \\
MUSE-7125 & $53.128100$ & $-27.789826$ & $-19.31\pm0.06$ & $5.0313$ & $216\pm38$ & $12\pm2$ & $295\pm62$ & $0.082\pm0.038$ & $0.063\pm0.029$ \\
MUSE-7205 & $53.162998$ & $-27.760258$ & $-17.86\pm0.10$ & $4.9222$ & $286\pm30$ & $58\pm8$ & $147\pm138$ & $0.075\pm0.056$ & $0.057\pm0.043$ \\
MUSE-7225 & $53.161663$ & $-27.763103$ & $-18.81\pm0.03$ & $5.5589$ & $557\pm49$ & $63\pm5$ & $215\pm23$ & $0.192\pm0.054$ & $0.147\pm0.041$ \\
MUSE-7319 & $53.186211$ & $-27.787121$ & $-17.93\pm0.15$ & $5.1444$ & $324\pm54$ & $68\pm14$ & $210\pm60$ & $0.249\pm0.268$ & $0.190\pm0.204$ \\
MUSE-7337 & $53.169041$ & $-27.787611$ & $-17.93\pm0.08$ & $5.4715$ & $193\pm15$ & $47\pm5$ & $222\pm111$ & $0.038\pm0.017$ & $0.029\pm0.013$ \\
MUSE-7605 & $53.165718$ & $-27.784885$ & $-18.09\pm0.08$ & $5.4725$ & $1096\pm24$ & $230\pm17$ & $278\pm83$ & $0.392\pm0.160$ & $0.299\pm0.122$ \\
MUSE-7922 & $53.180718$ & $-27.776484$ & $-17.19\pm0.52$ & $5.3851$ & $235\pm45$ & $109\pm56$ & $216\pm40$ & $0.303\pm0.159$ & $0.231\pm0.121$ \\
\hline
\end{tabular}
\caption{Information and Ly$\alpha$ properties of the $79$ galaxies with both Ly$\alpha$ and H$\alpha$ detections at $4.9<z<6.5$ in our sample. We provide the Ly$\alpha$ fluxes measured from VLT/MUSE and Keck/DEIMOS spectra. For galaxies with DEIMOS observations, we first list the Ly$\alpha$ fluxes converted to MUSE aperture values in order to be consistent with MUSE measurements (Section~\ref{sec:lya_flux}), and then the original Ly$\alpha$ fluxes measured from DEIMOS spectra in parentheses. Converting from MUSE measured Ly$\alpha$ fluxes to {\it JWST}/NIRSpec MSA measurements is to multiply MUSE fluxes by a factor of $\simeq0.8$ (Section~\ref{sec:lya_flux}). The Ly$\alpha$ velocity offset ($\Delta v_{{\rm Ly}\alpha}$) is calculated using the redshift measured from Ly$\alpha$ emission lines ($z_{{\rm Ly}\alpha}$) and the systemic redshift measured from H$\alpha$ emission lines (Section~\ref{sec:lya_measure}, see also Sun et al. in prep.). Note that the Ly$\alpha$ velocity offset measurement is also subject to an uncertainty of NIRCam grism wavelength calibration ($\simeq100$~km~s$^{-1}$). We compute the Ly$\alpha$ escape fraction for both case B ($f^{\rm case\ B}_{{\rm esc,Ly}\alpha}$) and case A recombination ($f^{\rm case\ A}_{{\rm esc,Ly}\alpha}$).}
\label{tab:lya_properties}
\end{table*}

\begin{table*}
\setcounter{table}{0}
\hspace*{-1cm}
\begin{tabular}{cccccccccc}
\hline
ID & R.A. & Decl. & M$_{\rm UV}$ & $z_{{\rm Ly}\alpha}$ & $F_{{\rm Ly}\alpha}$ & EW$_{{\rm Ly}\alpha}$ & $\Delta v_{{\rm Ly}\alpha}$ & $f^{\rm case\ B}_{{\rm esc,Ly}\alpha}$ & $f^{\rm case\ A}_{{\rm esc,Ly}\alpha}$ \\
 & (deg) & (deg) & (mag) & & ($10^{-20}$~erg~s$^{-1}$~cm$^{-2}$) & (\AA) & (km~s$^{-1}$) & & \\
\hline
MUSE-7934 & $53.146606$ & $-27.786125$ & $-19.77\pm0.03$ & $5.5270$ & $348\pm41$ & $16\pm1$ & $133\pm55$ & $0.111\pm0.048$ & $0.085\pm0.037$ \\
MUSE-7984 & $53.179279$ & $-27.773256$ & $-18.88\pm0.03$ & $6.1071$ & $320\pm53$ & $42\pm7$ & $232\pm26$ & $0.090\pm0.031$ & $0.069\pm0.024$ \\
MUSE-8124 & $53.156754$ & $-27.809155$ & $-19.54\pm0.04$ & $5.5887$ & $867\pm52$ & $50\pm3$ & $100\pm45$ & $0.503\pm0.241$ & $0.384\pm0.184$ \\
DEIMOS-43\_7167 & $189.085312$ & $ 62.212612$ & $-19.66\pm0.08$ & $4.9196$ & $2379\pm68$ ($1830$) & $92\pm7$ & $238\pm54$ & $0.232\pm0.045$ & $0.177\pm0.034$ \\
DEIMOS-42\_11827 & $189.143036$ & $ 62.166759$ & $-18.49\pm0.18$ & $4.9265$ & $1674\pm55$ ($1288$) & $190\pm32$ & $250\pm64$ & $0.170\pm0.045$ & $0.130\pm0.034$ \\
DEIMOS-23\_31399 & $189.362015$ & $ 62.249187$ & $-19.60\pm0.10$ & $5.0503$ & $739\pm63$ ($568$) & $32\pm4$ & $429\pm63$ & $0.228\pm0.127$ & $0.174\pm0.097$ \\
DEIMOS-25458 & $189.272995$ & $ 62.267040$ & $-18.43\pm0.23$ & $5.2012$ & $1548\pm89$ ($1191$) & $212\pm46$ & $160\pm52$ & $0.378\pm0.094$ & $0.289\pm0.071$ \\
DEIMOS-25824 & $189.301392$ & $ 62.268780$ & $-20.69\pm0.06$ & $5.2393$ & $623\pm125$ ($479$) & $10\pm2$ & $725\pm56$ & $0.149\pm0.052$ & $0.114\pm0.040$ \\
DEIMOS-11505 & $189.265076$ & $ 62.199612$ & $-18.77\pm0.10$ & $5.2260$ & $2007\pm121$ ($1544$) & $203\pm22$ & $233\pm75$ & $0.364\pm0.226$ & $0.277\pm0.172$ \\
DEIMOS-19842 & $189.363861$ & $ 62.239918$ & $-19.32\pm0.09$ & $5.2611$ & $462\pm151$ ($356$) & $28\pm9$ & $385\pm55$ & $0.135\pm0.058$ & $0.103\pm0.044$ \\
DEIMOS-43\_13063 & $189.156342$ & $ 62.210011$ & $-19.92\pm0.05$ & $5.1895$ & $358\pm97$ ($275$) & $12\pm3$ & $351\pm49$ & $0.052\pm0.015$ & $0.040\pm0.012$ \\
DEIMOS-33\_19970 & $189.230789$ & $ 62.263546$ & $-19.56\pm0.05$ & $5.1950$ & $395\pm92$ ($304$) & $19\pm4$ & $169\pm54$ & $0.055\pm0.015$ & $0.042\pm0.012$ \\
DEIMOS-42\_9127 & $189.110382$ & $ 62.202064$ & $-19.70\pm0.10$ & $5.2022$ & $394\pm58$ ($303$) & $16\pm2$ & $297\pm56$ & $0.040\pm0.009$ & $0.031\pm0.007$ \\
DEIMOS-33\_20014 & $189.231232$ & $ 62.259106$ & $-20.14\pm0.06$ & $5.1937$ & $2923\pm111$ ($2249$) & $82\pm5$ & $242\pm49$ & $0.078\pm0.004$ & $0.060\pm0.003$ \\
DEIMOS-33\_17440 & $189.205124$ & $ 62.260712$ & $-21.05\pm0.03$ & $5.1947$ & $4586\pm77$ ($3528$) & $56\pm1$ & $237\pm49$ & $0.229\pm0.016$ & $0.175\pm0.012$ \\
DEIMOS-vdrop\_225 & $189.183884$ & $ 62.179928$ & $-18.70\pm0.08$ & $5.2378$ & $1371\pm112$ ($1055$) & $148\pm16$ & $286\pm56$ & $0.932\pm0.482$ & $0.712\pm0.368$ \\
DEIMOS-vdrop\_375 & $189.285294$ & $ 62.251411$ & $-19.40\pm0.08$ & $5.2834$ & $1301\pm71$ ($1001$) & $75\pm6$ & $408\pm49$ & $0.253\pm0.045$ & $0.193\pm0.034$ \\
DEIMOS-33\_17034 & $189.200592$ & $ 62.259441$ & $-20.02\pm0.05$ & $5.6068$ & $3081\pm149$ ($2370$) & $116\pm7$ & $232\pm47$ & $0.354\pm0.051$ & $0.270\pm0.039$ \\
DEIMOS-42\_11693 & $189.141540$ & $ 62.190292$ & $-19.84\pm0.30$ & $5.6129$ & $2368\pm77$ ($1822$) & $106\pm29$ & $342\pm46$ & $0.162\pm0.015$ & $0.124\pm0.012$ \\
DEIMOS-vdrop\_167 & $189.142960$ & $ 62.196960$ & $-19.13\pm0.60$ & $5.6149$ & $953\pm86$ ($733$) & $82\pm46$ & $210\pm48$ & $1.097\pm0.488$ & $0.837\pm0.373$ \\
DEIMOS-33\_17705 & $189.208191$ & $ 62.232136$ & $-19.63\pm0.05$ & $5.8061$ & $650\pm41$ ($500$) & $38\pm3$ & $589\pm83$ & $0.092\pm0.033$ & $0.070\pm0.025$ \\
DEIMOS-32\_16773 & $189.197845$ & $ 62.199963$ & $-19.11\pm0.06$ & $5.9752$ & $525\pm38$ ($404$) & $53\pm4$ & $242\pm48$ & $0.103\pm0.032$ & $0.078\pm0.024$ \\
\hline
\end{tabular}
\caption{Continued.}
\label{tab:lya_properties}
\end{table*}


\begin{table*}
\centering
\hspace*{-0.5cm}
\begin{tabular}{ccccccc}
\hline
ID & JADES ID & $\log{(M_{\star}/M_{\odot})}$ & $\log{({\rm age}/{\rm yr})}$ & $A$(H$\alpha$) (mag) & [O~{\scriptsize III}]+H$\beta$ EW (\AA) & $\log{(\xi_{\rm ion}/{\rm erg}^{-1}\ {\rm Hz})}$ \\
\hline
MUSE-102049176 & JADES-GS+53.07450-27.82028 & $7.28^{+0.08}_{-0.07}$ & $6.24^{+0.17}_{-0.16}$ & $0.005^{+0.009}_{-0.003}$ & $5206^{+622}_{-598}$ & $25.91^{+0.10}_{-0.08}$ \\
MUSE-107041159 & JADES-GS+53.08803-27.81320 & $8.26^{+0.41}_{-0.32}$ & $7.76^{+0.51}_{-0.41}$ & $0.019^{+0.032}_{-0.016}$ & $610^{+292}_{-301}$ & $25.36^{+0.09}_{-0.08}$ \\
MUSE-116039142 & JADES-GS+53.13174-27.84712 & $8.79^{+0.13}_{-0.22}$ & $8.42^{+0.18}_{-0.27}$ & $0.003^{+0.007}_{-0.002}$ & $252^{+144}_{-80}$ & $25.45^{+0.07}_{-0.07}$ \\
MUSE-117039091 & JADES-GS+53.15625-27.83617 & $7.44^{+0.08}_{-0.07}$ & $7.05^{+0.09}_{-0.10}$ & $0.003^{+0.008}_{-0.002}$ & $1496^{+100}_{-100}$ & $25.57^{+0.05}_{-0.05}$ \\
MUSE-118034094 & JADES-GS+53.17529-27.84117 & $7.81^{+0.05}_{-0.03}$ & $6.64^{+0.09}_{-0.06}$ & $0.054^{+0.014}_{-0.012}$ & $2096^{+560}_{-434}$ & $25.81^{+0.10}_{-0.22}$ \\
MUSE-119039078 & JADES-GS+53.17747-27.83105 & $7.29^{+0.08}_{-0.09}$ & $7.12^{+0.10}_{-0.10}$ & $0.003^{+0.004}_{-0.001}$ & $1469^{+125}_{-112}$ & $25.54^{+0.05}_{-0.04}$ \\
MUSE-122002035 & JADES-GS+53.11984-27.83136 & $8.39^{+0.27}_{-0.34}$ & $8.22^{+0.38}_{-0.46}$ & $0.007^{+0.021}_{-0.005}$ & $347^{+263}_{-177}$ & $25.29^{+0.12}_{-0.04}$ \\
MUSE-123027133 & JADES-GS+53.14468-27.83103 & $7.13^{+0.26}_{-0.04}$ & $6.37^{+0.85}_{-0.29}$ & $0.002^{+0.002}_{-0.001}$ & $2109^{+176}_{-975}$ & $25.74^{+0.02}_{-0.02}$ \\
MUSE-124033068 & JADES-GS+53.16604-27.82394 & $7.33^{+0.09}_{-0.06}$ & $6.83^{+0.11}_{-0.08}$ & $0.007^{+0.020}_{-0.006}$ & $1267^{+126}_{-140}$ & $25.60^{+0.09}_{-0.07}$ \\
MUSE-125051124 & JADES-GS+53.18845-27.81950 & $8.49^{+0.14}_{-0.12}$ & $7.52^{+0.18}_{-0.17}$ & $0.077^{+0.021}_{-0.043}$ & $796^{+124}_{-111}$ & $25.58^{+0.10}_{-0.10}$ \\
MUSE-125052125 & JADES-GS+53.17350-27.82507 & $7.23^{+0.05}_{-0.05}$ & $6.28^{+0.18}_{-0.19}$ & $0.012^{+0.021}_{-0.009}$ & $1757^{+125}_{-136}$ & $25.75^{+0.01}_{-0.07}$ \\
MUSE-128044246 & JADES-GS+53.12103-27.81599 & $7.34^{+0.08}_{-0.07}$ & $7.09^{+0.10}_{-0.08}$ & $0.004^{+0.011}_{-0.003}$ & $1429^{+118}_{-142}$ & $25.55^{+0.05}_{-0.04}$ \\
MUSE-128045247 & JADES-GS+53.11988-27.82207 & $8.14^{+0.11}_{-0.10}$ & $7.46^{+0.19}_{-0.13}$ & $0.005^{+0.011}_{-0.003}$ & $787^{+174}_{-157}$ & $25.54^{+0.09}_{-0.07}$ \\
MUSE-134036056 & JADES-GS+53.13135-27.80687 & $7.88^{+0.06}_{-0.06}$ & $7.03^{+0.07}_{-0.07}$ & $0.004^{+0.011}_{-0.003}$ & $1496^{+134}_{-144}$ & $25.54^{+0.08}_{-0.06}$ \\
MUSE-134037057 & JADES-GS+53.13271-27.79899 & $7.22^{+0.07}_{-0.07}$ & $6.91^{+0.09}_{-0.09}$ & $0.003^{+0.007}_{-0.002}$ & $1446^{+126}_{-124}$ & $25.86^{+0.04}_{-0.03}$ \\
MUSE-135049239 & JADES-GS+53.12487-27.78413 & $8.68^{+0.12}_{-0.14}$ & $8.31^{+0.18}_{-0.19}$ & $0.038^{+0.061}_{-0.034}$ & $364^{+64}_{-56}$ & $25.33^{+0.06}_{-0.03}$ \\
MUSE-139047301 & JADES-GS+53.14289-27.75896 & $7.55^{+0.09}_{-0.09}$ & $7.43^{+0.14}_{-0.12}$ & $0.003^{+0.007}_{-0.002}$ & $963^{+143}_{-144}$ & $25.48^{+0.08}_{-0.06}$ \\
MUSE-68 & JADES-GS+53.17123-27.77852 & $8.00^{+0.07}_{-0.08}$ & $7.17^{+0.11}_{-0.08}$ & $0.005^{+0.013}_{-0.004}$ & $1297^{+133}_{-133}$ & $25.52^{+0.06}_{-0.03}$ \\
MUSE-313 & JADES-GS+53.17095-27.78251 & $7.35^{+0.09}_{-0.06}$ & $7.25^{+0.10}_{-0.08}$ & $0.002^{+0.006}_{-0.001}$ & $1260^{+120}_{-134}$ & $26.03^{+0.09}_{-0.15}$ \\
MUSE-417 & JADES-GS+53.15788-27.78007 & $7.39^{+0.09}_{-0.08}$ & $7.29^{+0.15}_{-0.11}$ & $0.005^{+0.013}_{-0.004}$ & $1131^{+140}_{-158}$ & $25.52^{+0.07}_{-0.06}$ \\
MUSE-547 & JADES-GS+53.16062-27.77161 & $7.65^{+0.13}_{-0.10}$ & $7.30^{+0.15}_{-0.15}$ & $0.007^{+0.019}_{-0.006}$ & $1110^{+152}_{-129}$ & $25.50^{+0.08}_{-0.06}$ \\
MUSE-1478 & JADES-GS+53.15321-27.76623 & $8.49^{+0.24}_{-0.13}$ & $7.18^{+0.40}_{-0.24}$ & $0.112^{+0.026}_{-0.055}$ & $967^{+171}_{-178}$ & $25.85^{+0.14}_{-0.29}$ \\
MUSE-1670 & JADES-GS+53.16674-27.80425 & $8.97^{+0.15}_{-0.15}$ & $7.48^{+0.19}_{-0.21}$ & $0.101^{+0.019}_{-0.023}$ & $840^{+103}_{-98}$ & $25.46^{+0.06}_{-0.05}$ \\
MUSE-2069 & JADES-GS+53.15105-27.78294 & $7.98^{+0.15}_{-0.11}$ & $7.44^{+0.22}_{-0.15}$ & $0.007^{+0.015}_{-0.005}$ & $915^{+196}_{-169}$ & $25.49^{+0.08}_{-0.08}$ \\
MUSE-2071 & JADES-GS+53.14615-27.77786 & $7.54^{+0.09}_{-0.08}$ & $7.06^{+0.10}_{-0.07}$ & $0.008^{+0.021}_{-0.006}$ & $1434^{+131}_{-133}$ & $25.54^{+0.05}_{-0.05}$ \\
MUSE-2168 & JADES-GS+53.13600-27.79849 & $7.62^{+0.05}_{-0.03}$ & $6.55^{+0.05}_{-0.03}$ & $0.072^{+0.017}_{-0.019}$ & $3122^{+269}_{-237}$ & $25.77^{+0.05}_{-0.04}$ \\
MUSE-2296 & JADES-GS+53.16836-27.80420 & $7.56^{+0.08}_{-0.07}$ & $7.18^{+0.11}_{-0.09}$ & $0.004^{+0.010}_{-0.003}$ & $1306^{+142}_{-148}$ & $25.54^{+0.07}_{-0.05}$ \\
MUSE-2302 & JADES-GS+53.18044-27.77066 & $7.07^{+0.05}_{-0.04}$ & $6.30^{+0.14}_{-0.20}$ & $0.005^{+0.010}_{-0.003}$ & $4401^{+477}_{-544}$ & $25.80^{+0.06}_{-0.03}$ \\
MUSE-2307 & JADES-GS+53.18335-27.79602 & $7.12^{+0.08}_{-0.07}$ & $6.31^{+0.16}_{-0.19}$ & $0.006^{+0.009}_{-0.004}$ & $3764^{+412}_{-355}$ & $25.94^{+0.15}_{-0.17}$ \\
MUSE-2350 & JADES-GS+53.15638-27.80966 & $7.12^{+0.06}_{-0.06}$ & $6.50^{+0.43}_{-0.32}$ & $0.002^{+0.004}_{-0.001}$ & $2321^{+436}_{-976}$ & $25.60^{+0.03}_{-0.03}$ \\
MUSE-2449 & JADES-GS+53.16470-27.76965 & $7.34^{+0.17}_{-0.09}$ & $6.96^{+0.23}_{-0.11}$ & $0.012^{+0.029}_{-0.010}$ & $1125^{+184}_{-280}$ & $25.59^{+0.06}_{-0.05}$ \\
MUSE-2481 & JADES-GS+53.14988-27.81073 & $8.39^{+0.15}_{-0.19}$ & $8.06^{+0.24}_{-0.31}$ & $0.044^{+0.023}_{-0.030}$ & $539^{+141}_{-106}$ & $25.40^{+0.19}_{-0.08}$ \\
MUSE-2502 & JADES-GS+53.15807-27.81801 & $7.74^{+0.12}_{-0.11}$ & $6.97^{+0.15}_{-0.19}$ & $0.007^{+0.021}_{-0.006}$ & $1409^{+224}_{-138}$ & $25.57^{+0.06}_{-0.05}$ \\
MUSE-2873 & JADES-GS+53.17252-27.76436 & $7.38^{+0.05}_{-0.04}$ & $6.73^{+0.09}_{-0.06}$ & $0.057^{+0.027}_{-0.024}$ & $1677^{+218}_{-200}$ & $26.06^{+0.11}_{-0.15}$ \\
MUSE-2964 & JADES-GS+53.14946-27.80980 & $7.15^{+0.14}_{-0.10}$ & $6.96^{+0.16}_{-0.11}$ & $0.020^{+0.037}_{-0.017}$ & $1457^{+155}_{-180}$ & $25.58^{+0.08}_{-0.07}$ \\
MUSE-3090 & JADES-GS+53.15491-27.76255 & $7.59^{+0.13}_{-0.08}$ & $7.39^{+0.19}_{-0.13}$ & $0.007^{+0.017}_{-0.005}$ & $963^{+135}_{-136}$ & $25.51^{+0.09}_{-0.07}$ \\
MUSE-3093 & JADES-GS+53.13969-27.79649 & $7.41^{+0.15}_{-0.17}$ & $7.16^{+0.17}_{-0.17}$ & $0.088^{+0.026}_{-0.034}$ & $1149^{+193}_{-168}$ & $25.53^{+0.09}_{-0.07}$ \\
MUSE-3203 & JADES-GS+53.17655-27.77112 & $7.58^{+0.09}_{-0.08}$ & $7.09^{+0.12}_{-0.10}$ & $0.002^{+0.005}_{-0.001}$ & $1363^{+124}_{-123}$ & $25.55^{+0.08}_{-0.04}$ \\
MUSE-3238 & JADES-GS+53.17560-27.79589 & $7.66^{+0.03}_{-0.02}$ & $6.55^{+0.02}_{-0.01}$ & $0.233^{+0.019}_{-0.018}$ & $1197^{+104}_{-108}$ & $25.71^{+0.03}_{-0.01}$ \\
MUSE-4405 & JADES-GS+53.17051-27.81249 & $7.42^{+0.10}_{-0.10}$ & $6.87^{+0.14}_{-0.16}$ & $0.006^{+0.019}_{-0.005}$ & $1574^{+242}_{-144}$ & $25.54^{+0.06}_{-0.05}$ \\
MUSE-6231 & JADES-GS+53.17834-27.80097 & $7.67^{+0.07}_{-0.07}$ & $7.19^{+0.09}_{-0.09}$ & $0.004^{+0.007}_{-0.002}$ & $1229^{+100}_{-123}$ & $25.83^{+0.06}_{-0.06}$ \\
MUSE-6294 & JADES-GS+53.16611-27.78574 & $8.10^{+0.07}_{-0.07}$ & $6.82^{+0.11}_{-0.09}$ & $0.136^{+0.020}_{-0.018}$ & $1620^{+201}_{-177}$ & $25.73^{+0.10}_{-0.08}$ \\
MUSE-6462 & JADES-GS+53.16407-27.79972 & $8.69^{+0.08}_{-0.10}$ & $7.31^{+0.12}_{-0.12}$ & $0.113^{+0.019}_{-0.022}$ & $1054^{+123}_{-117}$ & $25.59^{+0.09}_{-0.07}$ \\
MUSE-7125 & JADES-GS+53.12813-27.78987 & $8.72^{+0.16}_{-0.20}$ & $8.29^{+0.25}_{-0.32}$ & $0.004^{+0.010}_{-0.003}$ & $249^{+83}_{-55}$ & $25.42^{+0.08}_{-0.08}$ \\
MUSE-7205 & JADES-GS+53.16298-27.76031 & $7.19^{+0.09}_{-0.10}$ & $7.03^{+0.13}_{-0.10}$ & $0.005^{+0.013}_{-0.004}$ & $1452^{+165}_{-190}$ & $25.64^{+0.06}_{-0.05}$ \\
MUSE-7225 & JADES-GS+53.16167-27.76318 & $7.71^{+0.06}_{-0.06}$ & $7.35^{+0.09}_{-0.07}$ & $0.002^{+0.003}_{-0.001}$ & $1211^{+89}_{-96}$ & $25.48^{+0.04}_{-0.03}$ \\
MUSE-7337 & JADES-GS+53.16904-27.78769 & $7.39^{+0.15}_{-0.16}$ & $7.09^{+0.18}_{-0.19}$ & $0.069^{+0.025}_{-0.035}$ & $1236^{+132}_{-140}$ & $25.55^{+0.10}_{-0.07}$ \\
MUSE-7605 & JADES-GS+53.16577-27.78490 & $7.13^{+0.09}_{-0.07}$ & $6.83^{+0.11}_{-0.09}$ & $0.005^{+0.013}_{-0.003}$ & $1351^{+162}_{-133}$ & $25.62^{+0.07}_{-0.06}$ \\
MUSE-7922 & JADES-GS+53.18071-27.77656 & $8.58^{+0.25}_{-0.27}$ & $8.02^{+0.37}_{-0.35}$ & $0.111^{+0.041}_{-0.039}$ & $524^{+186}_{-163}$ & $25.38^{+0.11}_{-0.08}$ \\
MUSE-7934 & JADES-GS+53.14667-27.78621 & $8.72^{+0.12}_{-0.11}$ & $7.91^{+0.21}_{-0.25}$ & $0.006^{+0.023}_{-0.004}$ & $410^{+84}_{-75}$ & $25.52^{+0.11}_{-0.08}$ \\
MUSE-7984 & JADES-GS+53.17929-27.77331 & $7.77^{+0.09}_{-0.11}$ & $7.17^{+0.12}_{-0.14}$ & $0.061^{+0.019}_{-0.028}$ & $1125^{+121}_{-104}$ & $25.55^{+0.06}_{-0.05}$ \\
MUSE-8124 & JADES-GS+53.15677-27.80921 & $8.51^{+0.09}_{-0.10}$ & $7.79^{+0.15}_{-0.14}$ & $0.063^{+0.017}_{-0.020}$ & $429^{+74}_{-58}$ & $25.53^{+0.06}_{-0.05}$ \\
DEIMOS-11505 & JADES-GN+189.26510+62.19963 & $7.88^{+0.10}_{-0.15}$ & $6.25^{+0.16}_{-0.15}$ & $0.024^{+0.020}_{-0.016}$ & $3662^{+694}_{-442}$ & $26.21^{+0.08}_{-0.09}$ \\
DEIMOS-43\_13063 & JADES-GN+189.15632+62.21000 & $8.76^{+0.17}_{-0.13}$ & $7.51^{+0.24}_{-0.18}$ & $0.155^{+0.028}_{-0.027}$ & $853^{+124}_{-129}$ & $25.54^{+0.11}_{-0.11}$ \\
DEIMOS-33\_19970 & JADES-GN+189.23078+62.26355 & $7.88^{+0.08}_{-0.05}$ & $7.20^{+0.08}_{-0.07}$ & $0.004^{+0.008}_{-0.003}$ & $1283^{+108}_{-140}$ & $25.54^{+0.05}_{-0.03}$ \\
DEIMOS-33\_20014 & JADES-GN+189.23125+62.25912 & $8.05^{+0.35}_{-0.07}$ & $6.53^{+0.04}_{-0.17}$ & $0.030^{+0.015}_{-0.015}$ & $3261^{+280}_{-273}$ & $25.96^{+0.32}_{-0.16}$ \\
DEIMOS-33\_17440 & JADES-GN+189.20512+62.26072 & $8.65^{+0.08}_{-0.07}$ & $7.34^{+0.12}_{-0.09}$ & $0.006^{+0.013}_{-0.004}$ & $1052^{+137}_{-125}$ & $25.53^{+0.07}_{-0.07}$ \\
DEIMOS-vdrop\_225 & JADES-GN+189.18385+62.17992 & $7.36^{+0.09}_{-0.07}$ & $6.43^{+0.07}_{-0.27}$ & $0.002^{+0.004}_{-0.001}$ & $4042^{+556}_{-520}$ & $25.79^{+0.12}_{-0.18}$ \\
DEIMOS-33\_17034 & JADES-GN+189.20058+62.25945 & $8.13^{+0.06}_{-0.07}$ & $7.25^{+0.09}_{-0.08}$ & $0.002^{+0.004}_{-0.001}$ & $1318^{+91}_{-92}$ & $25.97^{+0.08}_{-0.08}$ \\
DEIMOS-33\_17705 & JADES-GN+189.20819+62.23212 & $7.99^{+0.10}_{-0.09}$ & $7.13^{+0.12}_{-0.13}$ & $0.021^{+0.027}_{-0.018}$ & $1319^{+118}_{-124}$ & $25.73^{+0.06}_{-0.16}$ \\
DEIMOS-32\_16773 & JADES-GN+189.19779+62.19996 & $7.71^{+0.07}_{-0.06}$ & $7.19^{+0.08}_{-0.09}$ & $0.005^{+0.012}_{-0.004}$ & $1332^{+102}_{-100}$ & $25.96^{+0.09}_{-0.09}$ \\
\hline
\end{tabular}
\caption{Physical properties (median values and marginalized $68$~per~cent credible intervals) derived from \textsc{beagle} models for the $61$ Ly$\alpha$ emitters with H$\alpha$ detections at $4.9<z<6.5$ with JADES NIRCam SEDs.}
\label{tab:sed}
\end{table*}


\bsp	
\label{lastpage}
\end{document}